\documentclass{aa}
\usepackage[varg]{txfonts}

\usepackage{graphicx}
\usepackage{natbib}
\bibpunct{(}{)}{;}{a}{}{,} 
\usepackage{color}
\usepackage{float}
\usepackage{subfigure}
\usepackage{hyperref}
\usepackage{longtable}
\usepackage{rotating}
\usepackage{lscape}
\usepackage{soul}

\newcommand{\lya}{\mbox{Ly$\alpha$}}

\newcommand{\flcgs}{\mbox{erg s$^{-1}$ cm$^{-2}$}}
\newcommand{\sbl}{\mbox{erg s$^{-1}$ cm$^{-2}$ arcsec$^{-2}$}}

\newcommand{\udft}{\textsf{udf-10}}
\newcommand{\mosaic}{\textsf{mosaic}}

\begin{document}

\title{The MUSE Hubble Ultra Deep Field Survey}
\subtitle{VIII: Extended Lyman-$\rm \alpha$ haloes around high-$z$ star-forming galaxies\thanks{MUSE Ultra Deep Field $\lya$ haloes catalog (Table B.1) is only available in electronic form at the CDS via anonymous ftp to \url{cdsarc.u-strasbg.fr} (130.79.128.5) or via \url{http://cdsweb.u-strasbg.fr/cgi-bin/qcat?J/A+A/}}}

\author{Floriane Leclercq\inst{\ref{inst1}\thanks{e-mail: floriane.leclercq@univ-lyon1.fr}}
\and Roland Bacon\inst{\ref{inst1}}
\and Lutz Wisotzki\inst{\ref{inst2}}
\and Peter Mitchell\inst{\ref{inst1}}
\and Thibault Garel\inst{\ref{inst1}}
\and Anne Verhamme\inst{\ref{inst1},\ref{inst4}}
\and J\'er\'emy Blaizot\inst{\ref{inst1}}
\and Takuya Hashimoto\inst{\ref{inst1}}
\and Edmund Christian Herenz\inst{\ref{inst7}}
\and Simon Conseil\inst{\ref{inst1}}
\and Sebastiano Cantalupo\inst{\ref{inst5}}
\and Hanae Inami\inst{\ref{inst1}}
\and Thierry Contini\inst{\ref{inst6}}
\and Johan Richard\inst{\ref{inst1}}
\and Michael Maseda\inst{\ref{inst8}}
\and Joop Schaye\inst{\ref{inst8}}
\and Raffaella Anna Marino\inst{\ref{inst5}}
\and Mohammad Akhlaghi\inst{\ref{inst1}}
\and Jarle Brinchmann\inst{\ref{inst8},\ref{inst9}}
\and Marcella Carollo\inst{\ref{inst5}}
}

\institute{Univ Lyon, Univ Lyon1, Ens de Lyon, CNRS, Centre de Recherche Astrophysique de Lyon UMR5574, F-69230, Saint-Genis-Laval, France\label{inst1}
\and Leibniz-Institut fur Astrophysik Potsdam (AIP), An der Sternwarte 16, 14482 Potsdam, Germany\label{inst2}
\and Observatoire de Geneve, Universite de Geneve, 51 Ch. des Maillettes, 1290 Versoix, Switzerland \label{inst4}
\and Institute for Astronomy, ETH Zurich, Wolfgang-Pauli-Strasse 27, 8093 Zurich, Switzerland \label{inst5}
\and Institut de Recherche en Astrophysique et Planétologie (IRAP), Université de Toulouse, CNRS, UPS, F-31400 Toulouse, France \label{inst6}
\and Department of Astronomy, Stockholm University, AlbaNova University Centre, SE-106 91, Stockholm, Sweden \label{inst7}
\and Leiden Observatory, Leiden University, P.O. Box 9513, 2300 RA Leiden, The Netherlands \label{inst8}
\and Instituto de Astrof{\'\i}sica e Ci{\^e}ncias do Espaço, Universidade do Porto, CAUP, Rua das Estrelas, PT4150-762 Porto, Portugal \label{inst9}
}

\date{Accepted September 25, 2017}


\abstract
{
We report the detection of extended $\lya$ haloes around 145 individual star-forming galaxies at redshifts $3\le z\le 6$ in the Hubble Ultra Deep Field observed with the Multi-Unit Spectroscopic Explorer (MUSE) at ESO-VLT. 
Our sample consists of continuum-faint ($-15\geq M_{\rm UV}\geq-22$) $\lya$ emitters (LAEs).
Using a 2D, two-component (continuum-like and halo) decomposition of $\lya$ emission assuming circular exponential distributions, we measure scale lengths and luminosities of $\lya$ haloes. We find that 80 \% of our objects having reliable $\lya$ halo measurements show $\lya$ emission that is significantly more extended than the UV continuum detected by HST (by a factor $\approx$4 to
>20). 
The median exponential scale length of the $\lya$ haloes in our sample is $\approx$4.5 kpc with a few haloes exceeding 10 kpc. By comparing the maximal detected extent of the $\lya$ emission with the predicted dark matter halo virial radii of simulated galaxies, we show that the detected $\lya$ emission of our selected sample of $\lya$ emitters probes a significant portion of the cold circum-galactic medium of these galaxies (>50\% in average).
This result therefore shows that there must be significant HI reservoirs in the circum-galactic medium and reinforces the idea that $\lya$ haloes are ubiquitous around high-redshift $\lya$ emitting galaxies. 
Our characterization of the $\lya$ haloes indicates that the majority of the $\lya$ flux comes from the halo ($\approx$65\%) and that their scale lengths seem to be linked to the UV properties of the galaxies (sizes and magnitudes).
We do not observe a significant $\lya$ halo size evolution with redshift, although our sample for $z>5$ is very small. 
We also explore the diversity of the $\lya$ line profiles in our sample and we find that the $\lya$ lines cover a large range of full width at half maximum (FWHM) from 118 to 512 km s$^{-1}$. While the FWHM does not seem to be correlated to the $\lya$ scale length, most compact $\lya$ haloes and those that are not detected with high significance tend to have narrower $\lya$ profiles (<350 km s$^{-1}$).
Finally, we investigate the origin of the extended $\lya$ emission but we conclude that our data do not allow us to disentangle the possible processes, i.e. scattering from star-forming regions, fluorescence, cooling radiation from cold gas accretion, and emission from satellite galaxies.
}

\maketitle

\section{Introduction}

Observing the circum-galactic medium (CGM) represents an important challenge for
understanding how galaxies form and evolve. 
Galaxy evolution is driven primarily by the flows of gas that surround galaxies. 
Moreover, the CGM contains a large amount of the baryonic matter in galaxies and as such observations of this gas provide crucial information.
A powerful tracer of this gas is Lyman alpha ($\lya$) emission, which allows circum-galactic gas to be observed around high-redshift galaxies as a $\lya$ halo. 
A number of physical mechanisms can contribute to spatially extended $\lya$ emission, including fluorescence, cooling radiation, or the scattering of $\lya$ photons produced in star-forming HII regions \citep{G96,K96,H00,H01,C05,D06,K10,Bar10,L15}.

Such extended $\lya$ emission has been detected around nearby galaxies (e.g. \citealt{K03,H05,H15}). By selecting 14 nearby galaxies that cover the same range of far-UV luminosities as high-$z$ galaxies, the LARS ($\lya$ Reference Sample; \citealt{O14}) collaboration constructed a sample that is comparable to high-$z$ samples. 
Most of their galaxies show $\lya$ emission that is more extended than both the stellar UV continuum and the H$\alpha$ emission showing the rich gas content of the CGM \citep{H13,H14,He16}.

At high redshift, the mapping of the extended $\lya$ haloes around galaxies (non-AGN) is however a lot more difficult because of sensitivity and resolution limitations.
Detections of extended Lyman alpha emission at high redshift have been obtained in the past. 
While some large $\lya$ blobs have been observed (e.g. \citealt{S00,M04,M11}), most of these studies were forced to employ stacking analyses because of sensitivity limitations.
The first tentative detections of $\lya$ haloes around normal star-forming galaxies emitting $\lya$ emission using narrowband (NB) imaging methods were reported by \cite{MW98} and \cite{F01}. 
Later, \cite{H04} observed 22 Lyman break galaxies (LBG) and detected extended $\lya$ emission by stacking the NB images. 
These authors were followed six years later by \cite{O10} who detected $\lya$ haloes in their composite NB images of 401 $\lya$ emitters (LAEs) at $z$=5.7 and 207 at $z$=6.6. \cite{Ma12} and \cite{Mo14} significantly increased the size of LAEs samples used by stacking $\approx$2000 and $\approx$4500 LAEs at redshift $z\simeq$3 and 2.2$\leq z \leq$6.6, respectively. 
\cite{Mo14} found typical $\lya$ halo exponential scale lengths of 5-10 physical kpc. \cite{Ma12} found that $\lya$ halo sizes are dependent on environmental density; these halo sizes extend from 9 to up to 30 physical kpc towards overdense regions. 
More recently, \cite{X17} studied $\approx$1500 galaxies in two overdense regions at $z\approx$3 and 4. Using stacking methods these authors reported $\lya$ halo exponential scale lengths of 5-6 physical kpc and found that $\lya$ halo sizes correlate with the UV continuum and $\lya$ luminosities, but not with overdensity.
\cite{S11} stacked 92 brighter (R$_{AB}$ $\simeq$ 24.5) and more massive LBGs at $z$=2.3$-$3, finding large $\lya$ extents of $\approx$80 physical kpc beyond the mean UV continuum size at a surface brightness level of $\sim$10$^{-19}$ erg s$^{-1}$ cm$^{-2}$ arcsec$^{-2}$. Put together, all these studies showed that Lyman alpha emission is on average more spatially extended than the UV stellar continuum emission from galaxy counterparts. 

Meanwhile, other studies have found conflicting results. \cite{F13} argued that the observed extended emission is artificially created by an underestimation of the stacking procedure systematics. After carrying out an error budget analysis, they did not find evidence for significant extended $\lya$ emission. \cite{B10} also reported compact $\lya$ emission in their stack of eight star-forming galaxies at $z$=3. 

Over a similar period and using a different approach, \cite{R08} performed an ultra-deep (92h) long-slit observation and identified 27 faint LAEs (few $\times$10$^{-18}$ $\flcgs$) at redshift 2.67< $z$ < 3.75. 
This observation enabled the individual detections of extended $\lya$ emission along the slit for most of their objects although with large uncertainties owing to slit losses and the high errors on the continuum size measurements. Some other detections of extended $\lya$ emission around high-redshift star-forming galaxies were obtained using the magnification power of gravitational lensing (e.g. \citealt{S07}, \citealt{P16}). 

Recently, a significant step forward has been taken thanks to the substantial increase in sensitivity provided by the Multi-Unit Spectroscopic Explorer (MUSE) at the ESO-VLT \citep{Ba10}. 
\citeauthor{W16} (2016; hereafter W16) reported the detection of 21 $\lya$ haloes around relatively continuum-faint (m$_{AB}$ $\gtrsim$ 27) star-forming galaxies at redshift 3<$z$<6 within the Hubble Deep Field South (HDFS) observed with MUSE. Their data reach an unprecedented limiting surface brightness (SB) of $\sim$10$^{-19}$ $\sbl$ (1$\sigma$) enabling the study of the CGM on a galaxy-by-galaxy basis. 
The $\lya$ haloes from the W16 study have exponential scale lengths ranging from $\approx$1 kpc to $\approx$7 kpc and appear to be on average 10 times larger than their corresponding UV galaxy sizes. These new observational data also enable the direct comparison of the $\lya$ halo properties with the stellar properties of the host galaxies and the investigation of the origin of the $\lya$ haloes. This pioneering study was however limited to a small sample and therefore the results need to be confirmed with better statistics.

Here, we extend the W16 LAE sample by one order of magnitude using the \textit{Hubble} Ultra Deep Field (UDF) data obtained with MUSE \citep{B15}. The significant effort on the data reduction of this data set improves the limiting SB sensitivity by one order of magnitude over previous narrowband studies.
First, we follow a similar approach as W16 to quantitatively characterize the spatial extent of the $\lya$ emission around high-redshift galaxies in the UDF ($-15\geq M_{\rm UV}\geq-22$). We then analyse the sizes and $\lya$ luminosities of our $\lya$ haloes as a function of the UV properties of their HST counterparts and compare our results to W16. 
In addition to its spatial distribution, the $\lya$ line profile encodes crucial information that can help shed light on the origin of the $\lya$ emission and constrain the gas opacity and kinematics \citep{H00,D06,V06,K10,G16}.
Taking advantage of the spectral information of MUSE data cubes, we also investigate how $\lya$ emission relates to various line properties, such as the line width and equivalent width.

The paper is organized as follows: we describe our data and our sample construction in section~\ref{sec2}. Section~\ref{sec3} presents our procedure for the extraction of the images and construction of radial SB profiles needed for the detection of extended $\lya$ emission. Section~\ref{sec4} explains the $\lya$ spatial distribution modelled that we use to determine the characteristics of the $\lya$ haloes that are presented in section~\ref{sec5}. Section~\ref{sec5} also includes the analysis of the $\lya$ line profile. In section~\ref{sec:hostgal} we investigate the relation between the $\lya$ haloes and their host galaxies. Finally, we discuss our results in section~\ref{sec7} and present our summary and conclusions in section~\ref{sec8}. Appendix~\ref{app1} gives a comparison of the $\lya$ haloes detected around galaxies, which are both in the deep \udft\ data cube and in the shallower \mosaic\ data cube.  

For this paper, we use AB magnitudes, physical distances, and assume a $\Lambda$CDM  cosmology  with $\Omega_{m}$=0.3, $\Omega_\Lambda$=0.7, and $H_{0}$=70 km s$^{-1}$ Mpc$^{-1}$.


\section{Data and sample definition}
\label{sec2}
\subsection{Observations and data reduction}

The UDF data were taken using the MUSE instrument between September 2014 and February 2016 under the MUSE consortium Guarantee Time Observations. 
A number of $1\arcmin \times 1\arcmin$ pointings (corresponding to the MUSE field of view) were completed at two levels of depth. The medium-deep data consist of a \mosaic\ of 9 deep, 10 hour pointings denoted \textsf{udf-0[1-9]}. 
The ultra-deep data, denoted \udft,\, consist of a single $\approx$20 hour pointing that overlaps with the \mosaic\ reaching a total of 30 hours depth.
During the observations the sky was clear with good seeing conditions (full width at half maximum (FWHM) of $0\farcs6$ at 7750$\AA$). More details about the data acquisition can be found in \citeauthor{B17} (2017; hereafter B17). 

The data reduction of both the \udft\ and \mosaic\ data cubes is described in B17. 
The two resulting data cubes contain $323\times322$ and $945\times947$ spatial pixels for the \udft\ and \mosaic\ field, respectively.
The number of spectra match the number of spatial pixels in each data cube with a wavelength range of 4750$\AA$ to 9350$\AA$ (3681 spectral pixels) with medium spectral resolution R$\sim$3000. 
The spatial sampling is $0\farcs2\times0\farcs2$ per spaxel and the spectral sampling is 1.25$\AA$ per pixel. The data cubes also contain the estimated variance for each pixel.
The data reach a limiting SB sensitivity (1$\sigma$) of 2.8 and 5.5$\times$10$^{-20}$ erg s$^{-1}$ cm$^{-2}$ $\AA^{-1}$ arcsec$^{-2}$ for an aperture of $1''\times1''$ in the 7000-8500$\AA$ range for the \udft\ and \mosaic\ data cubes, respectively (see B17 for more details).

Based on these reduced data cubes we constructed two catalogues corresponding to each data cube. 
The source detection and extraction were performed using HST priors, imposing a magnitude cut at 27 in the F775W band for the \mosaic\ field only, and the {\tt ORIGIN} (\citeauthor{Ma17}, in prep.) detection software. 
A complete description of the strategy used for the catalogue construction can be found in \citeauthor{I17} (2017; hereafter I17). 
{\tt The ORIGIN} software (see B17 for technical details) is designed to detect emission lines in 3D data sets. 
This software enables the discovery of a large number of LAEs that are barely seen or even undetectable in the HST images. 
Photometric magnitudes for these new objects were calculated following the method described in B17.

\subsection{Lyman alpha emitters sample}
\label{sec:sample}

\begin{figure}[t]
   \resizebox{\hsize}{!}{\includegraphics{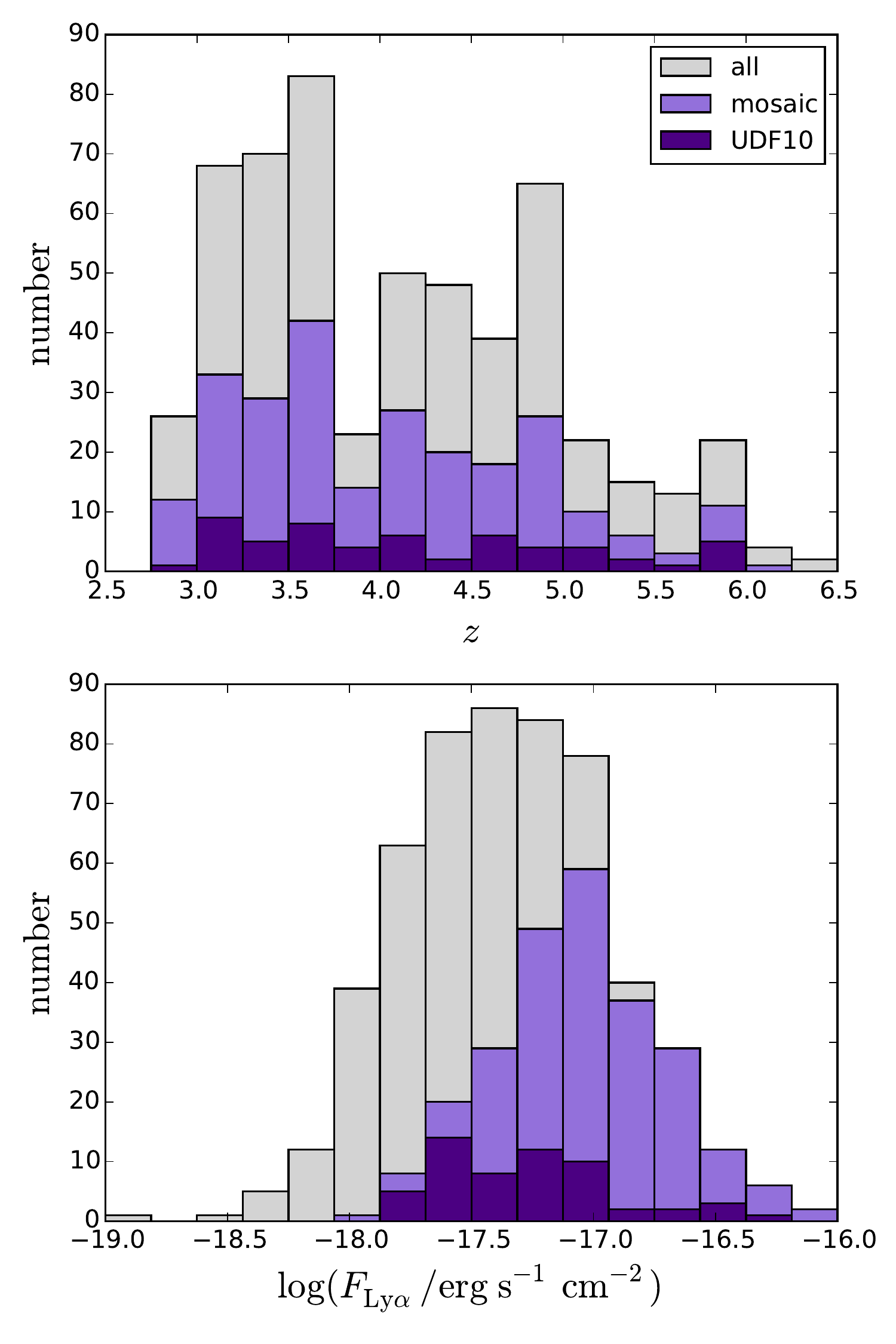}}
     \caption{Redshift distribution (upper panel) and total $\lya$ flux (measured using a CoG method, see section~\ref{sec:fluxmes} -- lower panel) histograms of our \udft\ (dark purple) and \mosaic\ (light purple) samples. The grey histograms show the distributions of the total sample (\udft\ and \mosaic) without applying the S/N cut.}
     \label{figz}
\end{figure}

Our parent sample was constructed from UDF catalogues (see I17) and according to the following criteria:
\begin{enumerate}
\item
We selected the LAEs ("TYPE=6" in the catalogues) with a reliable redshift ("CONFID=2 and 3"). 
This yields a sample of 155 and 620 objects for the \udft\ and \mosaic,\ respectively.
\item
Our primary objective being the study of individual galaxies, we removed galaxies in pairs closer than 50 kpc in projected transverse separation and with velocity differences of less than 1000 km s$^{-1}$  , which was estimated using the peak of the $\lya$ line or the red peak if the line was double peaked. 
We found 28 and 64 such objects in the \udft\ and \mosaic,\ respectively.
The study of the $\lya$ haloes of such LAE pairs will be part of another study. The analysis of merger rates from the MUSE UDF data is detailed in \cite{V17}.
\item
We also excluded 20 and 25 objects that are closer than 3\arcsec\ and 4\arcsec\ to the edges of the \udft\ and \mosaic\ data cubes, respectively. 
This is necessary to ensure we can analyse extended $\lya$ emission over a large spatial window for our entire sample. 
Objects from the \udft\ data cube are allowed to be closer to the edges because of the higher quality of the data given that the \udft\ data cube is combined with the wider \mosaic\ data cube.
\item
Among the remaining objects, we manually removed 7 and 29 objects in the \udft\ and \mosaic\ fields, respectively, which are contaminated by emission lines from foreground sources, skyline residuals, or by continuum remnants visible in the NB image (see section~\ref{sec:nbimage} for the continuum subtraction method). 
\item
Finally, following the procedure described in section~\ref{sec:nbimage}, we created NB images around the $\lya$ emission line and imposed a minimal signal-to-noise ratio (S/N) of 6 in a fixed and large aperture set using a curve of growth (CoG) method (see section~\ref{sec:fluxmes}).
The S/N is defined as the $\lya$ flux divided by the standard deviation from the data cube. 
This cut is motivated by our detection limit estimation described in section~\ref{sec:snrlim}. 
It eliminates 43 LAEs in the \udft\ field and 282 in the \mosaic\ field. 
The S/N cut introduces a selection bias towards brighter haloes. This bias is noticeable in the lower panel of Figure~\ref{figz}, where the total $\lya$ flux distribution before and after the S/N cut is shown.
\end{enumerate}
In total 26 galaxies are in both \udft\ and \mosaic\ fields. 
For these objects, we only show results from the \udft\ data cube because of the higher S/N; a comparison of the results from the two data cubes is given in Appendix~\ref{sec:limdetec_both}.
Our final sample consists of 252 galaxies: 57 in the \udft\ field and 195 that are uniquely in the \mosaic\ field.
The sample spans a redshift range from 2.93 to 6.04 and a total $\lya$ flux ranging from $\approx1.6\times10^{-18}$ $\flcgs$ to $\approx1.1\times10^{-16}$ $\flcgs$. 
$\lya$ fluxes are measured using a CoG method (see section~\ref{sec:fluxmes}).
The redshift and flux distributions of our sample (purple) and the sample without the S/N cut (grey) are shown in Figure~\ref{figz}. The flux distribution shows the selection bias towards the brighter LAEs.

\section{Detection of diffuse Ly$\alpha$ emission}
\label{sec3}

Our detection of extended $\lya$ emission employs a circularly symmetric analysis, following a similar approach to W16. 
The method uses the radial SB profiles of both the $\lya$ and UV continuum emission. 
In this section we first describe the methods used to create the $\lya$ NB and UV continuum images, then we explain how the radial surface brightness profiles are constructed, and finally we present some LAE radial SB profiles as examples. 

\subsection{Image construction}
\subsubsection{Ly$\alpha$ narrowband images}
\label{sec:nbimage}

We constructed a $10\arcsec\times10\arcsec$ Ly$\alpha$ NB image for each object from the MUSE data cube. We used a wide, fixed spatial aperture to ensure that we included all of the detectable $\lya$ emission around our galaxies.  
In order to remove the continuum, we first performed a spectral median filtering on the MUSE data cube in a wide spectral window of 200 spectral pixels, in effect removing any emission lines (see \citealt{HW17} for the validation of this continuum subtraction method on MUSE data cubes). A continuum-free data cube was then computed by subtracting the filtered data cube from the original. 
In some cases the continuum of very bright objects was not well subtracted. As specified in section~\ref{sec:sample}, the 21 affected objects were removed from the sample.

The spectral bandwidths of the Ly$\alpha$ NB images were defined to maximize the S/N in a fixed spatial aperture (radius of 2\arcsec). Following this procedure, we obtained NB images with spectral bandwidths ranging from 2.5 to 20$\AA$ (i.e. 2 and 16 MUSE pixels, respectively). The largest spectral bandwidths correspond to double-peaked lines (some examples can be seen in Figures~\ref{fig2a} and \ref{fig2b}). The average spectral width is 6.25$\AA$.

\subsubsection{Ultraviolet continuum images}
\label{sec:contim}

We constructed UV continuum images for our sample using one of three different HST images of the UDF \citep{I13}, depending on the redshift of the object. The F814W ACS/WFC, F105W WFC3/IR, and F125W WFC3/IR HST images are used for objects at $z$<4, 4$\leq$ $z$<5, and $z$ $\geq$5, respectively.
We chose these filters because they are not contaminated by the $\lya$ emission or by intergalactic medium (IGM) absorption. These filters also probe UV continuum over similar rest-frame wavelength ranges, which are approximately 1400-2300$\AA$, 1500-2400$\AA,$ and 1570-2300$\AA$ for the F814W ACS/WFC, F105W WFC3/IR, and F125W WFC3/IR HST filters, respectively.

For each object in our sample, we constructed UV continuum images with the HST counterparts from the I17 catalogue.
After masking the pixels outside the segmentation map for each HST counterpart, we resampled the masked HST images to MUSE resolution and convolved them with the MUSE PSF. 
The HST PSF is not taken into account here because, first, the HST PSF (FWHM of 0.09\arcsec\ for the F814W band and 0.19\arcsec\ for the F105W and F125W bands -- \citealt{R15}) is much smaller than the MUSE PSF ($\approx$0.7\arcsec) and, second, the constructed UV continuum images are only used to compare visually Ly$\alpha$ and UV spatial extents. Our UV continuum modelled based on the HST data (see section~\ref{sec:model})considers the HST PSF.

The method used to estimate the wavelength-dependent PSF of the \udft\ and \mosaic\ data cubes is detailed in B17. It is best described as a two-dimensional Moffat distribution with a fixed beta parameter of 2.8 and a wavelength-dependent FWHM that we evaluate at the wavelength of each $\lya$ line.

Twenty-one of our objects are not in the \cite{R15} HST catalogue and instead   were discovered by {\tt ORIGIN}; however, these 21 objects are visible in the HST image. Magnitudes and segmentation maps for these galaxies were calculated using {\tt NoiseChisel} \citep{A15} and added to the MUSE UDF catalogues (see B17 and I17). Thirteen other $\lya$ emitters of our sample discovered by MUSE do not show any HST counterpart (e.g. object \#6498 in Figure~\ref{fig2a}). When comparing to the corresponding $\lya$ radial SB profiles, we treat these galaxies as point-like sources convolved with the MUSE PSF. 

We could have constructed continuum images directly from the MUSE data cubes. However, this is only possible for the brightest objects as most of our objects have poor continuum S/N in the MUSE data cubes. In addition, source blending is important at the MUSE spatial resolution while at HST resolution most of our sources are well separated.

\subsection{Surface brightness radial profiles}
\label{SBprof}

To visually compare the spatial extents of the UV and $\lya$ emission, we constructed radial SB profiles. We performed aperture photometry on the $\lya$ NB images and UV continuum images by averaging the flux in successive, concentric, one-pixel-wide annuli centred on the $\lya$ emission centroid. 
For the objects in our sample without HST counterparts, we compared $\lya$ radial SB profiles to the MUSE PSF radial SB profiles. 
The $\lya$ centroid was measured by fitting a simple 2D Gaussian to the $\lya$ NB image. In some cases, the centroid measured from the MUSE data is offset from the coordinates from the HST catalogue. The offsets are relatively small: less than 0.3$\arcsec$ for $\approx$95\% of our sample (median value $\lesssim$0.1$\arcsec$). We therefore ignored these offsets when constructing SB profiles and assumed the UV and $\lya$ emission to be concentric. 
Errors on $\lya$ radial SB profiles were measured in each annulus using the estimated variance from the MUSE data cubes.

\begin{figure*}[p]
\centering
   \includegraphics[height=0.88\textheight]{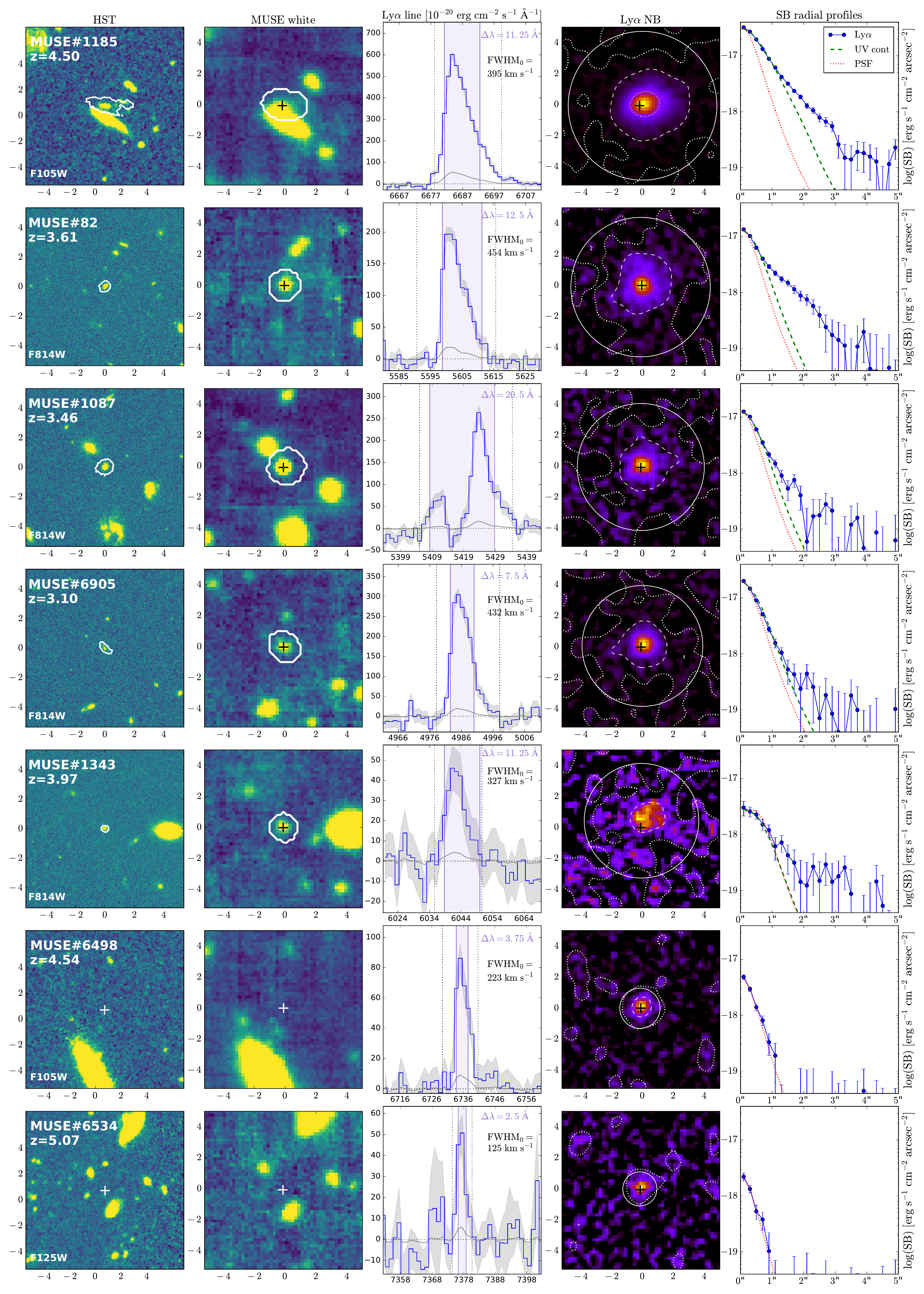}
     \caption{Representative sample of 7 LAEs from the MUSE UDF \mosaic\ field. Each row shows a different object. \textit{First column}: HST image (see section~\ref{sec:contim}) of the LAE indicated by the contour of its HST segmentation mask or by a white cross if it is not detected in the HST images (axis in arcsec). The MUSE ID, $z$ and the HST band are indicated. \textit{Second column}: MUSE white-light image summed over the full MUSE spectral range (axis in arcsec). The white contours correspond to the HST segmentation mask convolved with the MUSE PSF. The HST coordinates \citep{R15} are indicated by the cross. \textit{Third column}: $\lya$ line extracted in the HST segmentation mask convolved with the MUSE PSF. The purple area shows the NB image spectral width (indicated in purple). The two vertical black dotted lines indicate the bandwidth (in $\AA$) used to integrate the total $\lya$ flux (see section~\ref{sec:fluxmes}). The rest-frame FWHM of the single-peaked lines is also indicated. \textit{Fourth column}: $\lya$ narrowband image with SB contours at $10^{-17}~\sbl$ (central dotted white), $10^{-18}~\sbl$ (dashed white), and $10^{-19}~\sbl$ (outer dotted white). The radius of the solid white circle corresponds to the measured CoG radius $r_{\rm CoG}$ (see section~\ref{sec:fluxmes}). \textit{Last column}: radial SB profiles of $\lya$ emission (blue), UV continuum (green), and the PSF (red).}
     \label{fig2a}
\end{figure*}

\begin{figure*}[p]
\centering
   \includegraphics[width=17cm]{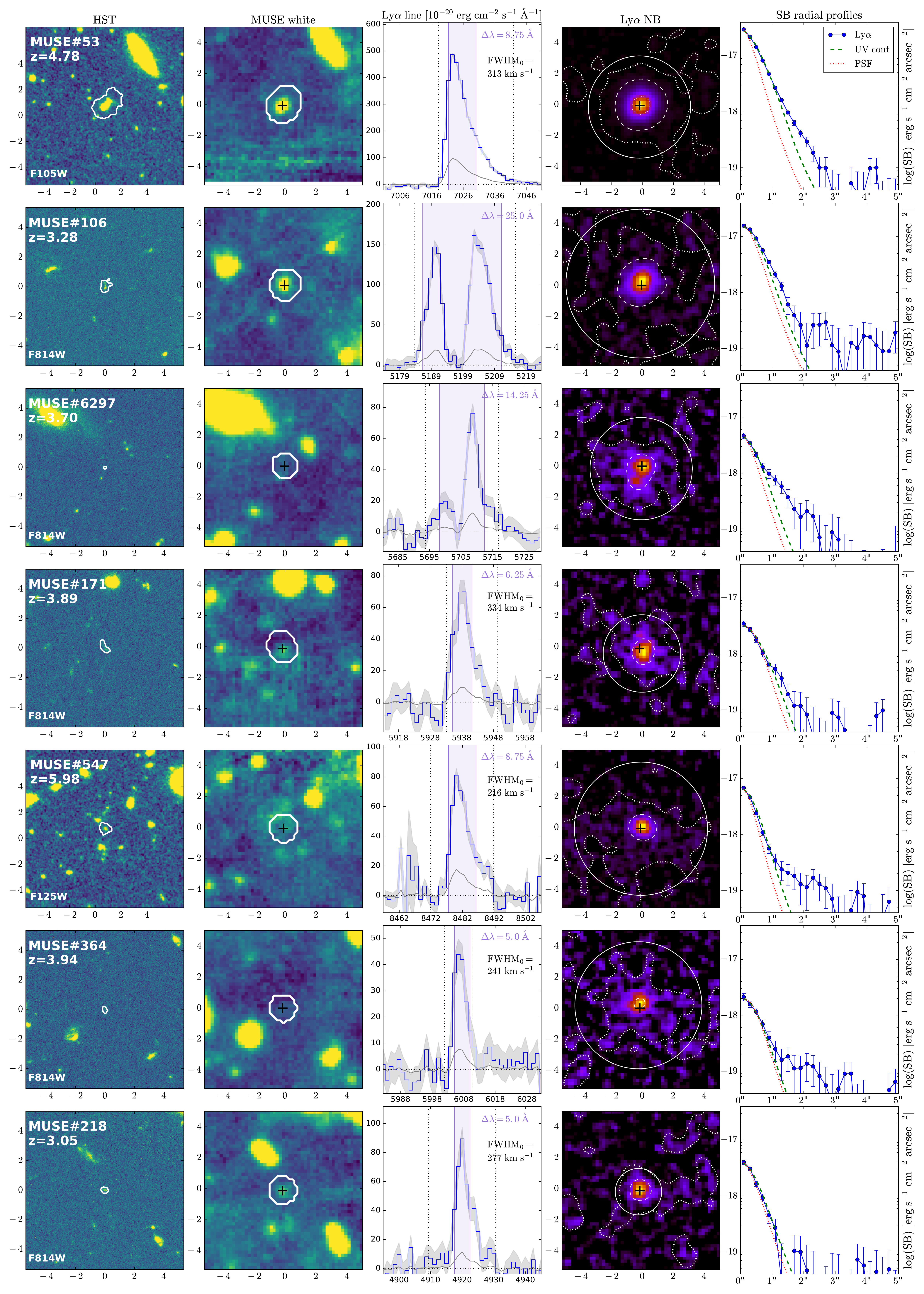}
     \caption{Same as Figure~\ref{fig2a} but for 7 representative objects in the MUSE UDF \udft\ field. Similar illustrations for all the objects in our sample are available at \url{http://muse-vlt.eu/science/udf/}.}
     \label{fig2b}
\end{figure*}

Figures~\ref{fig2a} and \ref{fig2b} show a representative subsample of 14 objects from the \mosaic\ and \udft\ fields. These objects were chosen to exhibit the diversity of the LAEs in terms of luminosity, line profile, and spatial extent. 
For each object we show the corresponding HST image we used for the study (see section~\ref{sec:contim}); the MUSE white light image, summed over the full MUSE data cube spectral range; the $\lya$ line, which is integrated in the HST counterparts mask convolved with the MUSE PSF (see the white contours on the white light image); the $\lya$ NB image (see section~\ref{sec:nbimage}); and the radial SB profiles. The UV continuum and PSF profiles have been re-scaled to the $\lya$ emission profile to aid the visual comparison.

Most of the objects show $\lya$ emission that is more spatially extended than the UV continuum. Some objects display a clear $\lya$ halo (e.g. objects \#1185, \#82, \#1087, \#53, and \#6297) but for other objects the extended $\lya$ emission is not as obvious (objects \#6498, \#6534, or \#218). 
Further analysis of the statistical significance of the detected $\lya$ haloes is presented in section \ref{sec:statsigni}.


\section{Ly$\alpha$ halo modelling}
\label{sec4}

\subsection{Two-dimensional two-component fits}
\label{sec:model}

Here, we describe how we characterize the spatial distribution of extended $\lya$ emission. 
Following W16, we fit $\lya$ emission with a two-dimensional, two-component exponential distribution using the Python/photutils package \citep{photu}.
Specifically, we decomposed the observed 2D $\lya$ distribution into central and extended exponential components using the HST morphological information as prior. 
The W16 work demonstrated that this decomposition is appropriate for characterizing $\lya$ haloes for a similar LAE sample. 
Adopting the same approach allows us to compare to their results directly.

The modelling is performed in two distinct steps:
\begin{enumerate}
\item First, the UV continuum is fit with a circular, 2D exponential distribution. 
We directly fit the HST image, chosen depending on the redshift of the object, (see section~\ref{sec:contim}) taking into account the corresponding PSF \citep[Table 1]{R15}. The HST counterpart of a given object was isolated by masking its surroundings using the HST segmentation mask. 
This first fit yields the continuum spatial scale length of each host galaxy. 
The \cite{R15} HST segmentation mask was created by combining the detection maps of the object in several HST bands. It is therefore supposed to delimit the galaxy in a rather large area and thus include most of the UV flux. If the object is located in a crowded region, the mask can be smaller to allow the separation of the sources. This is however very rare because of the high resolution of the HST images. Moreover, we find that the extent of the HST segmentation mask has a small impact on the resulting scale length. This is because the fit is mainly driven by the central emission of the galaxy. Consequently, even if there are potential faint UV counterparts surrounding the galaxy, the scale lengths are not drastically different.

\item Second, the $\lya$ NB image is fit by a sum of two circular, 2D exponential distributions, fixing the scale length of the first component to the continuum distribution value.
Hence, the first component corresponds to central, core emission and the second to emission from an extended halo. 
The fit takes into account the MUSE PSF by convolving the model with the PSF and the variance of each pixel in the image. 
\end{enumerate}

We thus have three parameters in total to fit the $\lya$ distribution: the halo scale length and fluxes of both the $\lya$ core and halo. 

\begin{figure*}
   \centerline{\includegraphics[scale=0.28]{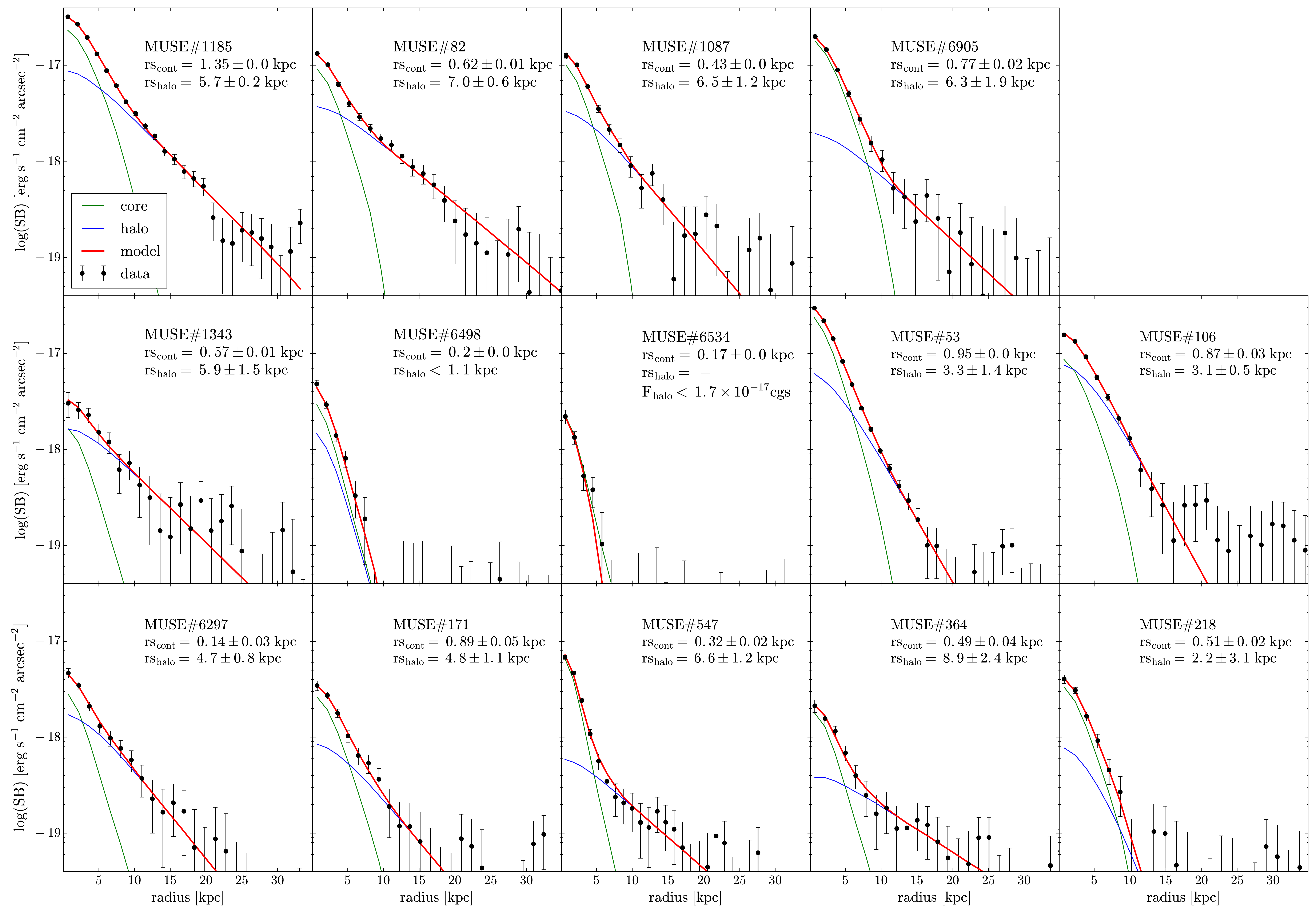}}
     \caption{Radial SB profiles of the modelled $\lya$ distribution decomposed into central (green lines) and extended (blue lines) exponential components.
     These are the same objects as in Figures~\ref{fig2a} and \ref{fig2b}.
     The total radial SB profiles of the modelled $\lya$ emission are shown in red. 
     For comparison, the observed radial SB profiles are overplotted as black points. 
     The fit is performed on the 2D $\lya$ NB image and not on the 1D radial SB profiles shown.
     The best-fit scale lengths are indicated in physical kpc. Upper limits and detection limits are also indicated (see section~\ref{sec:rhlim}). }

     \label{fig3}
\end{figure*}

Figure~\ref{fig3} shows the best-fit model radial SB profiles, decomposed into core emission (green line), extended halo emission (blue line), and with the total emission shown in red. 
We overplot the $\lya$ SB radial profiles (black dots with error bars).  For most of our objects, the modelled radial SB profiles are a good representation of the observed profiles. The 2D, two-component decomposition model therefore appears to be a good description of the $\lya$ distribution around LAEs. The W16 authors also found this decomposition to be consistent with their observed data.

\subsection{Error estimation}

We estimated errors on the best-fit halo scale length measurements.
First, we generated 100 realizations of each best-fit model $\lya$ image by combining the noise-free model image with realizations of the estimated noise.
The noise was assumed to follow a normal distribution with the variance at each pixel set equal to the variance of the corresponding pixel in the MUSE data cube.
Each realization of a given object was then fit and the final error on the halo scale length was given by the standard deviation across the recovered scale lengths.

To estimate the error on the core scale length (which is instead fit to the HST image) we followed a similar procedure using 100 empty regions of the HST image as artificial noise.

\subsection{Detection limit}
\label{sec:limdetec}
\subsubsection{Signal-to-noise limit}
\label{sec:snrlim}

To estimate the limitations of our 2D decomposition, we fit a range of simulated $\lya$ distributions combined with random realizations of the noise again using the variance from the MUSE data cubes. 
The variance used here was estimated around 6000$\AA$ in a 6$\AA$ spectral window corresponding to the median NB image spectral bandwidth of our sample.
This allows us to assess the S/N needed to measure $\lya$ halo properties reliably from our observed sample.

We considered simulated $\lya$ distributions with a fixed core scale length, $rs_{\rm cont}$, a fixed core flux, $F_{\rm cont}$, a range of 5 halo scale lengths, $rs_{\rm halo}$ and a broad range of halo fluxes, $F_{\rm halo}$. The fixed core values were set to the averages of our sample, $F_{\rm cont}=4.0\times10^{-18}$ $\flcgs$ and $rs_{\rm cont}=$ 0.06\arcsec\ (i.e. 0.3 MUSE pixel). We considered halo fluxes ranging from $1\times10^{-20}$ to $2\times10^{-18}$ $\flcgs$ and a set of halo sizes, $rs_{\rm halo}=$[0.2\arcsec, 0.4\arcsec, 0.6\arcsec, 1.0\arcsec , 1.4\arcsec, 1.8\arcsec, 2.2\arcsec] (i.e. [1, 2, 3, 5, 7, 9, 11] MUSE pixels). For each model $\lya$ distribution we generated 100 noise realizations and assessed the success rate for reliably recovering the halo size.

We find that halo sizes are reliably recovered above a S/N $\approx 6$. The S/N is measured inside an aperture corresponding to the CoG radius $r_{\rm CoG}$, which represents the radius for which the averaged flux in a concentric 1-pixel annulus reaches the noise value (see section~\ref{sec:fluxmes}). The smaller $\lya$ haloes are therefore less penalized by a S/N cut estimated within this aperture than if we had used a wide aperture that is identical for every object.

We also considered a range of core values ($rs_{\rm cont}$, $F_{\rm cont}$, not shown here) and find that this value of the S/N limit is still appropriate.
This S/N cut was thus adopted for the sample construction (see section \ref{sec:sample}).

\subsubsection{Size and flux limit}
\label{sec:rhlim}

In addition to being limited in sensitivity by S/N, we are also limited in our ability to measure the sizes of very compact $\lya$ haloes by the MUSE PSF.
To estimate the $\lya$ halo scale length below which we cannot trust our measurements, we again ran our modelled routine on several model $\lya$ distributions (with artificial noise based on the variance data cube). 
For each model object, we incrementally decreased the $\lya$ halo scale length until we could no longer recover the input value.
We find that the resulting scale length limit corresponds to one quarter of the MUSE PSF FWHM.
In practice, the halo scale length limit is a function of wavelength due to the PSF dependence on wavelength and thus on redshift.
As such, we calculated the limit separately for each object in our observed sample, yielding values ranging from 0.85 kpc to 1.48 kpc. 
If the best-fit $\lya$ scale length was below the scale length limit, we considered the limit value as an upper limit.

We also tested our ability to detect the faint $\lya$ haloes reliably. We performed the same exercise as our S/N limit procedure (see previous subsection) and find as expected that our halo flux limit increases with the halo scale length. This is because the surface brightness shrinks as the total $\lya$ flux is preserved.
Fixing the core component to the averages of our sample, we deduced that the halo flux limit corresponds to 9$\times$10$^{-19}$ $\flcgs$  and 5$\times$10$^{-19}$ $\flcgs$ times the halo scale length for the \mosaic\ and \udft,\ respectively.
\newline 
\\$\lya$ halo measurements and fitting results are given in Table~\ref{bigtab} of Appendix~\ref{ap:table}.


\section{Ly$\alpha$ halo characteristics}
\label{sec5}

In this section, we present the characteristics of our sample of $\lya$ haloes in terms of sizes, fluxes, spatial, and spectral shapes.


\subsection{Statistical significance of the detected $\lya$ haloes}
\label{sec:statsigni}

\begin{figure}[t]
   \resizebox{\hsize}{!}{\includegraphics{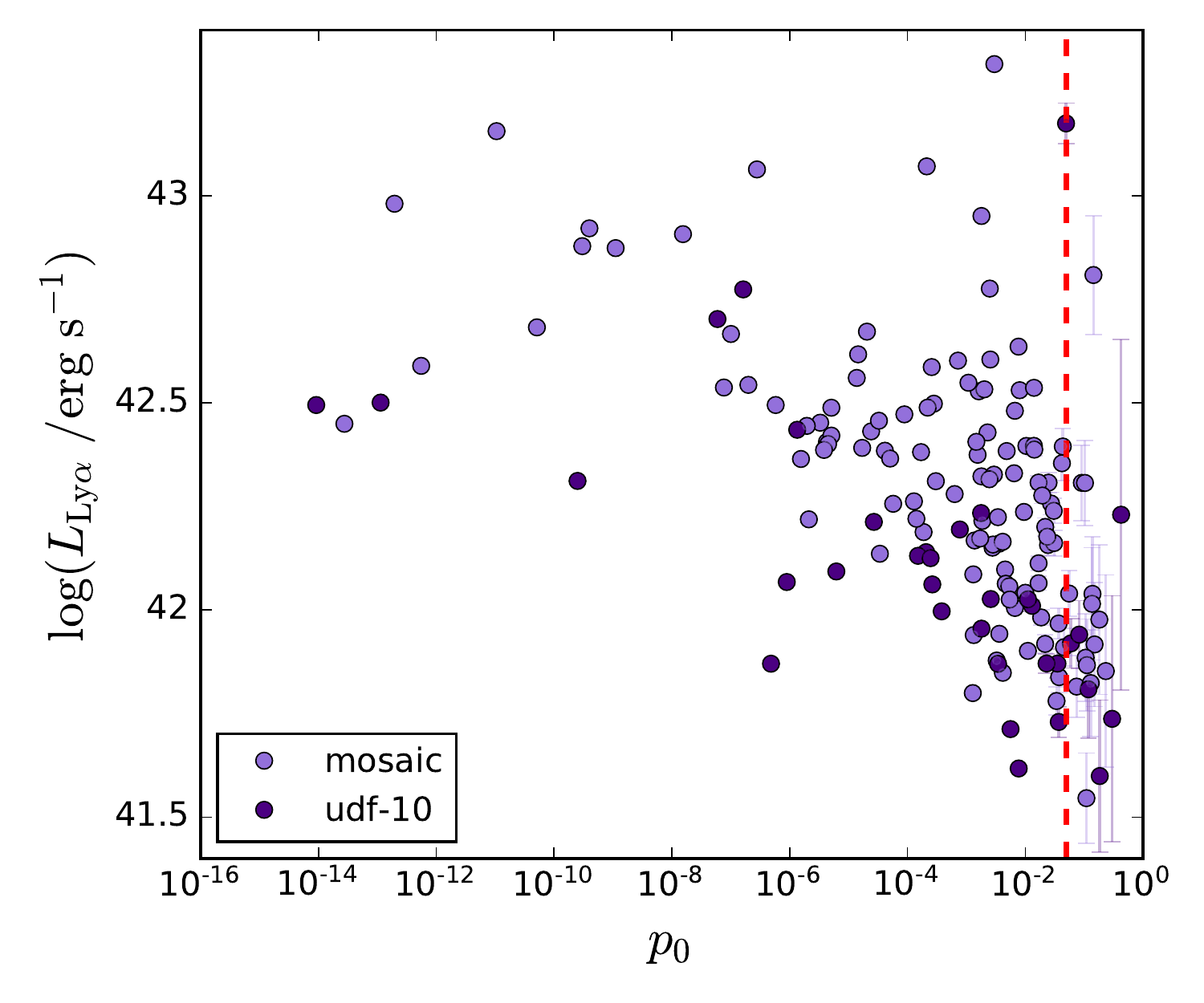}}
     \caption{$\lya$ luminosity vs. the statistical evidence for the presence ($p_{0}\leq0.05$) or absence ($p_{0}>0.05$) of a $\lya$ halo. Our $p_{0}$ threshold for a detection is indicated by the dotted red line. The value $p_{0}$ represents the probability of the two scale lengths to be identical (which would imply there is no $\lya$ halo). Light and dark purple symbols indicate the objects from the \udft\ and the \mosaic\ data cubes, respectively.}
     \label{pval}
\end{figure}

Of the 252 galaxies, 87 objects either have $\lya$ halo fluxes (62 objects), scale lengths (19 objects), or both (6 objects) below the detection limits (see section~\ref{sec:rhlim}). 
For the 19 objects with only the $\lya$ halo scale length below the detection limit, we use this value as an upper bound.
The objects with $\lya$ halo fluxes below the detection limit are ignored in the rest of this section.
This leaves us with a sample of 184 LAEs at this stage.

Following W16, in order to evaluate the statistical significance of the detected extended $\lya$ emission, we calculated the probability $p_{0}$ of the two scale lengths (galaxy and $\lya$ halo) to be identical by considering normal distribution. We consider a $\lya$ halo as detected if $p_{0} \leq 0.05$. 
Figure~\ref{pval} shows $p_{0}$ as a function of $\lya$ luminosity.
Out of the 184 objects for which we have a $\lya$ halo scale length measurement, 20 do not show statistical evidence for extended $\lya$ emission, mainly due to the large errors on their size measurements. Interestingly, some of these are relatively bright in $\lya$ (e.g. \#218 in Figure~\ref{fig2b}). Thirty objects have a $\lya$ halo with a very high significance ($p_{0}$ $\leq$ 10$^{-5}$).

Finally, out of the objects for which we have reliable $\lya$ halo measurements and excluding the objects with only upper limits to their $\lya$ halo scale length, 145 galaxies have a statistically significant $\lya$ halo (116 and 29 from the \mosaic\ and \udft\ field, respectively), which represents $\approx$80\% of the sample (145 out of 184 objects).


\subsection{Halo sizes}
\label{sec:halosize}
 
 \begin{figure}
\centering
   \resizebox{\hsize}{!}{\includegraphics{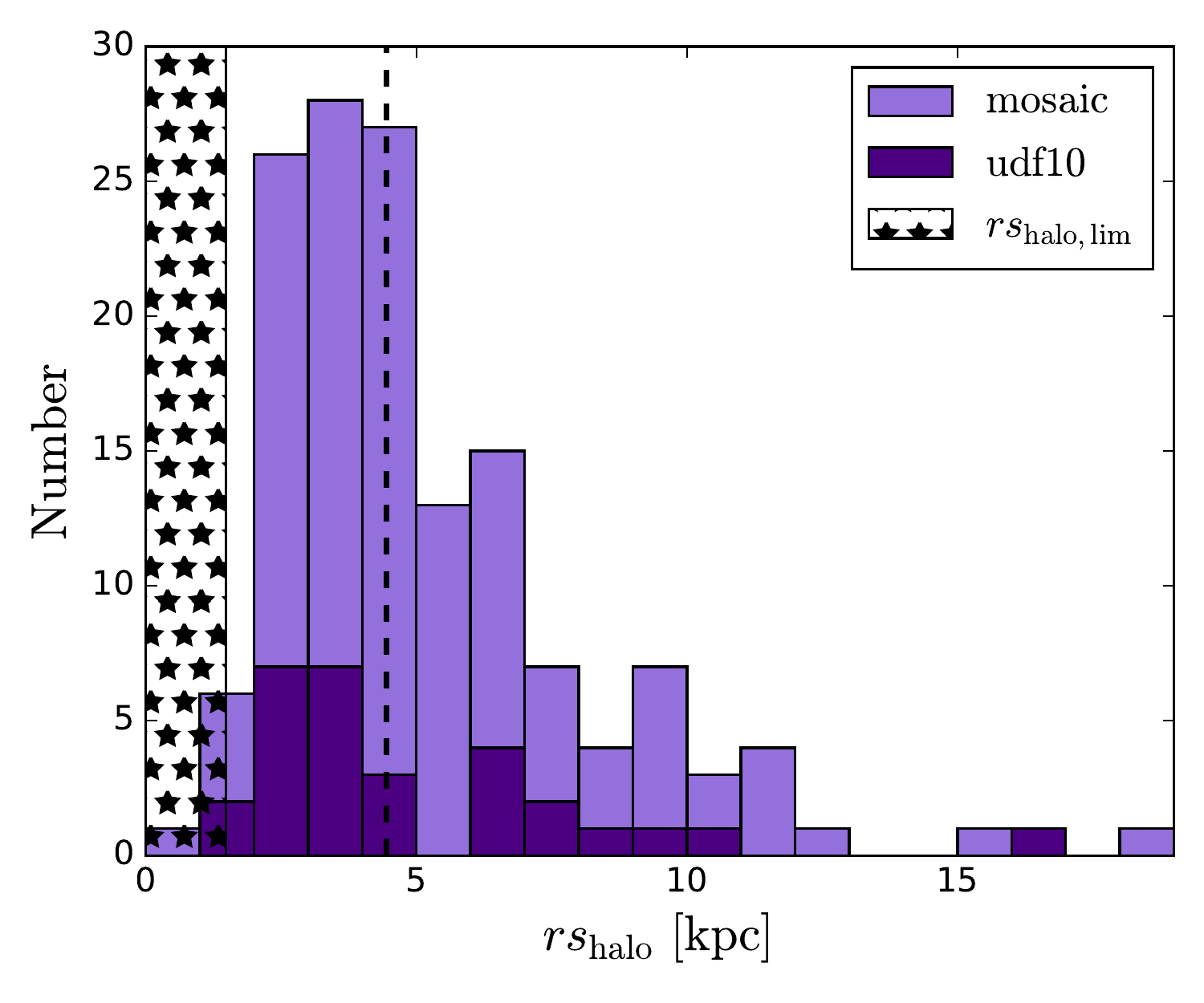}}
     \caption{Histogram of halo scale lengths resulting from the two-component model (see section~\ref{sec:model}). The dashed line indicates the median values (4.5 kpc) of the total distribution. The star-filled area shows the detection limit range (see section \ref{sec:rhlim}).}
     \label{figrh}
\end{figure}

Figure~\ref{figrh} shows the distribution of $\lya$ halo scale lengths for the 145 haloes that are considered to be statistically significant.
This distribution contains scale lengths ranging from 1.0 to 18.7 kpc with a median value of 4.5 kpc.
For reference, \citeauthor{X17} (2017; hereafter X17) measured similar halo scale lengths as our median scale length value; i.e. 5-6 kpc from their median stacking images of all LAEs.
The W16 authors measured slightly smaller scale lengths, with a median value of 3.4 kpc, but still in good agreement with our results.

Figure~\ref{figrh} also shows an extended tail of large halo scale lengths (> 7 kpc).
These large haloes represent less than 12\% (29 galaxies) of our total sample, plausibly explaining why they are not present in the smaller W16 sample.


\subsection{Fluxes and equivalent widths}

\subsubsection{Ly$\alpha$ flux and equivalent width fractions}
\label{sec:fluxfrac}

\begin{figure*}[t]
\centering
   \includegraphics[width=\textwidth]{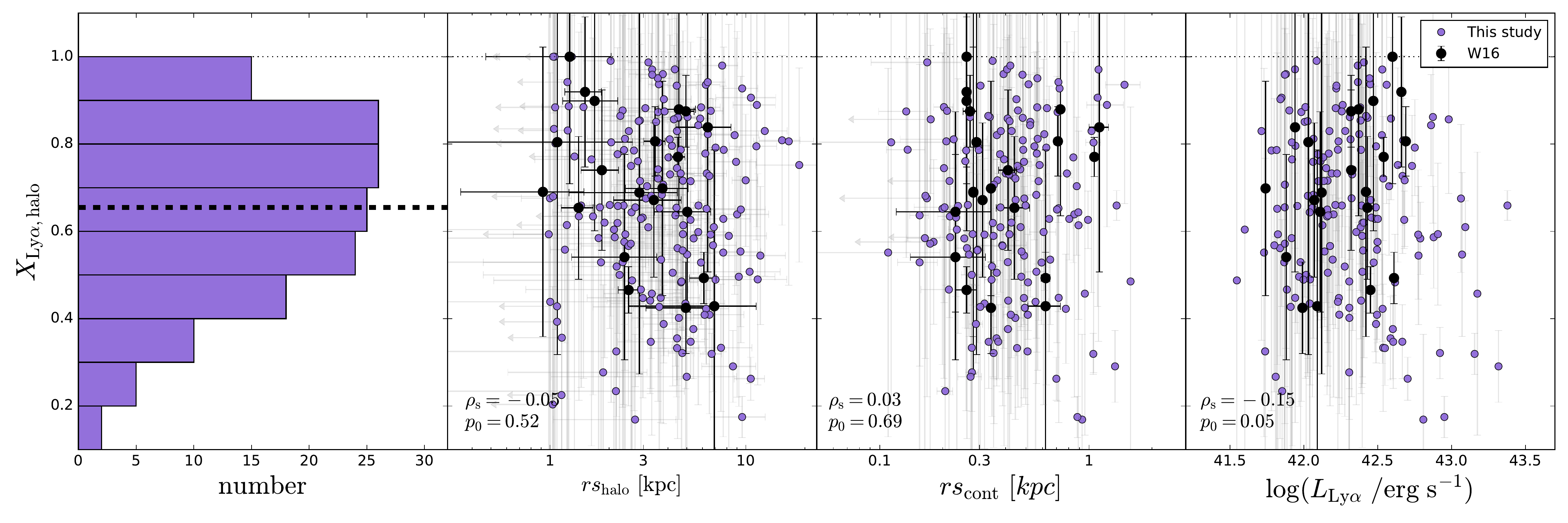}
        \caption{$\lya$ halo flux fraction $X_{\rm Ly\alpha, halo}$ as a function of $\lya$ halo and UV continuum scale lengths (second and third panel, respectively) and against total $\lya$ luminosity (right panel). 
       The first panel shows the $X_{\rm Ly\alpha, halo}$ distribution (objects without HST detection and with upper limit on their $\lya$ halo scale length are not included). The median value (0.65) is indicated by the black dashed line. 
       Upper limits on the scale lengths are indicated by arrows. 
       W16 measurements are indicated by the black points.
       Spearman rank correlation coefficients $\rho_{\rm s}$ and corresponding $p_{\rm 0}$ values for our results (excluding upper limits) and those of W16 are shown in each panel.}
   
     \label{figXhrlew}
\end{figure*}

\begin{figure}[t]
   \resizebox{\hsize}{!}{\includegraphics{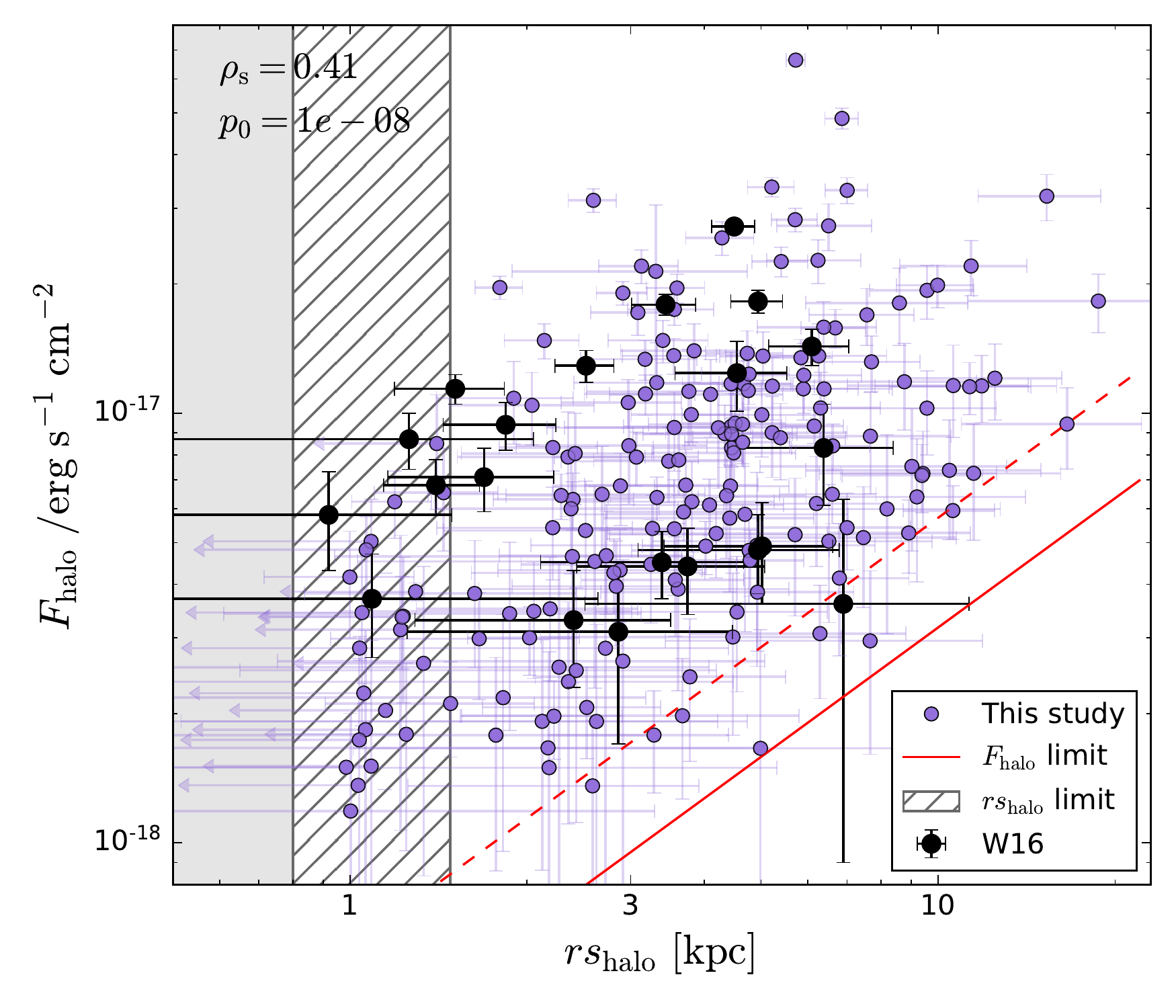}}
     \caption{$\lya$ halo flux as a function of halo scale length. 
       Our limiting $\lya$ halo flux and scale length are indicated by the red lines (dashed for the \mosaic\ and solid for \udft\ sample) and grey hashed area, respectively (see section~\ref{sec:rhlim}).
       Upper limits on the scale lengths are indicated by arrows. 
       W16 measurements are indicated by the black points.
       Spearman rank correlation coefficients $\rho_{\rm s}$ and corresponding $p_{\rm 0}$ values for our results and those of W16 (excluding upper limits) are shown in each panel. }
     \label{fhvsrh}
\end{figure}

Our 2D, two-component decomposition of the $\lya$ spatial distribution provides estimates of the decomposed flux from the core, $F_{\rm cont}$, and from the halo, $F_{\rm halo}$. 
The median halo flux is $\approx$5.4$\times$10$^{-18}$ $\flcgs$ and $\approx$9.4$\times$10$^{-18}$ $\flcgs$ for the \udft\ and \mosaic\ data cubes, respectively.
Following the same approach as W16, we defined the $\lya$ halo flux fraction, $X_{\rm Ly\alpha, halo}$, as $F_{\rm halo}/(F_{\rm halo}+F_{\rm cont})$, quantifying the contribution of the halo to the total $\lya$ flux. 
For galaxies with a UV continuum detection and with reliable $\lya$ halo scale length measurements, we find halo flux fractions ranging from $\approx$17 to $\approx$99\% (with an average of 65\%). 
Our results are in very good agreement with W16 (they find a mean value of 70\%), although we find more haloes with small $\lya$ halo flux fractions than W16. 
This difference is likely due to our larger sample and to the improved quality of the UDF data cubes with respect to the HDFS (see Figure~8 of B17).

Figure~\ref{figXhrlew} shows the $\lya$ halo flux fraction as a function of the core and halo scale lengths and as a function of the total $\lya$ luminosity measured by the CoG method (see section~\ref{sec:fluxmes}). 
In order to test for correlations, we calculated the Spearman rank correlation coefficients $\rho_{\rm s}$ (e.g. \citealt{W96}) and the corresponding p-values $p_{\rm 0}$, which correspond to the probability of the null hypothesis that no monotonic relation exists between the two variables. The value $\rho_{\rm s}$ varies between $-1$ and $+1$, with 0 implying no correlation. The test does not take the error bars into account.
We obtained ($\rho_{\rm s}$=$-$0.05, $p_{\rm 0}$=0.52), ($\rho_{\rm s}$=0.03, $p_{\rm 0}$=0.69), and ($\rho_{\rm s}$=$-$0.15, $p_{\rm 0}$=0.05) for the $X_{\rm Ly\alpha, halo}$-$rs_{\rm halo}$, $X_{\rm Ly\alpha, halo}$-$rs_{\rm cont}$, and $X_{\rm Ly\alpha, halo}$-$L_{\rm Ly\alpha}$, respectively.
We find no correlation between the fraction of $\lya$ emission in the halo and (i) the $\lya$ halo scale lengths (second panel), (ii) the UV continuum scale lengths (third panel), and (iii) the total $\lya$ luminosities (right panel).

Figure~\ref{fhvsrh} shows the $\lya$ flux in the halo as a function of its halo scale length. Our limiting $\lya$ halo flux and scale length (see section~\ref{sec:rhlim}) are indicated. We find a clear correlation between these two properties ($\rho_{\rm s}$=0.4, $p_{\rm 0}$<10$^{-8}$). While this correlation is partially created by our halo flux limit (red lines), the correlation is still readily apparent for brighter haloes. The correlation is positive, such that $\lya$ haloes with larger scale lengths have higher halo fluxes.

\subsubsection{Total Ly$\alpha$ flux and equivalent width}
\label{sec:fluxmes}

\begin{figure}[t]
   \resizebox{\hsize}{!}{\includegraphics{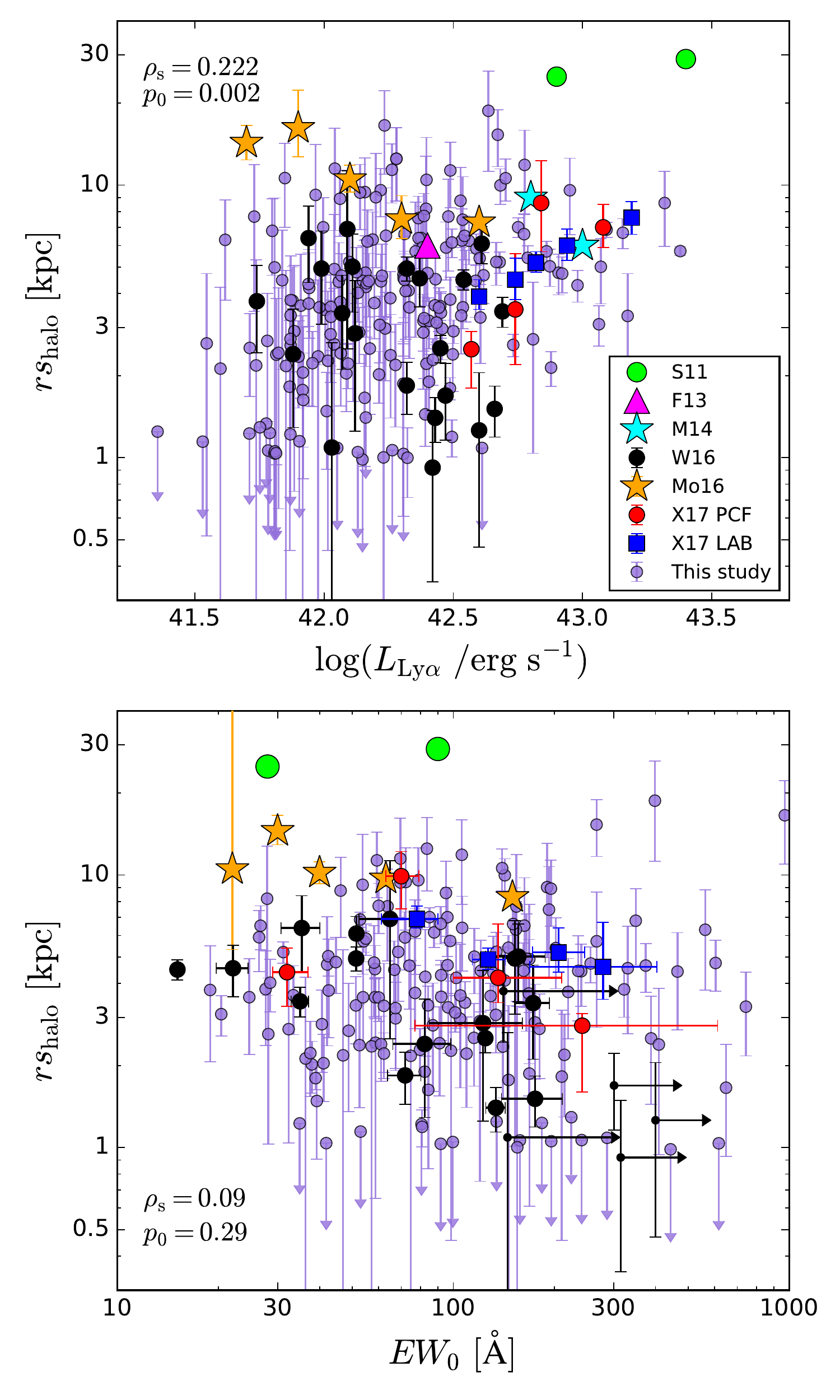}}
     \caption{Halo scale length plotted as a function of total $\lya$ luminosities (upper panel) and total rest-frame $\lya$ EW (lower panel).
       Only the 121 LAEs with the EWs from the H17 sample are included for the lower panel.
       Our results are shown by the purple dots while W16 results correspond to the black dots. Arrows show upper limits.
       Values from studies using stacking methods are indicated by coloured symbols: X17 (red circles for their stacked images of LAEs from a protocluster field (PCF), and blue squares around a $\lya$ blob (LAB)), \citealt{Mo16} (orange), \citealt{Mo14} (blue),
       \citealt{F13} (magenta),  and \citealt{S11} (green).
       Spearman rank correlation coefficients $\rho_{\rm s}$ and corresponding $p_{\rm 0}$ values for our results and those of W16 (without upper and lower limits) are shown in each panel.}   
     \label{figrlew}
\end{figure}

The $\lya$ flux was computed by integrating inside the circular aperture corresponding to the CoG radius. This radius ($r_{\rm CoG}$) was determined by averaging the flux in successive annuli of 1 pixel thickness around the $\lya$ emission centre until a certain annulus for which the averaged flux reaches the noise value. The centre of this last annulus corresponds to $r_{\rm CoG}$.
From this aperture, we extracted a spectrum and integrated the flux corresponding to the $\lya$ line width; the borders of the line are set when the flux goes under zero. These spectral bandwidths are indicated by the vertical black dotted lines in the third panel of Figs.~\ref{fig2a} and \ref{fig2b}\footnote{The spectra shown here were extracted using the HST segmentation map (for display purposes because of the higher S/N), whereas the bandwidth (indicated by dotted black lines) to measure the total $\lya$ flux were measured in spectra extracted in an aperture of radius r$_{\rm CoG}$. This explains why the dotted black lines do not cross zero exactly when the spectra do.}.

This method ensures that most of the $\lya$ flux is encompassed for each object, which is not the case if we use a single fixed aperture for all of the objects.
Figure~\ref{figz} (lower panel) shows the distribution of total $\lya$ fluxes for our sample, which spans 2 orders of magnitude from $1.74\times10^{-18}$ $\flcgs$ to $1.12\times10^{-16}$ $\flcgs$. 

Rest-frame $\lya$ equivalent widths (EWs) were calculated using the UV continuum measured by \citeauthor{H17} (2017; hereafter H17). The H17 authors performed careful UV continuum measurements using several HST bands (2 or 3 bands depending on the object).
After cross-matching the respective catalogues, we obtained $\lya$ EW measurements for 155 of the 184 galaxies for which we have a halo scale length measurement.

Figure~\ref{figrlew} shows the distribution of $\lya$ halo scale lengths as a function of total $\lya$ luminosity and rest-frame $\lya$ EW. 
With respect to previous studies that employed stacking, our sample goes much deeper and we probe much smaller $\lya$ haloes. 
The Spearman rank correlation test provides $\rho_{\rm s}$ = 0.222 ($p_{\rm 0}$=0.002) for the $rs_{\rm halo}$-$L_{\rm \lya}$ relation and $\rho_{\rm s}$=0.09 ($p_{\rm 0}$ = 0.29) for the $rs_{\rm halo}$-$EW_{\rm 0}$ relation.

We find a suggestion of a correlation between the $\lya$ scale length and total $\lya$ luminosity, albeit with very large scatter. In particular the bright LAEs tend to have large haloes ($rs_{\rm halo} \gtrsim$ 3 kpc), whereas there is more dispersion at lower $\lya$ luminosities. The 
X17 authors found  a clear correlation in their stacks corresponding to our bright LAEs whereas
W16 found no such correlation. By computing the Spearman rank correlation coefficients for our objects in their luminosity range (41.6< log($L_{\rm Ly\alpha}$)< 42.7), we find no correlation either ($\rho_{\rm s}$ = $0.07$, $p_{\rm 0}$ = 0.59).

We also do not find a correlation between halo sizes and rest-frame EWs. 
Both W16 and X17 found a similar result. 
As an aside, it is worth noting that some objects have very large EWs (exceeding 200$\AA$). 
The H17 work provides for a detailed analysis of these objects.


\subsection{Ly$\alpha$ line profiles}

\begin{figure}
   \resizebox{\hsize}{!}{\includegraphics{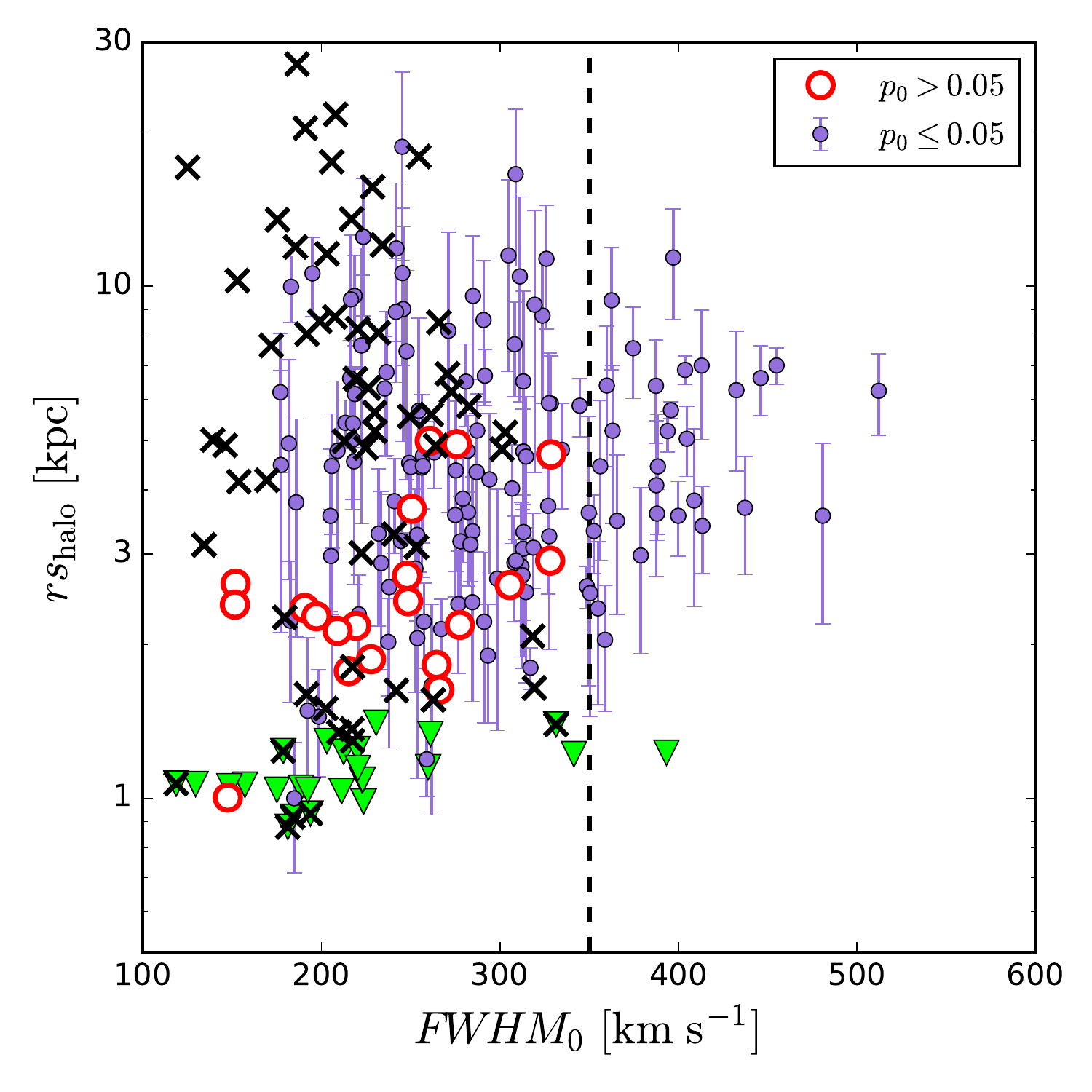}}
     \caption{$\lya$ halo scale length plotted as a function of rest-frame FWHM of the $\lya$ line. 
       The purple dots correspond to the objects that have a high significant $\lya$ halo ($p_{0}\leq0.05$).
       Most of the objects without a significant $\lya$ halo ($p_{0}>0.05$, see section~\ref{sec:statsigni}, red circles) or with upper limits on their halo properties (green triangles for scale lengths, black crosses for halo fluxes -- see section~\ref{sec:rhlim}) show a $\lya$ line narrower than 350 km s$^{-1}$ (black dashed line).} 
      \label{rhfwhm}
\end{figure}

\begin{figure}
   \resizebox{\hsize}{!}{\includegraphics{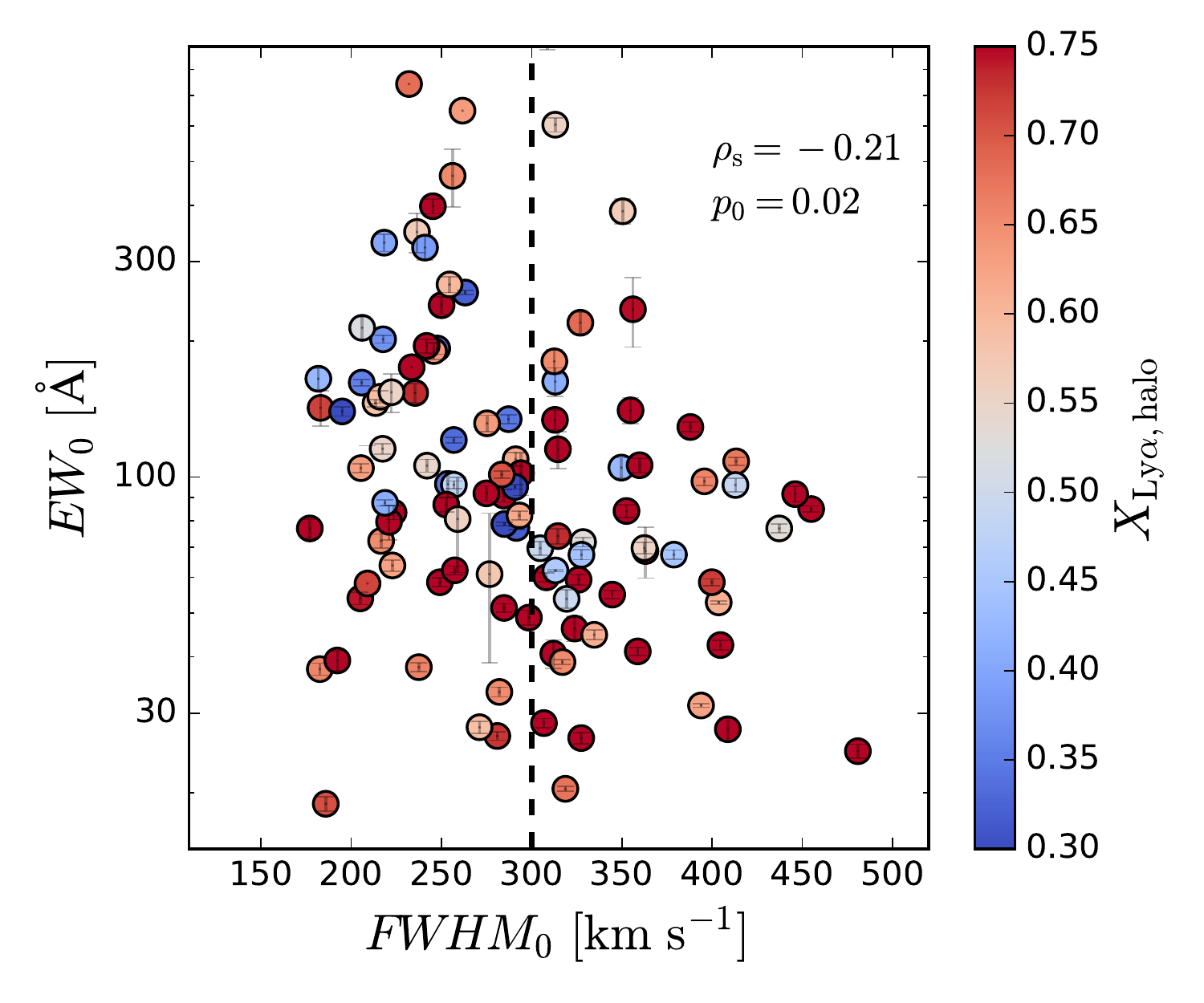}}
     \caption{$\lya$ equivalent width as a function of rest-frame FWHM of the $\lya$ line.
       Points are colour coded by $\lya$ halo flux fraction ($X_{\rm lya,halo}$). 
       We only show the objects with a statistically significant $\lya$ halo ($p_{0}\leq0.05$, see section~\ref{sec:statsigni}).} 
     \label{ewfwhmhff}
\end{figure}

Next, we explore the connection between the spectral and spatial properties of the $\lya$ emission. 
The diversity of $\lya$ line profiles can be appreciated by looking at Figures~\ref{fig2a} and \ref{fig2b}. 
While most of the lines are asymmetric and single peaked, others appear to be double peaked (see objects \#1087 and \#106 of Figure~\ref{fig2a} and Figure~\ref{fig2b}, respectively).
This diversity is directly reflected in the $\lya$ FWHM measurements, which span a large range from 118 km s$^{-1}$ to 512 km s$^{-1}$.

We performed the measurement of the $\lya$ FWHM only on the single-peaked $\lya$ lines so that objects with doubled-peaked profiles are excluded.
If the blue peak is comparable in flux to the red peak, the $\lya$ line is referred to as a $\lya$ doublet (see object \#106 on Figure~\ref{fig2b}), whereas if the blue peak is much fainter, the feature is referred as a blue bump (see object \#1087 on Figure~\ref{fig2a}). We carried out an inventory of the various line profiles encountered in our sample. 
Out of our 252 galaxies, 15 objects show a $\lya$ line with a blue bump and 8 objects have a $\lya$ doublet. 
Put together, the double-peaked profiles therefore represent a small fraction (<10\%) of our sample. 
The halo properties of such double-peaked line objects are not significantly different from those of the rest of the sample.

We connect the FWHM of single-peaked $\lya$ lines with $\lya$ halo sizes in Figure~\ref{rhfwhm}. 
The smallest $\lya$ haloes, for which we only have an upper limit on their scale length or halo flux (see section~\ref{sec:rhlim}) and the $\lya$ haloes with low statistical significance (see section~\ref{sec:statsigni}) appear to have a narrower $\lya$ line (<350 km s$^{-1}$), whereas the galaxies with significant extended $\lya$ emission span a wider range of FWHM values. 

Figure~\ref{ewfwhmhff} shows the $\lya$ EW plotted against the FWHM, colour coded by the $\lya$ halo flux fraction.
We show in this figure only the 121 objects with a statistically significant $\lya$ halo (see section~\ref{sec:statsigni}) and an EW measurement (see section~\ref{sec:fluxmes}). 
It is also apparent that we do not find evidence for a significant anti-correlation between the EWs and the FWHMs of the $\lya$ lines ($\rho_{\rm s}$=$-$0.21, $p_{\rm 0}$=0.02).

The objects that have less than 30\% of their total $\lya$ flux in the halo appear to have narrower $\lya$ lines than the rest of the sample (< 300 km s$^{-1}$). 
Certainly, the objects with a large $\lya$ line width (> 400 km s$^{-1}$) have small EW (<100$\AA$) and  >50\% of the $\lya$ flux is in the halo.


\section{Connecting host galaxies to $\lya$ haloes}
\label{sec:hostgal}

In this section, we investigate the connection between $\lya$ properties and the properties of the host galaxies.
First we consider the general UV properties. 
We then compare galaxy and $\lya$ halo sizes.
Finally, we explore the coevolution of UV and $\lya$ halo sizes with redshift.

\subsection{UV properties}

\begin{figure}
   \resizebox{\hsize}{!}{\includegraphics{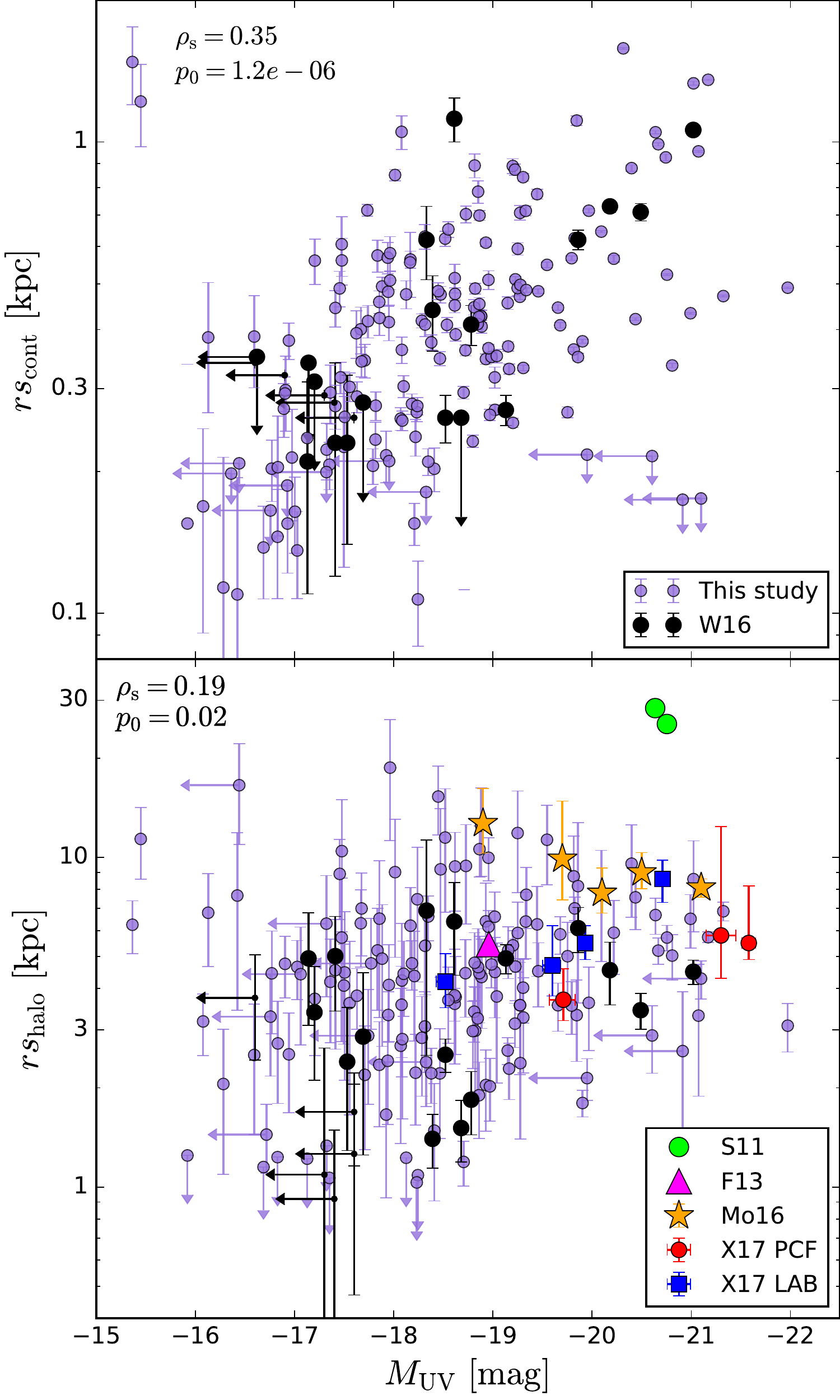}}
\caption{
UV continuum scale length (upper) and $\lya$ halo scale length (lower) as a function of absolute far-UV magnitude. 
Only the objects with a statistically significant $\lya$ halo are shown (see section~\ref{sec:statsigni}). 
Upper limits on the scale lengths and UV magnitudes are indicated by arrows. 
The W16 measurements are shown in black, \cite{S11} by green dots, \cite{F13} by magenta triangles, \citep{Mo16} by orange stars, and X17 by red points (LAEs from a protocluster field "PCF") and blue squares (LAEs around a $\lya$ blob "LAB").
Spearman rank correlation coefficients $\rho_{\rm s}$ and corresponding $p_{\rm 0}$ values for our results and those of W16 (without upper limits) are shown in each panel.}
\label{figrsfxmaguv}
\end{figure}

The upper panel of Figure~\ref{figrsfxmaguv} shows the expected correlation ($\rho_{\rm s}$=0.35, $p_{\rm 0}$$\sim$10$^{-6}$) between the UV sizes and UV magnitudes of galaxies \citep{S15}. 
The lower panel shows $\lya$ halo scale length as a function of absolute far-UV magnitude. 
According to the Spearman test coefficient ($\rho_{\rm s}$=0.19, $p_{\rm 0}$=0.02), there is a suggestion of a positive correlation between the $\lya$ halo size and UV magnitude for our selected sample of LAEs albeit the scatter is large. If the correlation is real, it would agree with the results of X17 who found that $\lya$ halo sizes are positively correlated with UV luminosities. 
Similarly, W16 found that UV-luminous galaxies ($M_{\rm UV}$ <$-$19) tend to have $\lya$ haloes with larger scale lengths ($rs_{\rm halo} \gtrsim$~3 kpc). This result is also observed in our larger sample.

\subsection{Comparison of sizes}
\label{sec:sizecomp}

\begin{figure*}[t]
\centering
   \includegraphics[width=13.5cm]{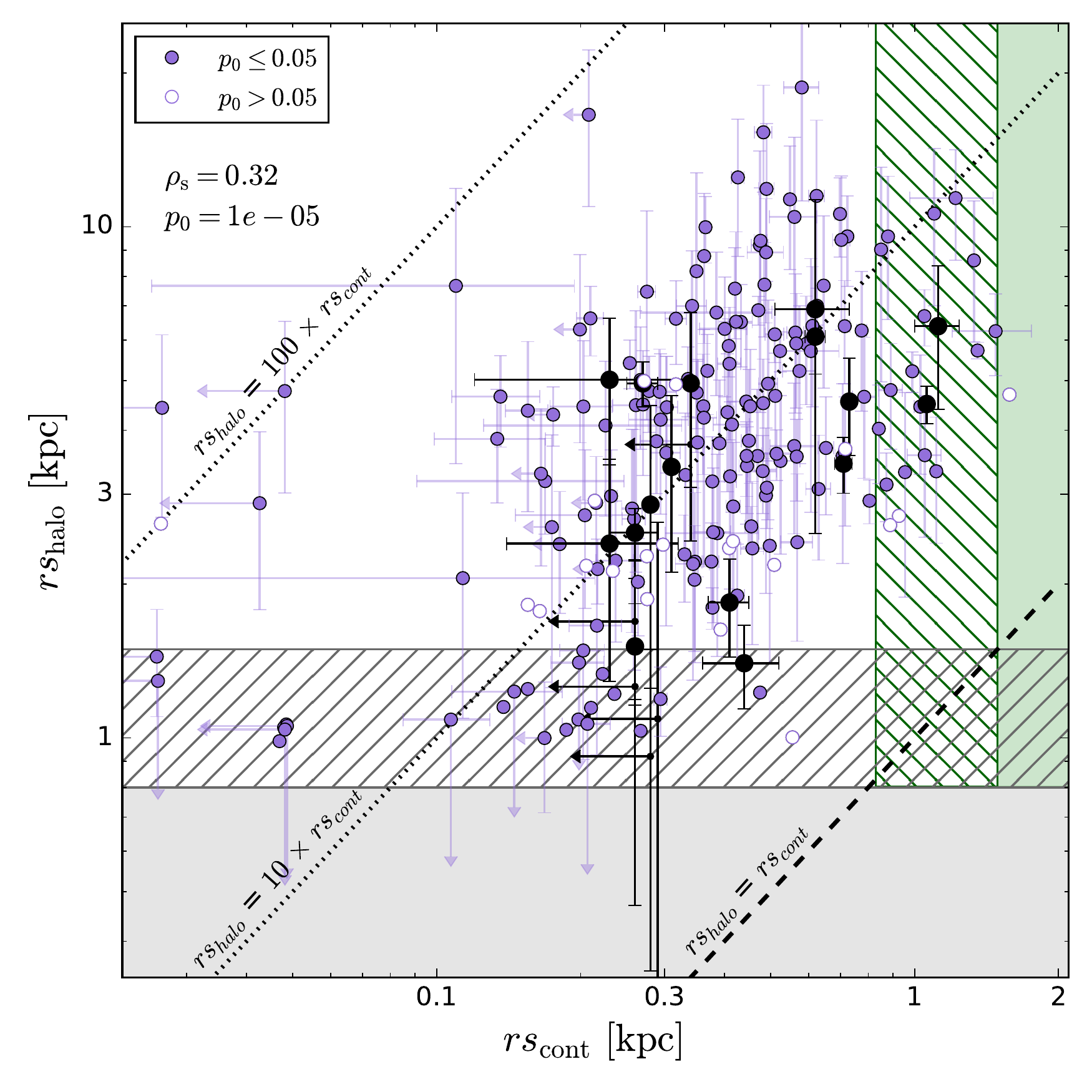}
\caption{
$\lya$ halo scale length as a function of UV continuum scale length. 
The grey area corresponds to the $\lya$ halo range for which we cannot reliably measure the $\lya$ halo size (see section~\ref{sec:rhlim}). 
This wavelength-dependent size limit spans from 0.85 kpc to 1.48 kpc and is represented by the grey hatched area (see section~\ref{sec:rhlim}). 
The green area shows the objects for which we would be able to detect the absence of a $\lya$ halo with our data. 
This limit also depends on the wavelength and is shown by the green dashed area. 
The black dashed line corresponds to a size ratio of 1 (meaning no halo). The two dotted lines indicate ratios of 10 and 100 as indicated in the figure. Upper limit scale lengths are indicated by arrows and objects without a statistically significant $\lya$ halo are shown by empty symbols. The W16 results are shown with black points.
Spearman rank correlation coefficients $\rho_{\rm s}$ and corresponding $p_{\rm 0}$ values for our results and those of W16 (without upper limits) are shown in each panel.}
\label{figrcrh}
\end{figure*}

Next we compare the UV continuum and $\lya$ emission scale lengths resulting from our 2D two-component model. 
Figure~\ref{figrcrh} shows the $\lya$ halo scale lengths plotted as a function of UV continuum scale length. 
First, according to the Spearman correlation test, $\lya$ scale lengths are positively correlated ($\rho_{\rm s}$=0.32, $p_{\rm 0}$ $\sim$10$^{-5}$) with galaxy UV sizes (albeit with large scatter). Indeed, the $\lya$ scale lengths are always between $\approx$4 and >20 times larger than the continuum scale length with a median size ratio of 10.8 (the lower quartile at 6.0 and the upper percentile at 19.1). This results is in very good agreement with W16, albeit this scatter in the ratio of scale lengths exceeds the range from their sample.
This plot also shows that we do not detect any LAEs without a $\lya$ halo. This is valid for 145 galaxies (80\%) of our sample given that the remaining 39 galaxies (20\%) only have either upper limits (grey area) or large error bars on their halo scale lengths (empty circles in Figure~\ref{figrcrh} upper panel).
As such, we can only confirm the absence of a $\lya$ halo around a galaxy larger than this size limit (i.e. $rs_{\rm cont}$ $\gtrsim$ 1~kpc; see section~\ref{sec:rhlim}). 
This condition is shown as the green area in Figure~\ref{figrcrh}. The dashed areas show the range of our wavelength-dependent detection size limit (see section~\ref{sec:rhlim}).

\subsection{Size evolution}
\label{sec:evored}

\begin{figure}[t]
\centering
\resizebox{\hsize}{!}{\includegraphics{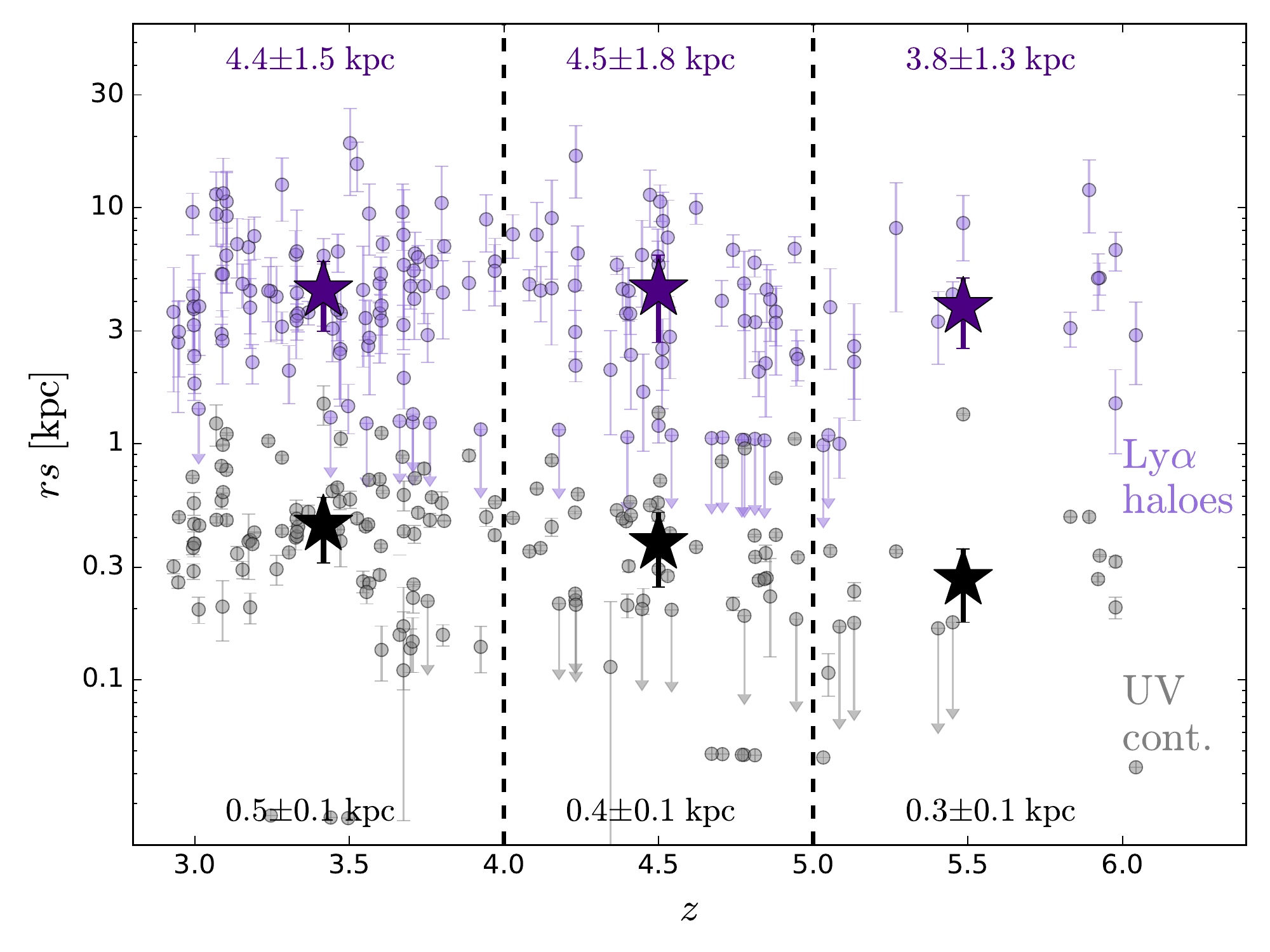}}
\caption{
$\lya$ halo (purple dots) and UV continuum (grey dots) scale lengths as a function of redshift. 
The median scale lengths in 3 redshift bins ($z$<4, 4$\leq$ $z$<5, $z$ $\geq$5) of both $\lya$ and UV continuum emission are indicated by the star symbols (error bars correspond to the median absolute deviation). The corresponding numerical values are given at the top and bottom of the figure for the $\lya$ and UV continuum emission, respectively. The objects with scale length upper limits are not taking into account for the median calculations.}
\label{figrsz}
\end{figure}

The evolution of both the UV and $\lya$ halo sizes is shown in Figure~\ref{figrsz}. 
While the UV size of galaxies decreases with redshift as expected \cite[][]{S15}, we do not find that $\lya$ halo sizes show significant evolution between redshifts 3 and 6. The W16 authors found that $\lya$ halo sizes decrease with increasing redshift. 
However, their sample at $z$>5 consists only of five galaxies and the dispersion is large. Because we do not have enough objects in the higher redshift bin, we conclude that there is currently no clear evidence supporting an evolution of $\lya$ halo sizes with redshift above $z$=3. \cite{Mo14} also investigated the size evolution of their stacked LAEs, finding no evidence for evolution of $\lya$ halo sizes from $z$=2.2 to $z$=5.7 and a possible but very uncertain increase from $z$=5.7 to $z$=6.6. 
This result implies a higher $\lya$/UV scale length ratio at high redshift and hence suggests that the fraction of CGM probed by the $\lya$ emission is actually increasing with redshift as the galaxies are known to be more compact and less massive as high redshift \citep{S15}.

We now compare our scale length measurements with 12 local starburst galaxies (0.028<$z$<0.18) from the LARS sample \citep{H13,Gu15}. These authors measured the spatial extent of the $\lya$ emission using the Petrosian 20 percent radius $R_{\rm p20}$ \citep{P76} and compared to the corresponding radius measured from H$\rm \alpha$. 
Similar to the high-redshift galaxies, some local galaxies show extended $\lya$ emission (seven galaxies according \citealt{H14}). 
Their resulting $\lya$/H$\rm \alpha$ size ratios range from 1 to 3.6, with an average of 2. 
For comparison, we also calculate the Petrosian radii of our galaxies. 
As in W16 (see their Figure 12), our galaxies at $z$>3 appear to have $\lya$ haloes with larger Petrosian radii as well as higher $\lya$/UV size ratios than local galaxies.


\section{Discussion}
\label{sec7}
\subsection{Probing the CGM}

\subsubsection{Ubiquity of $\lya$ haloes around LAEs}

The high fraction of LAEs with a significant $\lya$ halo ($\approx$80\%) demonstrates that $\lya$ haloes are a common property of LAEs at high redshift. 
As such, this also suggests that the CGM has a rich "cold" gas content.
Theoretical analyses and numerical simulations indeed predict that neutral hydrogen should be present around high-redshift star-forming galaxies (e.g. \citealt{K05,F11,FG15}). Our results therefore appear consistent with the canonical vision of the galaxy formation at high-$z$. For the remaining 20\% we have only upper limits or very uncertain $\lya$ halo size measurements, which prevents us from drawing firm conclusions. 

That such a result is not seen for local galaxies (cf. LARS sample; \cite{O14,H13}, see section~\ref{sec:evored}) underlines that there is a clear evolution of the CGM across cosmic time. 
In the LARS sample, $\lya$ haloes are only detected in 50\% of cases \citep{H14}. Moreover, \cite{H13} found the $\lya$ emission to be more extended than the UV continuum by a factor of 2.4 on average (interstellar medium scales), whereas we find a factor of $\approx$10 for our sample (CGM scales).
This difference could be due to an evolution of the $\lya$ escape fraction with redshift, possibly due to dust content evolution \citep{H11,D13} or because of sensitivity limitations. This can also suggest that the contribution of the mechanisms powering the $\lya$ haloes evolves with cosmic time.

This study is limited to $\lya$ emitters. 
In future work, we intend to search for extended $\lya$ emission around individual galaxies that were not detected based on their $\lya$ line (i.e. LBGs). 
For example, Steidel et al. (2011) detected extended $\lya$ emission around a stacked $\lya$ absorber galaxies sample and around massive LBGs. 
This promising result motivates us to search for $\lya$ haloes around all high-redshift galaxies with MUSE.

\subsubsection{$\lya$ halo size - host galaxy correlations}

Supposing $\lya$ haloes are a general property of star-forming galaxies, it is interesting to know how the halo sizes correlate with other properties of the host galaxies. 
We searched for such correlations in section \ref{sec:hostgal}. Both the $\lya$ halo flux and scale length seem to correlate with both UV magnitude and scale length, suggesting that the $\lya$ halo properties are actually linked to the UV properties of the host galaxy. Consequently, this may suggest that the star formation rate directly influences the powering of the $\lya$ haloes. Also, if the trends are real, the correlations suggest that the UV-brighter objects are associated with different physical conditions,  such as kinematics, gas content, and distribution, which favour the production of $\lya$ haloes compared to the fainter objects from our sample.

\subsubsection{$\lya$ spatial extent versus virial radius of DM haloes}

\begin{figure*}
\centering
   \includegraphics[width=\textwidth]{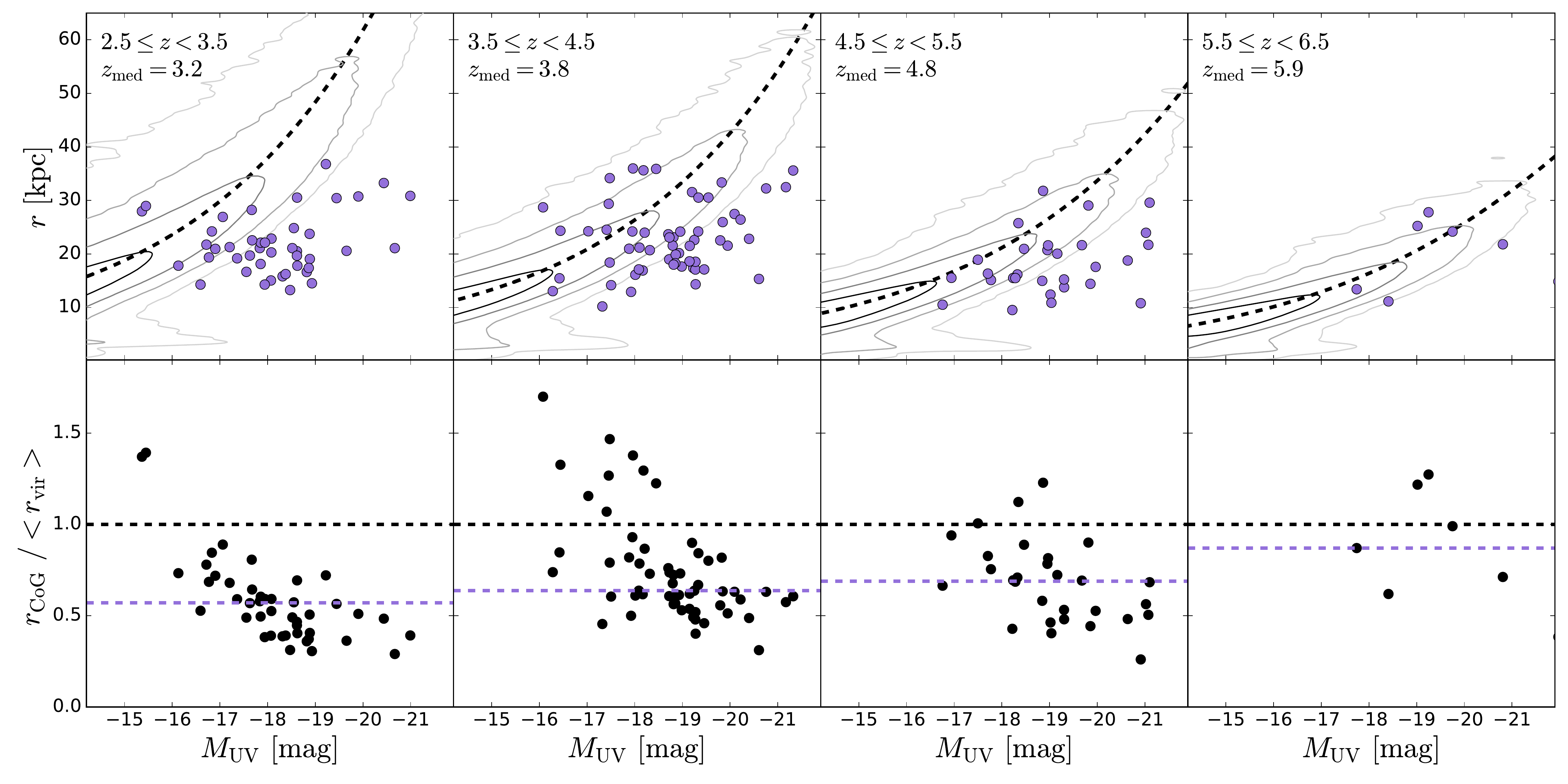}
     \caption{
     \textit{Upper panels}: Maximum radius of the $\lya$ haloes detected using the CoG method as a function of the absolute UV magnitude of their host galaxy. 
     The grey contours correspond to the predicted virial radius / UV magnitude relation predicted by a semi-analytic model \citep[contours at 10$^{-2}$, 10$^{-3}$, 10$^{-4}$, 10$^{-5}$ percent of the total number of modelled galaxies]{G15}. 
     The dashed black line corresponds to a polynomial fit of the distribution of the simulated galaxies.
     Each panel corresponds to a different redshift bin. 
     The plot aims to show what cold CGM scale we probe with Lyman alpha emission. 
     \textit{Lower panels}: Ratio of the predicted median virial radius at a given UV magnitude over the measured CoG radius of individual objects, plotted as a function of absolute UV magnitude in different redshift bins. 
     The median values are indicated by the dashed purple lines.}
     \label{rvir}
\end{figure*}

We now attempt to assess the maximum CGM scales that are traced by $\lya$ emission as a function of UV magnitude.
To do so, we compare the maximum detected extent of the $\lya$ haloes (measured using the CoG method) with the virial radius of the dark matter (DM) haloes of galaxies predicted by the semi-analytic model of \citeauthor{G15} (2015; see contours in the upper panels of Figure \ref{rvir}). 
We compare those two extents for four different redshift bins ($z$~$\simeq$[3,4,5,6]). 
In the lower panel, the purple dotted lines indicate the median value of the ratio of CoG radii over the mean virial radii $r_{\rm CoG}$ /<$r_{\rm vir}$> in each redshift bin. This ratio appears to increase with redshift (ratio median values of [57\%, 64\%, 69\%, 87\%] for median redshift bins of [3.2, 3.8, 4.8, 5.9]) suggesting that $\lya$ emission is probing a larger percentage of the CGM towards high redshift. This is not surprising because our measured $\lya$ halo sizes do not show any evolution with redshift while the \cite{G15} model predicts that, at fixed M$_{\rm UV}$, galaxies reside in smaller DM haloes at higher redshift.

While a weak anti-correlation can be guessed, the fraction of CGM probed by the $\lya$ emission is fairly constant with UV magnitude in each redshift bin.

Our $\lya$ haloes therefore reach on average more than $\gtrsim$50\% of the predicted virial radius of their host galaxy (irrespective of M$_{\rm UV}$) and go even beyond for higher redshift. 
This result demonstrates that the $\lya$ emission is a powerful tracer of the gas located inside the virial radius (e.g. the CGM) but not at larger scales (e.g. IGM) considering our current detection capacities.

\subsection{Origin of the $\lya$ haloes}
\label{sec:origin}

Taking advantage of our large statistical sample, we now attempt to assess the contribution of the different proposed $\lya$ emission processes that could be responsible for our observed $\lya$ haloes. 
For each process, we review the emission mechanism, discuss the expected observational signatures and, where possible, compare these expectations with our results.

\subsubsection{Stellar origin with scattering in an outflowing medium}

The scattering of $\lya$ photons produced in star-forming regions is one of the candidates to explain $\lya$ haloes. For this mechanism, $\lya$ photons are produced by recombination associated with the stellar UV radiation in the HII regions of galaxies. 
A fraction of those $\lya$ photons can be absorbed by interstellar dust but the escaping photons scatter into the surrounding neutral hydrogen gas and can be redirected towards the observer, leading to the observed $\lya$ haloes. 

The main question is therefore whether the stellar content of the galaxies produces enough ionizing photons to power the observed $\lya$ emission. In a similar approach to W16 (see their section 7.2 for more details), the condition to be tested is a condition on the $\lya$ EW as this quantity gives a direct comparison between the continuum and $\lya$ fluxes. The maximum dust-free $\lya$ EW estimated for a stellar origin ranges from $\approx$50 to 200$\AA$ \citep{C93}. While $\approx$17\% of our sample has $\lya$ EWs higher than 200$\AA,$ which suggests that $\lya$ photons do not only come from the HII regions, most of our galaxies have $\lya$ EWs lower than 200$\AA$. This suggests that the stellar UV continuum alone can power the haloes. 
In any case, given that the EW depends on stellar metallicity and initial mass function and can be affected by bursty star formation histories \citep{S03,R10}, the objects with EW>200$\AA$ values may be interpreted without invoking other $\lya$ production channels.
A more detailed discussion of the objects in our sample with large $\lya$ EWs is presented in H17.

Information is also encoded in the spectral shape of the $\lya$ line. 
Looking at our sample, most of our $\lya$ spectra show a single asymmetric line, as expected for $\lya$ scattering processes in outflowing media \citep{V06,DK12,Y16}. 
Outflows facilitate the escape of $\lya$ photons emitted in star-forming regions from the ISM \citep{D06,V12,Be14a,Be14b} and can therefore be responsible for the observed $\lya$ haloes. Some observational evidence has been found supporting this scenario in local galaxies (e.g. \citealt{Bi15,He16}).
However, the $\lya$ spectra do not indicate where the $\lya$ photons are produced. Indeed, they can be produced either in HII regions well within the galaxy and then scatter in the CGM or ISM (some very compact LAEs do show asymmetric $\lya$ line profiles) or  in the CGM and still scatter within the CGM producing asymmetric lines \citep{C05}.
Analyses of spatially resolved spectra along with $\lya$ transfer simulations should however be able to help disentangle between the different effects.

If the observed $\lya$ haloes are powered by $\lya$ radiation produced inside the galaxies and scattered outwards, we expect the spatial and spectral properties of these haloes to correlate (\citeauthor{Ve17}, in prep.). 
In particular, the halo flux fraction is predicted to increase with the spectral shift of the peak and the FWHM of the $\lya$ line. 
Figure~\ref{ewfwhmhff} however does not show such a trend.

Hence, our results indicate that the scattering of $\lya$ photons created in HII regions can contribute to the powering of $\lya$ haloes but it is difficult to quantify their contribution.

\subsubsection{Gravitational cooling radiation}
\label{sec:cool}

Another scenario to explain the extended $\lya$ emission is the so-called "cooling radiation" (e.g. \citealt{H00,fardal01,Fur05}). 
In this process, $\lya$ photons are emitted by collisionally excited circum-galactic gas, which converts gravitational energy into kinetic and thermal energy as it falls into the DM halo potential. 
Cooling radiation has been postulated to come into play at large radii, where the $\lya$ photons are less likely to be absorbed by dust \citep{F01,fardal01}. 
However, because the density is higher at the centre of the DM halo and the $\lya$ emissivity resulting from cooling is proportional to the density squared if the gas is warm enough, the $\lya$ emission is expected to be centrally concentrated \cite[as found in][]{RB12}. 
The bulk of the $\lya$ radiation that we would observe from such a geometry would therefore have scattered outwards through an infalling scattering medium. 

A number of other theoretical analyses have been carried out to predict the expected cooling radiation contribution \citep{Fur05,DL09,FG10,RB12}. 
Such numerical simulations are nevertheless difficult to perform as they require high resolution and expensive radiative transfer treatments. 
Recently, \cite{L15} performed hydrodynamic and radiative transfer simulations of LAEs and found that star formation accounts for the origin of the majority of diffuse $\lya$ emission but that cooling radiation can also have a significant contribution; i.e. 40-55\% of the total $\lya$ luminosity within distances up to the LAE virial radius.

According to theoretical predictions \citep{D06} and confirmed by numerical experiments \citep{T16}, for $\lya$ radiative transfer in a cooling gas, the resulting $\lya$ line is expected to be blueshifted with a blue tail with respect to the centre of the line. The effect of IGM absorption, however, even at the redshifts considered here, can have a significant impact on the blue side of the $\lya$ line profile \citep{L11}.
Most of our objects do not show any of the non-resonant lines needed to determine the systemic redshift of the galaxy and so to precisely measure such a blue shift. 
However, looking at our sample, none of our LAEs shows a clear single-peaked asymmetric line towards the blue. 
The predicted blueshifted feature could be manifested as the blue bumps observed for 10\% of our LAEs. 
Such lines indeed show a shifted blue peak and an enhanced red peak, suggesting that cooling radiation cannot fully account for the shape of the $\lya$ lines. 
However, some theoretical predictions are obtained by averaging over all directions, which is of course not the case for observed spectra. Moreover, the line of sight can strongly impact the line profiles \citep{V12,GD14}. As such, the spectral features of the $\lya$ line should be interpreted with caution here.

\begin{figure}
\resizebox{\hsize}{!}{\includegraphics{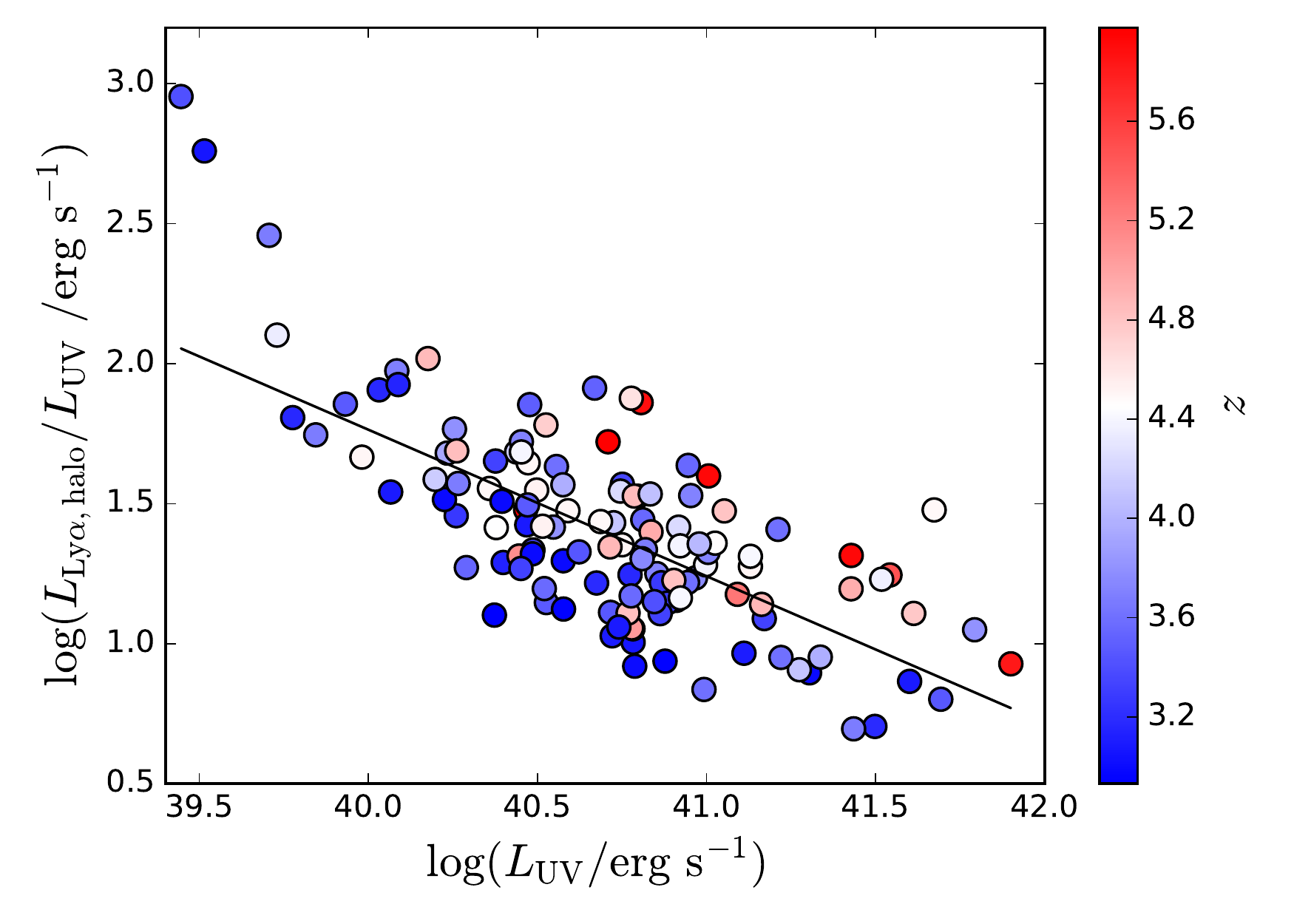}}
\caption{$\lya$ halo/UV luminosity ratios plotted against the UV luminosities at 1500$\AA$ colour coded by the redshift. The black solid line indicates a robust linear fit with a power-law exponent $-0.52\pm0.05$ leading to the relation $L_{\rm Lya,halo}$ $\propto$ $L_{\rm UV}^{0.45}$.}
\label{luvlya}
\end{figure}

Alongside $\lya$ spectral properties, $\lya$ luminosities also provide crucial information. Both \cite{RB12} and \cite{DL09} predict the $\lya$ luminosity produced by cooling radiation in a 10$^{11}$ M$_{\odot}$ DM halo to be 5$\times$10$^{41}$ erg s$^{-1}$. 
As our $\lya$ halo luminosities ($L_{\rm Lya,halo}$) are higher, this suggests that cooling is not the only process producing the $\lya$ halo emission or that the bulk of our LAEs reside in DM haloes more massive than $\approx 10^{11} \rm M_{\odot}$. This latter option is unlikely, however. 
In Figure~\ref{luvlya}, we plot the UV luminosity ($L_{\rm UV}$) to $\lya$ halo luminosity ratio as a function of $L_{\rm UV}$ for the sources in our sample\footnote{We plot here the UV/$\lya$ luminosity ratios on the y-axis to get rid of the luminosity distance on one of the axes. This ensures that the correlation is not artificially created by the redshift.}.
First, the anti-correlation between $L_{\rm Ly\alpha, halo}$/$L_{\rm UV}$ and $L_{\rm UV}$ shows that the halo component contributes more in UV-faint galaxies than in brighter UV sources. 
Interestingly, this trend may actually reflect the so-called "Ando effect", i.e. the fact that faint M$_{\rm UV}$ objects appear to have large $\lya$ EW\footnote{This is because most of the $\lya$ flux ($\approx$70\%) comes from the halo (see section~\ref{sec:fluxfrac})} , which is commonly observed at high redshift (\citealt{A06}, H17). 
Second, we perform a robust linear fit to the data and we measure a slope of $-0.52\pm0.05$ for the $L_{\rm Ly\alpha, halo}$/$L_{\rm UV}$ versus $L_{\rm UV}$ relation. 
This corresponds to $L_{\rm Ly\alpha, halo}$~$\propto$~$L_{\rm UV}^{0.45}$ and denotes that UV bright galaxies in our sample have more luminous $L_{\rm Ly\alpha, halo}$ haloes. 
This result can be directly compared to the predictions of \cite{RB12} who only considered cooling radiation at  $z$=3. These authors found a slope of 0.625 (if we assume UV luminosity proportional to the square of the DM halo mass), which is different from our result but not so dissimilar. Hence we cannot rule out this scenario. 
In the "scattering from HII regions" scenario, one would expect that the flux in the $\lya$ halo would globally scale with the number of $\lya$ photons that escape the galaxy, i.e. L$_{\rm UV}$ times the $\lya$ escape fraction from the ISM. We should therefore observe L$_{\rm Lya,halo}$ $\propto$ L$_{\rm UV}$; if all galaxies have the same $\lya$ escape fraction and if there are more neutral hydrogen atoms in the CGM than $\lya$ photons from the galaxies. It is worth noting however that varying dust content or ISM column density could strongly affect this relation by introducing a UV magnitude-dependent $\lya$ escape fraction.

Put together, our analysis does not allow us to give a firm conclusion about the contribution of cooling radiation in the production of $\lya$ haloes.

\subsubsection{$\lya$ fluorescence}

Another possible origin for the $\lya$ haloes is the $\lya$ fluorescence resulting from the recombinations of hydrogen that is photo-ionized by Lyman continuum (LyC) radiation generated by nearby quasars, young stars, or by the cosmic UV background (UVB) \citep{Fur05,C05,K10}.
This scenario is usually invoked for giant $\lya$ nebulae, within which quasars are known to reside (\citealt{C14,Bo16} and see \citealt{C17} for a review), as well as for compact dark galaxy sources (\citealt{C12,F16}; \citealt{M17}). 

According to the predictions of \cite{HM96} and \cite{C05}, the resulting $\lya$ SB produced by the diffuse ionizing background is significantly lower ($\sim$10$^{-20}$ erg s$^{-1}$ cm$^{-2}$ arcsec$^{-2}$ at $z$ $\approx$3). 
The expected effects of the UVB therefore appear to be negligible for the individual objects of our study.

According to the calculations of \citeauthor{G17} (2017), which uses the same MUSE UDF data as our study, the required LyC escape fraction from the ISM for stars to produce the observed $\lya$ halo in their stack of LAE pairs (SB of $\sim3\times10^{-20}$ erg s$^{-1}$ cm$^{-2}$ arcsec$^{-2}$) is extremely small ($f_{esc}\approx0.02$). This result suggests that it is not so difficult to have a large ionized fraction of gas in the inner parts of the haloes; the high gas densities and clumpiness of the medium moreover favour the $\lya$ fluorescence.

Consequently, we cannot rule out the contribution of the $\lya$ fluorescence for the powering of our observed $\lya$ haloes.

\subsubsection{Satellite galaxies}

\cite{Mo16} proposed another explanation where the $\lya$ halo would be powered by several satellite galaxies emitting $\lya$ emission around the central galaxy. 
\cite{SU10} and \cite{L15} have shown using cosmological simulations that $\lya$ haloes are indeed associated with such surrounding galaxies. 
Recently, \cite{MR17} applied an analytic formalism \citep{MR16} to investigate the plausibility of this scenario by using various satellite clustering conditions. 
These authors found that satellite sources can indeed play a role in the powering of $\lya$ haloes at large distances (20$\lesssim$ $r$ $\lesssim$ 40 physical kpc) from the galaxies. According to their modelling, such satellite galaxies would be very faint in the UV continuum (M$_{\rm UV}$> -17) so that they would be undetectable by any current instruments and may therefore be missed in current surveys.

Applied to the case of our data, we can expect that in the presence of satellites, which emit $\lya$ emission and are undetected in UV in the HST images, our $\lya$ haloes would appear clumpy and rather asymmetric. We do not observe such clumpiness in the central regions of our $\lya$ NB images. However the MUSE PSF acts to smooth out $\lya$ clumps, making their detection impossible.
Furthermore, at larger radii the S/N of the $\lya$ NB image drops significantly, making the detection of clumps or asymmetries very challenging. 
Moreover, a large contribution from the star formation in satellites is expected to provide similar UV and $\lya$ extended emission. Our measured UV continuum scale lengths however appear much smaller than the $\lya$ scale lengths (see section~\ref{sec:sizecomp}). In the presence of $\lya$-emitting satellites we might also expect the $\lya$ haloes would be offset from the host galaxy. Such an offset is not observed for most of our objects.

Put together with all these elements, it seems somewhat unlikely that there is a significant contribution from satellite galaxies to the powering of $\lya$ haloes. We cannot however completely rule out the possibility that unidentified satellites partly power the $\lya$ haloes.

\subsubsection{Future directions}

In summary, our results are suggestive of a scenario, in which the following range of processes can be responsible for the observed $\lya$ haloes: 
\begin{itemize}
\item
The scattering on CGM scales of $\lya$ photons that are produced by recombinations in HII regions.
\item
The cooling radiation triggered by gas inflowing onto the host galaxies.
\item
The $\lya$ fluorescence associated with hydrogen recombinations after ionization by Lyman continuum (LyC) radiation present in the CGM.
\end{itemize}
While those processes have to be considered together, their respective contributions cannot be constrained by our data. 
To try to disentangle the relative impact of the different processes, we need to know where the $\lya$ photons are produced, which is not straightforward because $\lya$ is a resonant line. 
As such, more observations are needed. In particular, H${\alpha}$ observations by the James Web Space Telescope (JWST) will directly tell us the origin of the $\lya$ emission as it is not a resonant line. 
H${\alpha}$ emission that is more extended than the UV continuum would be a direct piece of evidence that the $\lya$ emission is produced in the CGM (i.e. by fluorescence). On the other hand, compact H${\alpha}$ emission would indicate that $\lya$ photons are produced in the ISM and then propagate in the CGM by resonant scattering.
The adaptive optics (AO) on MUSE, currently in commissioning, also promises good progress as it will significantly improve the PSF and therefore enable the detection of smaller haloes and allow a precise characterization of the halo morphologies. Finally, the help of theoretical and numerical studies will be needed to fully understand the processes at play and their respective contributions.


\section{Summary and conclusions}
\label{sec8}

Thanks to the significant increase in sensitivity enabled by the MUSE instrument, we studied the CGM gas content of an unprecedentedly large sample of individual star-forming galaxies at redshift $z$=[3-6] in the Hubble Ultra Deep Field. Our LAE sample was selected to have a good S/N (>6) and to be isolated (see section~\ref{sec:sample}).
Our galaxy-by-galaxy based analysis allows us to characterize individual $\lya$ halo properties and to explore possible correlations with the UV properties of the host galaxies. Our major results are summarized as follows:
\begin{enumerate}
\item
We detect diffuse $\lya$ emission with high confidence around 145 individual LAEs. This represents 80\% of our objects for which we have reliable $\lya$ halo measurements. 
Among the objects for which we have reliable $\lya$ halo scale length measurements, 20 do not show a significant $\lya$ halo detection, mainly owing to large errors on their halo size measurement (see section~\ref{sec:statsigni}).
Put together, our data suggest that extended $\lya$ haloes are common around $\lya$ emitters at high redshift.
\item 
We find a large range of $\lya$ halo scale lengths, emphasizing the diversity of configurations of the cool CGM. 
The halo scale lengths in our sample range from 1.0 to 18.7 kpc with a median value of $\approx$4-5 kpc.
We also show that the $\lya$ emission probes the CGM out to large radii (Figure~\ref{rvir}), reaching on average $\approx$50\% of the virial radius according to the comparison of our data with predictions from a semi-analytic model.
This result shows that $\lya$ emission is a powerful tool to map the cold hydrogen  around high-redshift galaxies. 
\item
The $\lya$ haloes properties of our selected sample of LAEs appear to be dependent on the stellar content of the galaxies. 
Both $\lya$ halo spatial extents and fluxes are found to be positively correlated with UV magnitudes and spatial extents of the host galaxies, although the correlation with UV magnitude is not as clear.
\item 
While $\lya$ halo scale lengths appear to be considerably larger at $z$>3 than at $z\simeq$0 (from a comparison with the LARS sample), we do not observe any significant evolution of the $\lya$ scale lengths between redshift 3 and 6. 
This implies an evolution of the CGM content between $z\simeq$0 and $z$=3.
\item
The galaxies that are less likely to have a $\lya$ halo as well as those with small haloes ($rs_{\rm halo}$ $\lesssim$1 kpc, i.e. objects with upper limits on the halo size), have on average narrower $\lya$ lines than the rest of the sample (Fig~\ref{rhfwhm}). 
This suggests that the $\lya$ line is less broadened when the gas content in the CGM is low. However, it is worth noting that $\lya$ line of galaxies with a high confidence $\lya$ halo are not systematically broader.
\item 
With the information from our data we attempt to explore the origin of the $\lya$ haloes around star-forming galaxies. 
While we find no evidence for a dominant contribution from a single particular process, we are not able to rule out any of the scenarios we consider, i.e. scattering from star-forming regions, fluorescence, and cooling radiation from cold gas accretion, except maybe the scenario for which satellites significantly contribute. Indeed, while we do not find significant evidence for the "satellite scenario", our data cannot disentangle whether the $\lya$ photons are produced in the star-forming regions and then scatter in the CGM or "in-situ" in the CGM from gravitational cooling radiation and/or from fluorescence.
As a consequence, further observations and analysis will be needed to understand the powering process(es) of the $\lya$ haloes (JWST, MUSE with AO, and theoretical and numerical analyses). 
\end{enumerate}
The MUSE instrument has enabled us to extend the sample of measurements of $\lya$ haloes to fainter and smaller galaxies, which are more representative of the bulk of the galaxy population. 
Our study underlines the significant cold gas content of the Universe between redshifts 3 and 6, regardless of the nature of the $\lya$ halo emission mechanism.

This new study highlights that $\lya$ emission presents an exciting new opportunity to study the diffuse and low-SB gas in the vicinity of faint high-redshift galaxies.
In the coming years, adaptive optics mounted on MUSE/VLT will allow us to be even more precise in the detection and characterization of these $\lya$ haloes. 
By improving the PSF, it will be possible to detect smaller $\lya$ haloes and thus confirm if there are LAEs without any halo component. 
Within the context of the substantial recent progress in improving numerical simulations and theoretical models, detailed comparisons of models and observations of $\lya$ haloes around normal star-forming galaxies are now possible and promise to significantly expand our understanding of the mechanisms that regulate the gas that flows in and out of galaxies in the early Universe.


\begin{acknowledgements}
FL, RB, SC, HI, and MA acknowledge support from the ERC advanced grant 339659-MUSICOS. TG is grateful to the LABEX Lyon Institute of Origins (ANR-10-LABX-0066) of the Université de Lyon for its financial support within the programme "Investissements d'Avenir" (ANR-11-IDEX-0007) of the French government operated by the National Research Agency (ANR). RB, TC, and JR acknowledge support from the FOGHAR Project with ANR Grant ANR-13-BS05-0010. SC acknowledges support from Swiss National Science Foundation grant PP00P2 163824. TC acknowledges support from the OCEVU Labex (ANR-11-LABX-0060) and the A*MIDEX project (ANR-11-IDEX-0001-02) funded by the ``Investissements d'avenir'' French government programme managed by the ANR. JR acknowledges support from the ERC starting grant CALENDS. JS acknowledges support from ERC Grant agreement 278594-GasAroundGalaxies.
RAM acknowledges support by the Swiss National Science Foundation.
JB acknowledges support by Funda{\c c}{\~a}o para a Ci{\^e}ncia e a
Tecnologia (FCT) through national funds (UID/FIS/04434/2013) and Investigador FCT
contract IF/01654/2014/CP1215/CT0003., and by FEDER through COMPETE2020 (POCI-01-0145-FEDER-007672).
\end{acknowledgements}


\bibliographystyle{aa} 
\bibliography{biblio}

\begin{appendix}

\onecolumn

\section{Effect of the exposure time on the $\lya$ halo detection}
\label{sec:limdetec_both}
\label{app1}

Our sample is constructed from two overlapping data sets with various exposure times, where the \udft\ data is 3 times deeper on average than the \mosaic\ data. We are therefore able to investigate how our $\lya$ halo size measurements vary with depth. 

\begin{figure}[h]
\centering
\includegraphics[scale=0.6]{{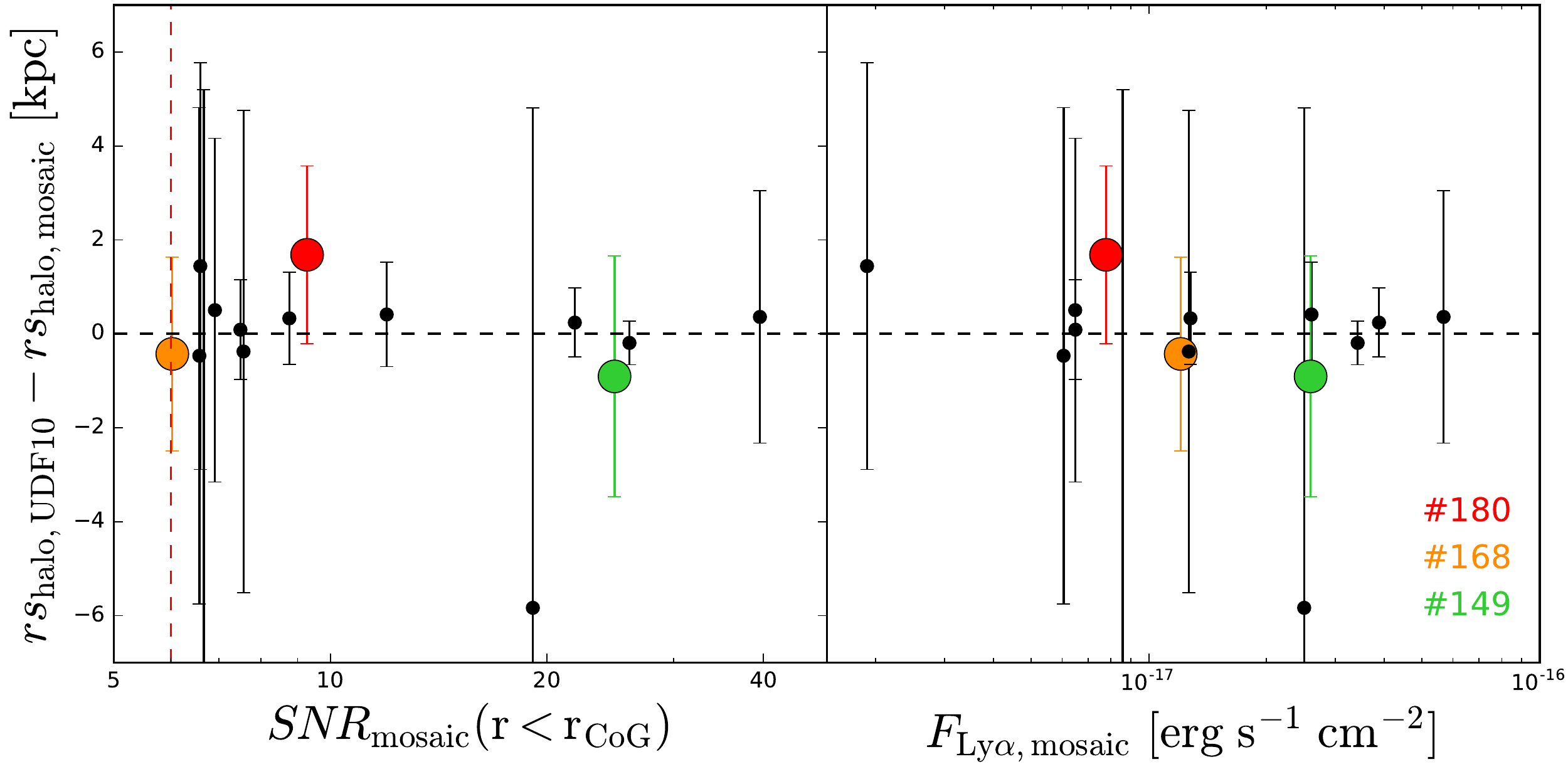}}
\caption{Comparison of the best-fit $\lya$ halo scale lengths of common objects from the deeper \udft\ data cube ($\approx$30 hours) and the shallower \mosaic\ data cube (10 hours) with reliable halo measurements (15 objects, see section~\ref{sec:rhlim}). \textit{Left}: Difference between the $\lya$ halo scale lengths, plotted as a function of the $\lya$ S/N from the \mosaic\ data cube. 
The S/N is calculated in a fixed aperture corresponding to the CoG radius $r_{\rm CoG}$ (see section~\ref{sec:fluxmes}). 
The median difference is $\lesssim$0.1 kpc. 
The dashed red line shows the S/N cut of 6 imposed on our sample (see section~\ref{sec:sample}). 
\textit{Right}: Difference between scale lengths, plotted against the total $\lya$ flux measured in the \mosaic\ data cube. The red, orange, and green points correspond to the objects shown as examples in Figure~\ref{ex_compudf10mosa}. The corresponding MUSE IDs are indicated.}
\label{compudf10mosa}
\end{figure}

Within our sample, 26 objects are both detected in the \udft\ and \mosaic\ fields but only 15 have a reliable $\lya$ halo measurement (see section~\ref{sec:rhlim}).
Figure \ref{compudf10mosa} shows the difference of the $\lya$ halo scale length measurements against the S/N (left panel) and the total $\lya$ flux (right panel) of the $\lya$ NB image constructed from the \mosaic\ data cube for these 15 objects. 
The median difference is small (<0.1 kpc) and the error bars encompass the measured offsets for every object.

\begin{figure*}
   \centerline{\includegraphics[scale=0.4]{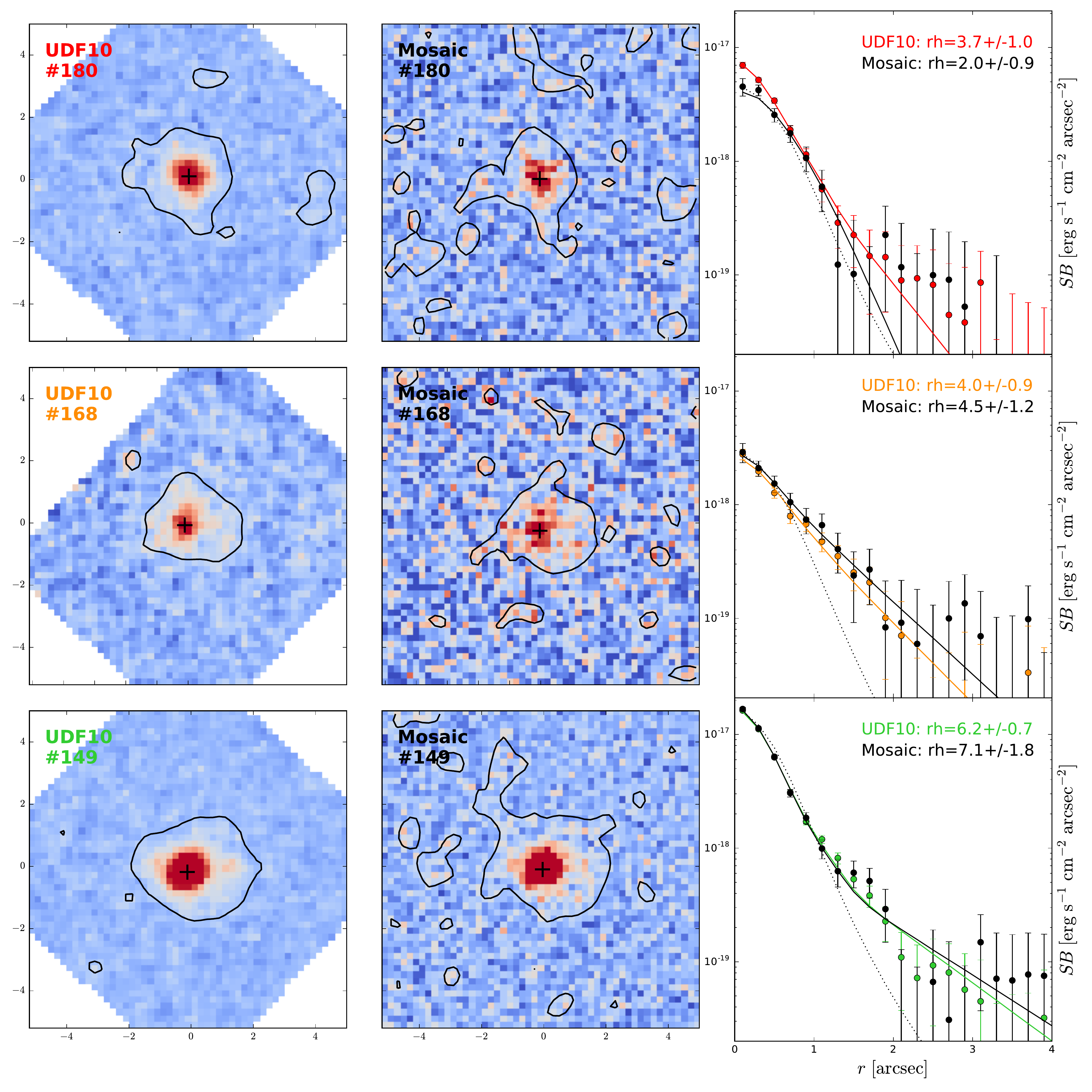}}
     \caption{Comparison of the $\lya$ halo detection for 3 representative objects from the deeper \udft\ data cube and the shallower \mosaic\ data cube.
     From top to bottom rows show objects \#180, \#168, and \#149, respectively. 
     \textit{Left and middle panels}: $\lya$ NB images constructed from the \udft\ and \mosaic\ data cubes, respectively. 
     The black contours show the $10^{-18.5}$ $\sbl$ SB level for the two images. 
     \textit{Right panels}: Comparison of the radial SB profiles measured on the $\lya$ NB images (data points) and from the modelled $\lya$ images (lines) in the \udft\ data cube (in colour) and the \mosaic\ data cube (in black). 
     The black dotted line shows the rescaled radial SB of the UV continuum. 
     The best-fit halo scale lengths from the two data cubes are indicated in the upper right corner of each panel.}
     \label{ex_compudf10mosa}
\end{figure*}

Figure \ref{ex_compudf10mosa} shows a more detailed comparison of the two data cubes for three representative objects. 
The S/N of $\lya$ NB images from the \udft\ is on average larger by a factor $\approx$2 compared to the \mosaic. This is consistent with the noise analysis of the UDF MUSE data cubes given in B17.
This is also clearly highlighted by the contours, which are clearer and less splintered in the \udft\ images. 
The NB images are optimized in terms of S/N from both data cubes. Hence, the spectral bandwidths are not always the same in the two data cubes for a given object. As a consequence the NB images and therefore the $\lya$ centroid measurement can be slightly different explaining that the SB profiles do not perfectly overlap in the inner region. 


The first object (\#180, top row) shows a discrepancy in the halo scale length measurements between the two data cubes (object indicated as red point in Figure~\ref{compudf10mosa}). 
Visual inspection of radial SB profiles (right panel) shows that the $\lya$ halo of this object is lost in the noise for the \mosaic\ data cube. 
We apply a S/N cut of 6 to define our sample (see section \ref{sec:snrlim}) 
The S/N of the object we are showing here is higher (9.3) than the S/N cut (6) and this poses questions about the reliability of our S/N cut. 

The next object \#168 (middle row of Figure~\ref{ex_compudf10mosa}, orange dot in Figure~\ref{compudf10mosa}) offers a counter example. 
While the S/N of the \mosaic\ $\lya$ NB image is lower than the previous example and very close to our S/N cut value (6.1), the scale lengths measured in the two data cubes are similar for this example. 
These two examples highlight that the S/N cut we define using simulated extended objects is not absolute and that for S/N<10, halo sizes can be underestimated in some cases. This uncertainty is however encompassed in the error bar.

The last example (\#149, last row of Figure~\ref{ex_compudf10mosa} and green dots in Figure~\ref{compudf10mosa}) shows the comparison of the detection for an object with a good S/N in the two data cubes. 
Reassuringly, the scale lengths measured in the NB images are similar with a larger error on the scale length fit to the \mosaic\ data.
\newline\\
Put together, this illustrates the importance of surface brightness sensitivity for the detection of extended $\lya$ emission.

\newpage

\section{Table of $\lya$ halo measurements}
\label{ap:table}

\begin{center}
\begin{small}
\begin{landscape}
\begin{longtable}{rclllccrlrrlcrrcc}
\caption{Table containing all the measurements from our analysis. ID: source identifier in the catalog by I17. F: Field of the source (1 for \udft\ and 2 for \mosaic\ ). $\alpha_{\rm 2000}$ and $\delta_{\rm 2000}$: coordinates in I17. $z$: redshift in I17. $m$: Continuum AB magnitude in the HST filter band used for the study (see section~\ref{sec:contim}) taken from \cite{R15} and I17. $M_{\rm UV}$: Absolute far-UV magnitude. $F_{\rm Ly\alpha}$: Total $\lya$ flux in 10$^{-18}$ erg s$^{-1}$ cm$^{-2}$, integrated over an aperture of radius $r_{\rm CoG}$ determined using the CoG method. $\rm log_{10}$ $L_{\rm Ly\alpha}$: logarithm of the $\lya$ luminosity in erg s$^{-1}$.  $EW_{\rm Ly\alpha}$: Total $\lya$ rest-frame equivalent width in $\AA$. $r_{\rm CoG}$: CoG method radius in physical kpc. $rs_{\rm cont}$: Exponential scale length of the UV continuum in physical kpc; upper limits are given if the object is considered as a point source in HST. $rs_{\rm halo}$: Exponential scale length of the $\lya$ halo from the decomposition into two components in physical kpc. $F_{\rm cont}$: Integrated flux of the ‘continuum-like’ $\lya$ component, in 10$^{-18}$ erg s$^{-1}$ cm$^{-2}$. $F_{\rm halo}$: Integrated flux of the $\lya$ halo, in 10$^{-18}$ erg s$^{-1}$ cm$^{-2}$. $p_{\rm 0}$: Probability of the two scale lengths (galaxy and $\lya$ halo) to be identical by considering a normal error distribution. $1-p_{\rm 0}$ is the confidence level that the scale lengths are not identical. $FWHM_{\rm 0}$: rest-frame full width at half maximum of the $\lya$ line in km~s$^{-1}$. An electronic version of the table is available at http://muse-vlt.eu/science/udf/}. \\
\hline \hline
$\rm ID$ & $\rm F$ & $\alpha_{\rm 2000}$ & $\delta_{\rm 2000}$ & $z$ & $m$ & $-M_{\rm UV}$ & $F_{\rm Ly\alpha}$ & $\rm log_{10} \ \it L_{\rm Ly\alpha}$ & $EW_{\rm Ly\alpha}$ & $r_{\rm CoG}$ & $rs_{\rm cont}$ & $rs_{\rm halo}$ & $F_{\rm cont}$ & $F_{\rm halo}$ & $ 1-p_{\rm 0}$ & $\rm FWHM_{\rm 0}$ \\ $ $ & $ $ & $\rm [ ^{\circ}]$ & $\rm [ ^{\circ}]$ & $ $ & $\rm [AB]$ & $\rm [AB]$ & $\rm [10^{-18}~cgs]$ & $\rm [erg~s^{-1}]$ &  $\rm [\AA]$ & $\rm [kpc]$ & $\rm [kpc]$ & $\rm [kpc]$ & $\rm [10^{-18}~cgs]$ & $\rm [10^{-18}~cgs]$ & $ $ & $\rm [km~s^{-1}]$ \\
\hline \\
\endfirsthead
\hline \hline
$\rm ID$ & $\rm F$ & $\alpha_{\rm 2000}$ & $\delta_{\rm 2000}$ & $z$ & $m$ & $-M_{\rm UV}$ & $F_{\rm Ly\alpha}$ & $\rm log_{10} \ \it L_{\rm Ly\alpha}$ & $EW_{\rm Ly\alpha}$ & $r_{\rm CoG}$ & $rs_{\rm cont}$ & $rs_{\rm halo}$ & $F_{\rm cont}$ & $F_{\rm halo}$ & $ 1-p_{\rm 0}$ & $\rm FWHM_{\rm 0}$ \\ $ $ & $ $ & $\rm [ ^{\circ}]$ & $\rm [ ^{\circ}]$ & $ $ & $\rm [AB]$ & $\rm [AB]$ & $\rm [10^{-18}~cgs]$ & $\rm [erg~s^{-1}]$ &  $\rm [\AA]$ & $\rm [kpc]$ & $\rm [kpc]$ & $\rm [kpc]$ & $\rm [10^{-18}~cgs]$ & $\rm [10^{-18}~cgs]$ & $ $ & $\rm [km~s^{-1}]$ \\
\hline \\
\endhead
53 & 1 & 53.158115 & -27.786372 & 4.78 & 25.3 & 21.07 & 60.8+/-0.8 & 43.17 & 62.1 & 21.7 & 0.95+/-0 & 3.31+/-1.43 & 25.4+/-4.1 & 21.4+/-9.1 & 0.9509 & 313 \\
68 & 2 & 53.171186 & -27.778461 & 4.94 & 25.8 & 20.64 & 53.9+/-1.3 & 43.16 & 77.2 & 18.8 & 1.05+/-0.01 & 6.69+/-0.84 & 33.7+/-0.5 & 15.8+/-1.6 & 1.0 & 291 \\
82 & 2 & 53.151551 & -27.785348 & 3.61 & 26.1 & 19.82 & 44.7+/-3.0 & 42.75 & 85.0 & 33.4 & 0.62+/-0.01 & 7.00+/-0.58 & 8.7+/-0.6 & 33.0+/-2.3 & 1.0 & 454 \\
106 & 1 & 53.163726 & -27.779076 & 3.28 & 26.5 & 19.22 & 37.6+/-2.5 & 42.44 & 101.3 & 36.8 & 0.87+/-0.02 & 3.13+/-0.48 & 9.3+/-2.3 & 22.0+/-2.0 & 1.0 & 283 \\
109 & 1 & 53.162561 & -27.772861 & 3.09 & -- & -- & 36.5+/-1.1 & 42.5 & -- & 19.5 & 0.80+/-0.02 & 2.91+/-0.29 & 11.3+/-1.4 & 19.0+/-1.2 & 1.0 & 309 \\
148 & 1 & 53.167601 & -27.774719 & 3.07 & 27.0 & 18.62 & 16.1+/-1.7 & 42.14 & 69.6 & 30.5 & 0.48+/-0.01 & 9.39+/-2.52 & 5.8+/-0.4 & 7.2+/-1.3 & 0.99979 & 362 \\
149 & 1 & 53.167866 & -27.778617 & 3.72 & 27.0 & 18.95 & 23.0+/-0.9 & 42.49 & 87.7 & 24.2 & 0.51+/-0.02 & 6.16+/-0.74 & 13.5+/-0.3 & 9.3+/-0.8 & 1.0 & 219 \\
153 & 2 & 53.168072 & -27.775265 & 3.55 & 27.1 & 18.81 & 23.3+/-1.7 & 42.45 & 108.4 & 23.1 & 0.45+/-0.02 & 3.40+/-0.66 & 7.2+/-1.4 & 14.8+/-1.7 & 1.0 & 413 \\
168 & 1 & 53.1702 & -27.77732 & 4.7 & 27.0 & 19.3 & 5.7+/-0.4 & 42.13 & 28.5 & 15.2 & 0.84+/-0.02 & 4.03+/-0.88 & 1.5+/-0.4 & 4.9+/-0.7 & 0.99985 & 307 \\
171 & 1 & 53.166029 & -27.782774 & 3.89 & 27.2 & 18.82 & 8.9+/-0.9 & 42.13 & 44.7 & 18.0 & 0.89+/-0.05 & 4.79+/-1.12 & 2.9+/-0.5 & 4.5+/-0.8 & 0.99975 & 335 \\
180 & 1 & 53.16396 & -27.779688 & 3.46 & 27.3 & 18.55 & 13.7+/-1.1 & 42.19 & 77.0 & 24.8 & 0.65+/-0.02 & 3.69+/-0.96 & 5.1+/-0.7 & 5.9+/-0.8 & 0.99923 & 437 \\
183 & 1 & 53.162393 & -27.771601 & 3.19 & 27.3 & 18.38 & 13.2+/-0.9 & 42.09 & 62.2 & 16.3 & 0.38+/-0.01 & 2.21+/-0.42 & 2.4+/-1.1 & 8.3+/-1.0 & 0.99999 & 257 \\
218 & 1 & 53.153465 & -27.776998 & 3.05 & 27.6 & 17.96 & 6.5+/-0.7 & 41.74 & 47.0 & 11.8 & 0.51+/-0.02 & 2.18+/-3.14 & 3.1+/-1.0 & 1.5+/-1.3 & 0.70282 & 278 \\
237 & 1 & 53.164516 & -27.77388 & 4.82 & 27.4 & 18.97 & 6.5+/-0.5 & 42.21 & 37.9 & 21.6 & 0.26+/-0.01 & 2.02+/-0.43 & 1.5+/-0.6 & 3.0+/-0.5 & 0.99997 & 237 \\
242 & 1 & 53.157457 & -27.77556 & 3.76 & 27.8 & 18.12 & 3.7+/-0.5 & 41.71 & 34.9 & 12.4 & 0.47+/-0.03 & < 1.23 & 0.2+/-1.1 & 3.4+/-0.8 & -- & 393 \\
279 & 1 & 53.159155 & -27.784108 & 3.61 & 28.1 & 17.82 & 3.2+/-0.4 & 41.6 & 36.3 & 11.1 & 0.23+/-0.02 & 2.12+/-2.09 & 1.3+/-1.0 & 1.9+/-1.1 & 0.81649 & 209 \\
324 & 2 & 53.170996 & -27.781188 & 3.0 & 28.4 & 17.2 & 14.3+/-1.8 & 42.06 & 219.8 & 21.3 & 0.56+/-0.06 & 3.72+/-1.22 & 3.1+/-1.3 & 6.8+/-1.4 & 0.99535 & 327 \\
364 & 1 & 53.15384 & -27.779243 & 3.94 & 28.6 & 17.45 & 7.4+/-0.9 & 42.06 & 195.3 & 29.4 & 0.49+/-0.04 & 8.91+/-2.43 & 1.7+/-0.2 & 5.3+/-1.0 & 0.99974 & 242 \\
385 & 1 & 53.154241 & -27.77493 & 4.77 & 28.7 & 17.69 & 2.7+/-0.4 & 41.81 & 41.9 & 15.2 & < 0.05 & < 1.04 & 0.0+/-0 & 2.8+/-0.2 & -- & 175 \\
400 & 1 & 53.163176 & -27.779113 & 3.09 & 28.8 & 16.83 & 8.5+/-1.2 & 41.87 & 180.5 & 24.2 & 0.20+/-0.06 & 2.73+/-0.93 & 2.5+/-1.3 & 4.7+/-1.3 & 0.99655 & 313 \\
417 & 1 & 53.157861 & -27.78 & 5.13 & 28.3 & 18.22 & 3.1+/-0.3 & 41.96 & 37.6 & 9.5 & 0.24+/-0.02 & 2.22+/-0.68 & 1.0+/-0.3 & 2.0+/-0.4 & 0.9982 & 183 \\
547 & 1 & 53.160608 & -27.771537 & 5.98 & 27.7 & 19.02 & 14.3+/-1.3 & 42.77 & 150.7 & 25.2 & 0.32+/-0.02 & 6.61+/-1.23 & 4.5+/-0.2 & 6.5+/-0.9 & 1.0 & 216 \\
559 & 1 & 53.153576 & -27.776791 & 4.51 & 29.3 & 16.94 & 2.4+/-0.4 & 41.71 & 115.3 & 15.5 & 0.38+/-0.03 & 2.53+/-0.85 & 0.4+/-0.3 & 2.1+/-0.5 & 0.99439 & 315 \\
590 & 1 & 53.170112 & -27.781033 & 3.67 & 29.5 & 16.42 & 4.1+/-0.7 & 41.73 & 154.2 & 15.5 & 0.11+/-0.08 & 7.66+/-4.23 & 2.4+/-0.3 & 3.0+/-1.3 & 0.96305 & 222 \\
605 & 1 & 53.156764 & -27.779805 & 3.44 & 29.6 & 16.24 & 5.0+/-0.5 & 41.75 & 224.6 & 14.3 & < 0.03 & < 1.29 & 0.0+/-0 & 3.8+/-0.2 & -- & 203 \\
666 & 1 & 53.162908 & -27.785182 & 4.67 & 29.7 & 16.63 & 2.8+/-0.3 & 41.81 & 195.7 & 12.6 & < 0.05 & < 1.06 & 0.0+/-0 & 2.2+/-0.1 & -- & 148 \\
837 & 1 & 53.157037 & -27.772649 & 5.98 & 28.3 & 18.41 & 2.5+/-0.4 & 42.01 & 39.2 & 11.1 & 0.20+/-0.02 & 1.48+/-0.57 & 0.3+/-0.6 & 2.1+/-0.5 & 0.987 & 192 \\
1059 & 2 & 53.153442 & -27.766119 & 3.81 & 24.7 & 21.32 & 86.4+/-3.3 & 43.09 & 52.8 & 35.6 & 0.47+/-0 & 6.87+/-0.45 & 31.1+/-0.7 & 48.5+/-2.7 & 1.0 & 403 \\
1087 & 2 & 53.1679 & -27.797953 & 3.46 & 24.8 & 20.99 & 40.7+/-3.8 & 42.54 & 26.7 & 30.9 & 0.43+/-0 & 6.51+/-1.20 & 10.3+/-1.1 & 27.3+/-3.4 & 1.0 & 281 \\
1113 & 2 & 53.169939 & -27.76834 & 3.09 & 25.0 & 20.67 & 54.9+/-1.8 & 42.68 & 31.2 & 21.1 & 0.99+/-0.01 & 5.22+/-0.47 & 20.1+/-1.2 & 33.6+/-1.7 & 1.0 & 393 \\
1185 & 2 & 53.161971 & -27.819086 & 4.5 & 25.1 & 21.17 & 112.0+/-2.4 & 43.38 & 97.9 & 32.5 & 1.35+/-0 & 5.73+/-0.21 & 34.3+/-1.0 & 66.4+/-1.7 & 1.0 & 395 \\
1226 & 2 & 53.16223 & -27.815769 & 3.1 & -- & -- & 8.0+/-1.4 & 41.85 & -- & 17.9 & 1.10+/-0.01 & 10.61+/-3.60 & 1.2+/-0.9 & 11.6+/-2.8 & 0.99585 & 245 \\
1253 & 2 & 53.178249 & -27.773996 & 3.67 & 25.5 & 20.4 & 67.9+/-2.1 & 42.95 & 78.7 & 22.8 & 0.88+/-0.01 & 9.57+/-2.99 & 48.5+/-0.6 & 10.3+/-2.3 & 0.9982 & 284 \\
1281 & 2 & 53.163474 & -27.81384 & 2.99 & -- & -- & 20.6+/-2.5 & 42.22 & -- & 29.2 & 0.72+/-0.01 & 9.58+/-1.92 & 1.5+/-0.8 & 19.3+/-2.8 & 1.0 & 218 \\
1283 & 2 & 53.186448 & -27.792795 & 4.36 & 25.5 & 20.76 & 36.6+/-3.0 & 42.86 & 54.2 & 32.2 & 0.52+/-0 & 5.71+/-0.51 & 5.2+/-0.7 & 28.2+/-1.8 & 1.0 & -- \\
1343 & 2 & 53.142007 & -27.797409 & 3.97 & 25.8 & 20.22 & 15.2+/-2.2 & 42.38 & 26.4 & 26.4 & 0.57+/-0.01 & 5.91+/-1.49 & 1.6+/-0.8 & 12.3+/-2.3 & 0.99983 & 327 \\
1423 & 2 & 53.160865 & -27.801122 & 3.6 & 26.1 & 19.84 & 27.0+/-2.6 & 42.53 & 51.3 & 26.0 & 1.11+/-0.03 & 3.32+/-0.92 & 0.4+/-2.1 & 11.8+/-2.8 & 0.99198 & 284 \\
1445 & 2 & 53.146042 & -27.791589 & 3.33 & 26.1 & 19.66 & 33.0+/-2.3 & 42.54 & 58.5 & 20.6 & 0.45+/-0.01 & 3.56+/-0.59 & 6.7+/-1.4 & 17.4+/-1.4 & 1.0 & 399 \\
1446 & 2 & 53.125488 & -27.788223 & 3.24 & -- & -- & 15.3+/-1.7 & 42.17 & -- & 20.8 & 1.03+/-0.03 & 4.45+/-1.17 & 2.4+/-1.5 & 11.7+/-2.2 & 0.99829 & 388 \\
1525 & 2 & 53.164051 & -27.819038 & 4.03 & 25.8 & 20.31 & 12.4+/-1.3 & 42.31 & 42.1 & 17.8 & 1.58+/-0.01 & 4.70+/-2.33 & 6.2+/-1.6 & 5.8+/-2.7 & 0.90911 & 328 \\
1593 & 2 & 53.184386 & -27.791262 & 4.11 & 26.0 & 20.09 & 17.6+/-2.5 & 42.48 & 63.7 & 27.5 & 0.65+/-0 & 7.67+/-2.84 & 5.2+/-0.7 & 8.8+/-2.3 & 0.99336 & 222 \\
1670 & 2 & 53.166721 & -27.804163 & 5.83 & 24.7 & 21.97 & 29.5+/-2.3 & 43.06 & 20.4 & 14.9 & 0.49+/-0 & 3.09+/-0.52 & 8.3+/-1.4 & 17.2+/-2.0 & 1.0 & 318 \\
1711 & 2 & 53.141656 & -27.774938 & 3.77 & 26.7 & 19.25 & 21.2+/-1.7 & 42.47 & 71.8 & 22.6 & 0.59+/-0.02 & 5.90+/-1.42 & 10.2+/-0.7 & 11.4+/-1.8 & 0.99991 & 328 \\
1723 & 2 & 53.152995 & -27.806005 & 3.6 & 26.7 & 19.15 & 36.9+/-2.0 & 42.67 & 134.4 & 21.5 & 0.37+/-0.01 & 5.23+/-0.93 & 21.8+/-0.7 & 11.6+/-1.9 & 1.0 & 287 \\
1724 & 2 & 53.158366 & -27.81343 & 3.7 & 26.7 & 19.2 & 8.1+/-1.2 & 42.04 & 28.1 & 15.4 & 0.89+/-0.03 & 2.61+/-1.58 & 2.5+/-2.5 & 4.5+/-2.2 & 0.86202 & 305 \\
1726 & 2 & 53.178697 & -27.798836 & 3.71 & 26.7 & 19.2 & 45.6+/-2.7 & 42.79 & 145.9 & 31.6 & 0.25+/-0.01 & 5.41+/-0.59 & 16.1+/-0.6 & 22.5+/-1.8 & 1.0 & 213 \\
1737 & 2 & 53.137692 & -27.781268 & 4.88 & 26.4 & 19.97 & 10.4+/-1.2 & 42.43 & 33.4 & 17.6 & 0.71+/-0.01 & 3.62+/-1.03 & 4.1+/-0.9 & 7.8+/-1.3 & 0.99772 & 282 \\
1756 & 2 & 53.17816 & -27.790191 & 3.3 & 26.8 & 18.93 & 11.9+/-0.9 & 42.09 & 41.1 & 14.5 & 0.35+/-0.01 & 2.04+/-0.56 & 0.1+/-2.2 & 10.4+/-2.0 & 0.99869 & 358 \\
1761 & 2 & 53.168317 & -27.813246 & 4.03 & 26.8 & 19.34 & 14.8+/-1.5 & 42.39 & 60.0 & 30.5 & 0.49+/-0 & 7.71+/-1.61 & 3.6+/-0.6 & 13.2+/-2.0 & 1.0 & 308 \\
1769 & 2 & 53.174588 & -27.777868 & 3.28 & 26.9 & 18.88 & 19.0+/-2.7 & 42.28 & 83.5 & 23.8 & 0.43+/-0.01 & 12.49+/-3.74 & 2.5+/-0.5 & 12.1+/-2.4 & 0.99937 & 223 \\
1775 & 2 & 53.182996 & -27.78046 & 4.38 & 26.8 & 19.46 & 13.5+/-1.6 & 42.43 & 58.5 & 17.1 & 0.48+/-0.01 & 4.52+/-0.99 & 1.5+/-0.7 & 9.5+/-1.6 & 0.99998 & 249 \\
1778 & 2 & 53.149925 & -27.801733 & 4.51 & 26.5 & 19.81 & 15.7+/-2.4 & 42.53 & 46.2 & 29.1 & 0.36+/-0 & 8.76+/-2.85 & 2.6+/-0.4 & 11.8+/-2.7 & 0.99838 & 323 \\
1817 & 2 & 53.156111 & -27.789788 & 3.42 & 26.9 & 18.86 & 32.0+/-1.4 & 42.55 & 120.8 & 17.4 & 0.45+/-0.01 & 4.45+/-1.30 & 18.0+/-1.0 & 8.9+/-1.5 & 0.99892 & 256 \\
1833 & 2 & 53.15249 & -27.797708 & 3.01 & 27.0 & 18.61 & 8.5+/-1.8 & 40.77 & 27.6 & 19.7 & 0.45+/-0.02 & 3.81+/-1.45 & 0.7+/-1.4 & 6.2+/-1.5 & 0.98987 & 408 \\
1835 & 2 & 53.162661 & -27.802301 & 4.81 & 26.7 & 19.68 & 15.6+/-1.4 & 42.59 & 54.9 & 21.7 & 0.41+/-0.01 & 5.85+/-0.76 & 2.2+/-0.4 & 13.5+/-1.4 & 1.0 & 344 \\
1843 & 2 & 53.179827 & -27.800877 & 4.81 & 27.1 & 19.31 & 16.1+/-0.9 & 42.6 & 96.6 & 13.8 & 0.33+/-0 & 3.27+/-1.05 & 10.1+/-0.7 & 5.4+/-1.1 & 0.99745 & 253 \\
1864 & 2 & 53.192881 & -27.786896 & 4.39 & 27.0 & 19.27 & 10.6+/-1.6 & 42.33 & 53.9 & 17.1 & 0.47+/-0.01 & 3.56+/-1.24 & 3.2+/-1.7 & 9.3+/-1.9 & 0.99356 & 205 \\
1931 & 2 & 53.164766 & -27.761537 & 3.09 & 27.1 & 18.52 & 12.6+/-1.6 & 42.04 & 69.6 & 21.1 & 0.62+/-0.02 & 11.49+/-4.67 & 7.5+/-0.7 & 7.2+/-2.2 & 0.99 & 304 \\
1939 & 2 & 53.17625 & -27.80269 & 3.74 & 27.1 & 18.85 & 12.2+/-1.7 & 42.22 & 74.0 & 18.3 & 0.78+/-0.04 & 4.65+/-1.43 & 3.4+/-1.0 & 9.4+/-1.8 & 0.99658 & 314 \\
1950 & 2 & 53.18085 & -27.774124 & 4.47 & 26.7 & 19.55 & 14.7+/-1.8 & 42.49 & 59.3 & 30.5 & 0.55+/-0.01 & 11.32+/-3.06 & 3.0+/-0.3 & 11.5+/-1.6 & 0.99978 & 326 \\
1961 & 2 & 53.181006 & -27.78112 & 4.41 & 27.0 & 19.28 & 8.9+/-1.2 & 42.26 & 52.6 & 14.3 & 0.50+/-0.01 & 2.38+/-0.98 & 3.1+/-1.8 & 6.0+/-1.8 & 0.97267 & -- \\
1969 & 2 & 53.167393 & -27.766818 & 4.08 & 27.1 & 18.99 & 49.3+/-1.7 & 42.92 & 256.4 & 17.7 & 0.35+/-0.01 & 4.74+/-0.71 & 29.1+/-0.6 & 13.8+/-1.8 & 1.0 & 263 \\
1971 & 2 & 53.146738 & -27.787648 & 3.56 & 27.1 & 18.73 & 11.8+/-1.8 & 42.16 & 72.2 & 23.1 & 0.70+/-0.03 & 9.43+/-3.15 & 3.9+/-0.5 & 7.2+/-1.8 & 0.99718 & 216 \\
2069 & 2 & 53.151038 & -27.782866 & 5.27 & 26.7 & 19.86 & 8.0+/-1.4 & 42.39 & 27.9 & 14.4 & 0.35+/-0.01 & 8.19+/-4.57 & 4.2+/-0.5 & 6.0+/-2.9 & 0.95676 & 271 \\
2134 & 2 & 53.154089 & -27.798788 & 3.52 & 27.4 & 18.45 & 39.4+/-4.4 & 42.67 & 267.0 & 35.9 & 0.48+/-0.02 & 15.31+/-3.61 & 7.6+/-0.7 & 32.0+/-4.0 & 0.99998 & -- \\
2168 & 2 & 53.135907 & -27.798357 & 5.78 & 25.9 & 20.74 & 16.7+/-1.0 & 42.81 & 32.4 & 11.4 & 0.93+/-0.01 & 2.72+/-1.68 & 14.0+/-2.1 & 2.8+/-2.5 & 0.85677 & 247 \\
2171 & 2 & 53.152636 & -27.789346 & 4.12 & 27.4 & 18.72 & 22.3+/-1.5 & 42.59 & 161.9 & 19.0 & 0.36+/-0.01 & 4.45+/-1.18 & 15.1+/-0.9 & 8.3+/-1.4 & 0.99974 & 205 \\
2178 & 2 & 53.148771 & -27.771703 & 2.95 & 27.5 & 18.07 & 7.8+/-1.1 & 41.78 & 48.8 & 15.1 & 0.26+/-0.01 & 2.68+/-1.33 & 1.8+/-1.9 & 6.5+/-1.9 & 0.96615 & 298 \\
2198 & 2 & 53.169812 & -27.803363 & 3.45 & 27.5 & 18.31 & 14.5+/-1.8 & 42.22 & 133.9 & 15.8 & 0.63+/-0.04 & 3.07+/-0.84 & 1.1+/-1.3 & 7.9+/-1.1 & 0.99814 & 312 \\
2296 & 2 & 53.168338 & -27.804127 & 4.95 & 27.3 & 19.16 & 9.9+/-1.1 & 42.42 & 79.6 & 20.1 & 0.33+/-0.01 & 2.29+/-0.44 & 1.0+/-0.7 & 6.4+/-0.8 & 0.99999 & 221 \\
2302 & 2 & 53.180422 & -27.770581 & 5.03 & 27.8 & 18.71 & 5.1+/-0.6 & 42.15 & 443.4 & 8.3 & < 0.05 & < 0.99 & 1.0+/-0.8 & 1.5+/-2.0 & -- & 223 \\
2350 & 2 & 53.156355 & -27.809589 & 5.05 & 28.2 & 18.24 & 14.6+/-1.2 & 42.61 & 287.0 & 14.7 & 0.11+/-0.02 & < 1.09 & 6.7+/-1.4 & 5.0+/-1.3 & -- & 222 \\
2365 & 2 & 53.162773 & -27.75828 & 3.6 & 27.7 & 18.18 & 64.5+/-19.5 & 42.91 & -- & 35.6 & 0.28+/-0.01 & 4.77+/-0.81 & 8.4+/-0.8 & 12.3+/-1.8 & 1.0 & 281 \\
2370 & 2 & 53.15048 & -27.765236 & 3.0 & 27.7 & 17.85 & 9.9+/-1.7 & 41.9 & 140.6 & 18.1 & 0.46+/-0.03 & 2.35+/-0.83 & 1.1+/-2.4 & 7.9+/-2.2 & 0.98896 & 354 \\
2426 & 2 & 53.168213 & -27.761765 & 3.8 & 27.8 & 18.21 & 10.2+/-1.7 & 42.16 & 131.6 & 24.0 & 0.16+/-0.02 & 4.37+/-1.59 & 3.7+/-0.8 & 6.4+/-1.8 & 0.99588 & 275 \\
2460 & 2 & 53.182485 & -27.794075 & 3.07 & 27.8 & 17.79 & 8.3+/-0.8 & 41.85 & 74.7 & 13.3 & 0.21+/-0.02 & 2.17+/-2.69 & 5.4+/-1.0 & 1.7+/-2.1 & 0.7678 & 219 \\
2495 & 2 & 53.184432 & -27.784364 & 3.0 & -- & -- & 30.0+/-2.2 & 42.38 & -- & 21.3 & 0.38+/-0.05 & 3.18+/-0.71 & 8.6+/-1.6 & 13.4+/-1.4 & 0.99996 & 277 \\
2515 & 2 & 53.137916 & -27.787741 & 3.47 & 27.9 & 17.94 & 10.1+/-1.0 & 42.06 & 91.0 & 14.3 & 0.57+/-0.04 & 2.41+/-0.87 & 1.9+/-2.2 & 8.1+/-1.9 & 0.98319 & 284 \\
2522 & 2 & 53.133274 & -27.78967 & 3.33 & 27.9 & 17.86 & 16.7+/-1.8 & 42.24 & 162.7 & 22.1 & 0.42+/-0.03 & 6.52+/-3.25 & 7.3+/-0.7 & 5.0+/-1.8 & 0.96959 & 313 \\
2547 & 2 & 53.148058 & -27.795 & 4.51 & 27.8 & 18.47 & 11.6+/-1.7 & 42.4 & 109.5 & 21.0 & 0.35+/-0.01 & 2.21+/-0.81 & 3.3+/-1.5 & 5.4+/-1.3 & 0.98959 & 291 \\
2582 & 2 & 53.155149 & -27.762365 & 2.93 & 28.0 & 17.55 & 10.6+/-1.7 & 41.91 & 105.0 & 16.7 & 0.30+/-0.02 & 3.61+/-1.96 & 5.2+/-1.7 & 3.9+/-1.6 & 0.95463 & 349 \\
2598 & 2 & 53.187891 & -27.788202 & 3.01 & -- & -- & 17.7+/-1.8 & 42.16 & -- & 15.0 & 0.20+/-0.03 & < 1.4 & 4.9+/-2.4 & 8.5+/-2.5 & -- & 230 \\
2687 & 2 & 53.191017 & -27.791646 & 4.84 & 28.2 & 18.23 & 8.0+/-0.9 & 42.31 & 91.7 & 13.7 & 0.27+/-0.01 & < 1.03 & 5.3+/-0.7 & 1.4+/-1.4 & -- & 211 \\
2727 & 2 & 53.162616 & -27.803606 & 4.53 & 28.1 & 18.24 & 15.9+/-0.9 & 42.54 & 192.7 & 15.5 & 0.28+/-0.01 & 7.47+/-3.27 & 10.3+/-0.3 & 5.1+/-1.6 & 0.98597 & 247 \\
2749 & 2 & 53.184182 & -27.797324 & 3.73 & 28.1 & 17.82 & 14.9+/-1.7 & 42.31 & 168.4 & 22.7 & 0.28+/-0.02 & 1.87+/-1.26 & 8.9+/-1.6 & 3.4+/-1.6 & 0.89709 & 227 \\
2755 & 2 & 53.160811 & -27.804489 & 4.5 & 28.1 & 18.17 & 6.9+/-1.0 & 42.17 & 77.0 & 16.9 & 0.56+/-0.02 & 6.21+/-1.89 & 1.8+/-0.6 & 6.2+/-1.8 & 0.99862 & 177 \\
2821 & 2 & 53.150033 & -27.776318 & 3.17 & 28.2 & 17.46 & 11.9+/-1.9 & 42.04 & 168.4 & 24.0 & 0.32+/-0.02 & 4.92+/-2.89 & 6.4+/-0.9 & 4.9+/-2.3 & 0.94458 & 275 \\
2848 & 2 & 53.178279 & -27.776469 & 3.77 & 28.2 & 17.74 & 4.8+/-0.8 & 41.82 & 58.7 & 10.9 & 0.42+/-0.03 & 2.42+/-1.77 & 1.7+/-1.2 & 2.5+/-1.4 & 0.87106 & 248 \\
2861 & 2 & 53.185756 & -27.783405 & 4.54 & 28.0 & 18.28 & 4.9+/-0.8 & 42.03 & 40.6 & 15.5 & 0.42+/-0.01 & 2.84+/-0.95 & 0.7+/-0.8 & 4.0+/-1.1 & 0.99457 & 312 \\
2974 & 2 & 53.182002 & -27.800434 & 3.26 & 28.4 & 17.36 & 8.8+/-1.6 & 41.94 & 102.0 & 19.2 & 0.29+/-0.04 & 4.19+/-1.45 & 1.3+/-0.8 & 5.3+/-1.1 & 0.99638 & 294 \\
2975 & 2 & 53.156412 & -27.806404 & 3.54 & 28.4 & 17.5 & 5.7+/-0.9 & 41.84 & -- & 14.2 & 0.26+/-0.03 & 4.48+/-2.37 & 2.4+/-0.7 & 3.0+/-1.2 & 0.96252 & 177 \\
3089 & 2 & 53.162774 & -27.760765 & 5.92 & 27.0 & 19.75 & 29.0+/-3.7 & 43.07 & 115.5 & 24.2 & 0.27+/-0.01 & 5.02+/-1.35 & 8.2+/-0.6 & 9.9+/-2.0 & 0.99979 & 217 \\
3138 & 2 & 53.135192 & -27.795094 & 5.08 & -- & -- & 7.4+/-0.9 & 42.32 & -- & 13.4 & < 0.17 & 1.00+/-0.29 & 2.0+/-1.3 & 4.2+/-1.2 & 0.9982 & 184 \\
3203 & 2 & 53.176528 & -27.771036 & 5.89 & 27.5 & 19.25 & 14.9+/-1.9 & 42.78 & 106.0 & 27.8 & 0.49+/-0.02 & 11.86+/-4.05 & 9.7+/-0.3 & 11.6+/-2.1 & 0.99751 & 242 \\
3234 & 2 & 53.133439 & -27.787097 & 4.14 & 28.7 & 17.41 & 5.9+/-0.9 & 42.01 & 98.4 & 11.9 & 0.28+/-0.03 & 2.27+/-1.81 & 2.6+/-1.4 & 2.6+/-1.9 & 0.86434 & 197 \\
3242 & 2 & 53.156797 & -27.769165 & 4.81 & 28.6 & 17.81 & 5.4+/-0.8 & 42.13 & 99.7 & 13.8 & < 0.05 & < 1.05 & 0.7+/-1.4 & 3.4+/-1.6 & -- & 188 \\
3281 & 2 & 53.142058 & -27.77978 & 5.93 & 25.9 & 20.81 & 18.4+/-3.1 & 42.87 & 42.5 & 21.8 & 0.34+/-0 & 5.04+/-0.79 & 2.2+/-0.5 & 13.6+/-1.9 & 1.0 & 404 \\
3286 & 2 & 53.160157 & -27.810876 & 3.71 & 28.6 & 17.32 & 4.4+/-0.8 & 41.78 & -- & 11.0 & 0.22+/-0.03 & < 1.33 & 0.8+/-1.2 & 2.6+/-1.8 & -- & 261 \\
3329 & 2 & 53.143178 & -27.796885 & 3.25 & 28.7 & 17.06 & 11.7+/-2.4 & 42.06 & 464.9 & 26.9 & < 0.03 & 4.42+/-1.72 & 3.0+/-1.1 & 5.7+/-1.9 & 0.99465 & 256 \\
3343 & 2 & 53.172141 & -27.806734 & 4.4 & 28.1 & 18.1 & 8.2+/-1.3 & 42.22 & 240.3 & 21.2 & 0.30+/-0.02 & 4.44+/-1.14 & 0.5+/-0.5 & 6.8+/-1.0 & 0.99986 & 250 \\
3464 & 2 & 53.138159 & -27.794645 & 3.56 & 28.7 & 17.13 & 6.1+/-0.9 & 41.87 & 80.4 & 12.7 & 0.24+/-0.03 & < 1.22 & 2.0+/-1.7 & 3.1+/-2.1 & -- & 341 \\
3475 & 2 & 53.153859 & -27.791684 & 3.15 & 28.8 & 16.9 & 27.5+/-2.5 & 42.4 & 603.5 & 21.0 & 0.29+/-0.02 & 4.76+/-1.01 & 9.0+/-0.9 & 11.3+/-2.0 & 1.0 & 313 \\
3520 & 2 & 53.178226 & -27.777924 & 3.38 & 28.8 & 17.0 & 7.6+/-0.9 & 41.92 & 146.0 & 11.4 & 0.16+/-0.03 & 1.77+/-1.55 & 1.3+/-0.9 & 1.8+/-0.9 & 0.85009 & 215 \\
3634 & 2 & 53.145494 & -27.788994 & 4.15 & 28.7 & 17.41 & 11.5+/-1.8 & 42.31 & 330.7 & 24.5 & 0.44+/-0.04 & 4.55+/-1.93 & 5.1+/-0.8 & 3.4+/-1.2 & 0.98322 & 218 \\
3699 & 2 & 53.137205 & -27.787321 & 4.4 & 28.9 & 17.35 & 9.1+/-1.3 & 42.26 & 241.0 & 17.1 & 0.21+/-0.02 & < 1.07 & 1.2+/-1.1 & 4.8+/-0.9 & -- & 129 \\
3707 & 2 & 53.152402 & -27.803915 & 3.18 & 28.9 & 16.77 & 15.2+/-2.1 & 42.15 & 235.4 & 19.4 & 0.20+/-0.03 & 4.45+/-1.53 & 3.2+/-0.9 & 9.3+/-1.8 & 0.99722 & 356 \\
3723 & 2 & 53.192662 & -27.790666 & 4.78 & 28.6 & 17.77 & 5.9+/-0.9 & 42.16 & 58.1 & 15.1 & < 0.05 & 4.77+/-1.76 & 1.9+/-0.4 & 4.8+/-1.2 & 0.99641 & 209 \\
4045 & 2 & 53.150585 & -27.801962 & 3.5 & 29.1 & 16.72 & 21.1+/-2.2 & 42.39 & -- & 21.7 & < 0.03 & 1.44+/-0.34 & 3.4+/-1.9 & 6.5+/-1.7 & 0.99998 & 198 \\
4231 & 2 & 53.172862 & -27.779775 & 3.47 & 29.2 & 16.59 & 13.1+/-1.3 & 42.18 & 388.1 & 14.3 & 0.39+/-0.09 & 2.52+/-1.07 & 4.1+/-1.6 & 5.3+/-1.6 & 0.97642 & 350 \\
4381 & 2 & 53.160523 & -27.768769 & 4.03 & 29.2 & 16.93 & 4.5+/-0.7 & 41.87 & 210.3 & 10.6 & 0.16+/-0.03 & 1.82+/-1.36 & 1.9+/-1.2 & 2.2+/-1.2 & 0.88896 & 264 \\
4430 & 2 & 53.1337 & -27.782743 & 3.92 & 29.4 & 16.69 & 5.2+/-0.7 & 41.9 & -- & 9.3 & 0.14+/-0.03 & < 1.15 & 3.7+/-1.0 & 2.0+/-2.2 & -- & 259 \\
4755 & 2 & 53.161439 & -27.798449 & 3.17 & 29.5 & 16.13 & 6.8+/-0.9 & 41.8 & 349.2 & 17.8 & 0.38+/-0.12 & 6.80+/-2.13 & 3.1+/-0.4 & 4.1+/-1.4 & 0.99872 & 236 \\
5742 & 2 & 53.168894 & -27.796529 & 4.71 & 30.0 & 16.3 & 2.6+/-0.5 & 41.78 & 157.9 & 7.3 & < 0.05 & < 1.06 & 0.2+/-0.7 & 1.8+/-1.3 & -- & 157 \\
6102 & 2 & 53.170131 & -27.80879 & 6.04 & 29.0 & 17.74 & 5.7+/-1.0 & 42.38 & 175.2 & 13.4 & < 0.04 & 2.88+/-1.10 & 0.7+/-0.7 & 4.3+/-1.0 & 0.99518 & 233 \\
6291 & 2 & 53.154597 & -27.770516 & 3.3 & 29.7 & 16.04 & 3.5+/-0.6 & 41.55 & -- & 11.5 & < 0.03 & 2.63+/-2.11 & 2.0+/-1.0 & 1.9+/-1.3 & 0.89106 & 152 \\
6292 & 1 & 53.169613 & -27.772612 & 3.33 & 27.8 & 17.94 & 7.1+/-0.8 & 41.87 & 84.1 & 22.1 & 0.48+/-0.03 & 3.33+/-0.58 & 0.3+/-0.6 & 6.4+/-0.7 & 1.0 & 353 \\
6296 & 1 & 53.160556 & -27.771091 & 3.73 & 28.3 & 17.62 & 6.1+/-0.7 & 41.92 & 84.3 & 16.8 & 0.39+/-0.05 & 1.63+/-0.79 & 1.2+/-1.4 & 3.8+/-1.2 & 0.9407 & 266 \\
6297 & 1 & 53.159262 & -27.784966 & 3.7 & 28.9 & 17.03 & 12.7+/-1.3 & 42.07 & 374.0 & 24.2 & 0.14+/-0.03 & 4.65+/-0.95 & 2.3+/-0.5 & 8.6+/-1.0 & 1.0 & -- \\
6298 & 1 & 53.169249 & -27.781255 & 3.14 & 28.0 & 17.68 & 11.0+/-1.0 & 42.0 & 96.1 & 22.6 & 0.34+/-0.02 & 7.00+/-1.98 & 5.7+/-0.3 & 5.4+/-1.1 & 0.99962 & 413 \\
6299 & 1 & 53.152594 & -27.775777 & 4.05 & 28.4 & 17.73 & 5.2+/-0.8 & 41.94 & 167.8 & 22.0 & 0.72+/-0.02 & 3.68+/-2.13 & 2.4+/-0.5 & 2.0+/-0.7 & 0.91794 & 251 \\
6304 & 1 & 53.15213 & -27.779299 & 3.23 & 28.8 & 16.89 & 6.7+/-1.0 & 41.81 & 152.9 & 19.3 & 0.27+/-0.04 & 4.99+/-3.98 & 4.6+/-0.5 & 1.7+/-1.5 & 0.88224 & 261 \\
6306 & 1 & 53.153037 & -27.786705 & 4.23 & 29.7 & 16.44 & 9.3+/-1.2 & 42.23 & 970.0 & 24.3 & < 0.21 & 16.56+/-5.61 & 2.3+/-0.3 & 9.4+/-2.0 & 0.99823 & 309 \\
6309 & 1 & 53.170063 & -27.775037 & 5.89 & 28.6 & 18.17 & 4.2+/-0.6 & 42.23 & 155.1 & 8.9 & 0.56+/-0.09 & 1.00+/-2.29 & 1.5+/-0.6 & 1.2+/-0.8 & 0.57732 & 148 \\
6316 & 1 & 53.170329 & -27.778353 & 4.45 & 28.9 & 17.32 & 3.2+/-0.4 & 41.62 & 562.2 & 10.2 & < 0.2 & 6.30+/-2.52 & 1.6+/-0.3 & 3.1+/-0.9 & 0.99229 & -- \\
6317 & 1 & 53.167677 & -27.777434 & 5.4 & 29.8 & 16.76 & 3.2+/-0.5 & 42.03 & 742.9 & 10.5 & < 0.17 & 3.29+/-1.12 & 0.8+/-0.3 & 1.8+/-0.5 & 0.99741 & 232 \\
6319 & 1 & 53.165703 & -27.781258 & 4.18 & 28.2 & 17.95 & 1.9+/-0.3 & 41.53 & 53.0 & 10.5 & < 0.21 & < 1.15 & 1.6+/-0.4 & 0.5+/-0.9 & -- & 220 \\
6324 & 1 & 53.162423 & -27.782167 & 5.13 & 25.6 & 20.91 & 2.6+/-0.3 & 41.87 & 4.0 & 10.8 & < 0.17 & 2.58+/-1.33 & 1.0+/-0.5 & 1.4+/-0.6 & 0.96492 & 238 \\
6341 & 1 & 53.15254 & -27.783798 & 4.78 & 29.5 & 16.93 & 2.6+/-0.3 & 41.81 & 617.0 & 9.9 & < 0.19 & < 1.04 & 0.8+/-0.7 & 1.7+/-1.0 & -- & 192 \\
6375 & 2 & 53.143203 & -27.787981 & 3.42 & 30.4 & 15.36 & 32.8+/-3.0 & 42.56 & -- & 28.0 & 1.48+/-0.28 & 6.25+/-1.14 & 1.6+/-1.6 & 22.7+/-2.6 & 0.99999 & 512 \\
6416 & 2 & 53.157596 & -27.79784 & 4.23 & 26.2 & 19.95 & 41.0+/-3.4 & 42.88 & -- & 21.6 & < 0.22 & 2.14+/-0.31 & 10.4+/-1.2 & 14.8+/-1.4 & 1.0 & 267 \\
6462 & 2 & 53.163975 & -27.799572 & 5.45 & 25.5 & 21.1 & 28.6+/-2.7 & 42.98 & 92.0 & 29.6 & < 0.18 & 4.29+/-0.57 & 4.3+/-0.9 & 25.6+/-2.3 & 1.0 & -- \\
6498 & 2 & 53.19483 & -27.784605 & 4.54 & 29.9 & 16.36 & 5.2+/-0.5 & 42.05 & -- & 8.8 & < 0.2 & < 1.09 & 2.3+/-0.9 & 1.5+/-1.5 & -- & 223 \\
6506 & 2 & 53.147566 & -27.80398 & 3.8 & 28.5 & 17.48 & 17.5+/-3.1 & 42.4 & -- & 34.2 & 0.56+/-0.06 & 10.45+/-4.50 & 7.1+/-0.5 & 7.4+/-2.3 & 0.98604 & 311 \\
6512 & 2 & 53.152062 & -27.792605 & 4.15 & 28.1 & 18.01 & 9.0+/-1.4 & 42.2 & 190.2 & 16.1 & 0.85+/-0.02 & 9.03+/-4.05 & 4.3+/-0.7 & 7.5+/-2.7 & 0.97823 & 245 \\
6518 & 2 & 53.143246 & -27.786831 & 3.75 & 25.4 & 20.61 & 9.9+/-1.2 & 42.14 & -- & 15.3 & < 0.22 & 2.88+/-0.67 & 2.7+/-1.0 & 6.8+/-1.3 & 0.99997 & 308 \\
6521 & 2 & 53.173838 & -27.800337 & 3.07 & 30.2 & 15.45 & 21.4+/-2.2 & 42.26 & -- & 29.0 & 1.22+/-0.24 & 11.38+/-2.78 & 2.7+/-0.9 & 22.0+/-3.2 & 0.99987 & 397 \\
6576 & 2 & 53.186435 & -27.794264 & 3.7 & 29.0 & 16.97 & 5.7+/-0.9 & 41.88 & 120.1 & 11.0 & 0.21+/-0.06 & 2.91+/-2.16 & 3.0+/-1.1 & 2.6+/-1.9 & 0.89425 & 328 \\
6581 & 2 & 53.151712 & -27.802763 & 3.6 & -- & -- & 14.3+/-1.6 & 42.26 & -- & 18.5 & 0.13+/-0.04 & 3.85+/-0.96 & 2.0+/-1.4 & 14.0+/-2.2 & 0.99994 & 279 \\
6658 & 1 & 53.15499 & -27.776767 & 3.47 & 27.8 & 18.08 & 9.2+/-1.2 & 42.03 & 92.1 & 20.3 & 1.05+/-0.09 & 3.57+/-1.10 & 1.0+/-1.0 & 4.1+/-1.1 & 0.98892 & 275 \\
6680 & 1 & 53.165731 & -27.771716 & 4.5 & 27.4 & 18.86 & 23.6+/-0.9 & 42.7 & 139.8 & 31.8 & 0.70+/-0.02 & 10.60+/-1.87 & 16.7+/-0.2 & 5.9+/-0.8 & 1.0 & 195 \\
6682 & 1 & 53.169899 & -27.776872 & 5.05 & 27.4 & 19.04 & 2.6+/-0.5 & 41.87 & 18.9 & 10.9 & 0.35+/-0.01 & 3.79+/-1.72 & 1.0+/-0.5 & 2.4+/-0.9 & 0.97707 & 186 \\
6684 & 1 & 53.16347 & -27.781441 & 4.74 & 28.0 & 18.35 & 8.5+/-1.2 & 42.31 & 91.7 & 25.8 & 0.21+/-0.01 & 6.62+/-1.03 & 1.2+/-0.3 & 8.4+/-1.0 & 1.0 & 446 \\
6694 & 1 & 53.158467 & -27.771203 & 3.66 & 30.0 & 15.92 & 1.7+/-0.4 & 41.36 & 134.6 & 11.1 & 0.16+/-0.18 & < 1.25 & 0.2+/-0.8 & 1.8+/-0.8 & -- & 220 \\
6700 & 1 & 53.168281 & -27.781056 & 3.0 & 25.7 & 19.9 & 37.9+/-1.6 & 42.49 & 38.9 & 30.7 & 0.38+/-0 & 1.80+/-0.17 & 10.3+/-1.4 & 19.6+/-1.2 & 1.0 & 317 \\
6751 & 2 & 53.16976 & -27.772501 & 3.33 & 26.9 & 18.89 & 8.4+/-1.1 & 41.94 & -- & 19.1 & 0.41+/-0.01 & 4.34+/-1.31 & 0.3+/-0.9 & 9.0+/-1.5 & 0.99867 & 286 \\
6883 & 2 & 53.176297 & -27.778913 & 3.19 & 25.2 & 20.44 & 24.6+/-3.2 & 42.36 & -- & 33.3 & 0.42+/-0 & 7.57+/-1.53 & 0.4+/-0.8 & 16.9+/-2.6 & 1.0 & 374 \\
6895 & 2 & 53.175969 & -27.792613 & 3.71 & 26.6 & 19.33 & 17.2+/-2.1 & 42.37 & -- & 24.2 & 0.71+/-0.02 & 6.39+/-1.46 & 1.0+/-1.0 & 15.9+/-2.3 & 0.99995 & 387 \\
6905 & 2 & 53.143035 & -27.799874 & 3.1 & 26.2 & 19.45 & 38.9+/-3.4 & 42.53 & -- & 30.4 & 0.77+/-0.02 & 6.27+/-1.91 & 18.6+/-1.0 & 13.6+/-2.5 & 0.99799 & 432 \\
6911 & 2 & 53.165732 & -27.81646 & 4.62 & 27.4 & 18.96 & 21.2+/-2.4 & 42.68 & 142.4 & 20.7 & 0.37+/-0.01 & 9.98+/-1.49 & 7.8+/-0.4 & 19.9+/-2.3 & 1.0 & 183 \\
6915 & 2 & 53.153093 & -27.807049 & 3.7 & 29.1 & 16.83 & 4.6+/-0.6 & 41.79 & 183.5 & 11.0 & 0.15+/-0.04 & < 1.23 & 0.7+/-1.3 & 3.4+/-1.3 & -- & 212 \\
7001 & 2 & 53.138579 & -27.790217 & 5.48 & 25.6 & 21.02 & 61.2+/-4.0 & 43.32 & 95.1 & 24.0 & 1.33+/-0.01 & 8.59+/-2.64 & 44.0+/-1.2 & 18.1+/-3.7 & 0.99702 & 290 \\
7047 & 2 & 53.154992 & -27.806988 & 4.23 & 26.9 & 19.23 & 22.5+/-1.5 & 42.62 & 96.2 & 17.4 & 0.51+/-0.01 & 4.67+/-0.99 & 12.5+/-0.8 & 11.7+/-1.5 & 0.99999 & 256 \\
7073 & 2 & 53.162333 & -27.793005 & 3.6 & 26.6 & 19.27 & 7.7+/-1.2 & 41.98 & 24.7 & 18.6 & 0.71+/-0.03 & 3.56+/-1.37 & 0.8+/-1.2 & 5.4+/-1.6 & 0.98143 & 480 \\
7085 & 2 & 53.13324 & -27.786724 & 2.95 & 26.7 & 18.82 & 22.2+/-1.7 & 42.24 & 67.3 & 16.6 & 0.49+/-0.01 & 2.98+/-1.06 & 10.4+/-2.3 & 8.4+/-2.3 & 0.99051 & 378 \\
7087 & 2 & 53.157569 & -27.764705 & 3.56 & 26.7 & 19.15 & 44.1+/-1.7 & 42.73 & -- & 18.6 & 0.46+/-0.01 & 2.59+/-0.24 & 10.4+/-1.8 & 31.3+/-2.0 & 1.0 & 348 \\
7091 & 2 & 53.173305 & -27.783819 & 4.41 & 26.4 & 19.79 & 13.7+/-1.4 & 42.44 & 60.0 & 22.5 & 0.57+/-0.01 & 3.55+/-0.65 & 0.7+/-0.7 & 13.6+/-1.3 & 1.0 & -- \\
7119 & 2 & 53.132734 & -27.795038 & 3.18 & 27.1 & 18.62 & 8.1+/-1.6 & 41.88 & 42.6 & 17.8 & 0.39+/-0.01 & 3.77+/-1.24 & 0.5+/-1.9 & 11.2+/-2.8 & 0.99674 & -- \\
7121 & 2 & 53.185076 & -27.783946 & 3.68 & 27.1 & 18.86 & 19.3+/-1.6 & 42.41 & 82.2 & 19.9 & 0.43+/-0.02 & 1.90+/-0.50 & 6.6+/-2.6 & 10.8+/-2.4 & 0.99853 & 293 \\
7131 & 2 & 53.135124 & -27.792277 & 3.1 & 27.2 & 18.47 & 10.6+/-1.3 & 41.97 & 53.7 & 13.3 & 0.47+/-0.02 & 9.20+/-4.88 & 6.5+/-0.7 & 6.4+/-2.3 & 0.9633 & 319 \\
7133 & 2 & 53.14691 & -27.790191 & 3.37 & 27.2 & 18.61 & 26.3+/-2.0 & 42.45 & 129.1 & 20.5 & 0.51+/-0.04 & 3.60+/-0.41 & 1.3+/-1.1 & 19.6+/-1.5 & 1.0 & 388 \\
7143 & 2 & 53.179101 & -27.786973 & 3.68 & 28.5 & 17.47 & 17.1+/-2.4 & 42.35 & 267.2 & 18.4 & 0.61+/-0.09 & 5.71+/-2.95 & 3.5+/-0.8 & 5.2+/-1.8 & 0.95819 & 254 \\
7157 & 2 & 53.158308 & -27.807171 & 4.23 & 27.4 & 18.8 & 15.5+/-1.6 & 42.46 & 104.7 & 21.6 & 0.23+/-0.01 & 2.97+/-0.69 & 6.2+/-1.3 & 10.6+/-1.4 & 0.99997 & 205 \\
7159 & 2 & 53.167929 & -27.812327 & 3.0 & 27.9 & 17.63 & 38.1+/-2.4 & 42.49 & 322.5 & 19.7 & 0.29+/-0.02 & 3.80+/-0.80 & 15.7+/-1.0 & 9.9+/-1.4 & 0.99999 & 240 \\
7161 & 2 & 53.135226 & -27.797719 & 4.24 & 27.2 & 18.93 & 12.8+/-2.2 & 42.37 & 106.3 & 20.2 & 0.61+/-0.01 & 6.40+/-1.96 & 2.5+/-0.7 & 11.4+/-2.4 & 0.99845 & 359 \\
7163 & 2 & 53.176588 & -27.768609 & 2.99 & 27.5 & 18.08 & 18.0+/-2.4 & 42.16 & 82.5 & 22.9 & 0.36+/-0.02 & 4.23+/-2.07 & 9.1+/-1.8 & 9.3+/-2.6 & 0.96919 & -- \\
7167 & 2 & 53.174189 & -27.789924 & 3.71 & 28.0 & 17.95 & 15.7+/-2.3 & 42.33 & 142.2 & 24.2 & 0.41+/-0.04 & 4.10+/-1.34 & 4.1+/-1.2 & 11.1+/-2.3 & 0.99707 & -- \\
7177 & 2 & 53.146248 & -27.800812 & 4.85 & 27.4 & 19.02 & 8.0+/-1.0 & 42.31 & 66.8 & 12.4 & 0.27+/-0.01 & 4.49+/-1.23 & 1.3+/-0.6 & 8.1+/-1.7 & 0.9997 & -- \\
7179 & 2 & 53.17318 & -27.7678 & 3.5 & 27.9 & 17.96 & 36.8+/-4.5 & 42.64 & 398.2 & 36.0 & 0.58+/-0.05 & 18.73+/-7.49 & 6.0+/-0.5 & 18.2+/-2.9 & 0.99233 & 245 \\
7183 & 2 & 53.146084 & -27.785423 & 4.88 & 27.6 & 18.85 & 9.4+/-0.8 & 42.39 & 67.3 & 15.0 & 0.41+/-0.01 & 3.25+/-1.29 & 5.6+/-0.9 & 4.4+/-1.2 & 0.98578 & 327 \\
7185 & 2 & 53.152803 & -27.79541 & 3.97 & 27.7 & 18.32 & 25.2+/-1.7 & 42.6 & 202.0 & 20.7 & 0.41+/-0.02 & 5.39+/-1.56 & 14.5+/-0.6 & 8.8+/-1.8 & 0.99928 & 217 \\
7189 & 2 & 53.150236 & -27.766438 & 4.14 & 27.6 & 18.54 & 5.4+/-1.0 & 41.98 & 57.1 & 13.3 & 0.41+/-0.02 & 2.35+/-2.13 & 2.1+/-1.1 & 2.4+/-1.3 & 0.81946 & 190 \\
7193 & 2 & 53.135782 & -27.795472 & 3.09 & 27.8 & 17.84 & 14.4+/-2.4 & 42.1 & 68.6 & 21.1 & 0.57+/-0.04 & 5.22+/-1.78 & 3.6+/-1.0 & 9.0+/-1.9 & 0.99546 & 363 \\
7195 & 2 & 53.167605 & -27.803793 & 3.56 & 27.8 & 18.09 & 10.6+/-1.5 & 42.11 & 86.9 & 17.1 & 0.26+/-0.01 & 2.81+/-1.20 & 1.3+/-0.9 & 4.3+/-0.8 & 0.98321 & 252 \\
7199 & 2 & 53.159299 & -27.760893 & 3.33 & 28.1 & 17.67 & 14.9+/-2.2 & 42.19 & 153.6 & 28.2 & 0.40+/-0.05 & 6.31+/-1.66 & 3.8+/-0.8 & 10.3+/-2.2 & 0.99981 & 235 \\
7201 & 2 & 53.151967 & -27.807977 & 4.5 & 27.6 & 18.7 & 14.7+/-1.3 & 42.49 & 80.7 & 23.7 & 0.29+/-0.01 & 1.19+/-0.18 & 4.9+/-1.2 & 6.2+/-1.1 & 1.0 & 259 \\
7215 & 2 & 53.156944 & -27.816141 & 4.86 & 28.9 & 17.5 & 8.1+/-1.3 & 42.32 & -- & 18.9 & 0.23+/-0.10 & 4.09+/-1.37 & 1.4+/-0.5 & 6.1+/-0.9 & 0.99754 & 387 \\
7223 & 2 & 53.138414 & -27.777219 & 4.94 & 28.1 & 18.33 & 11.8+/-0.9 & 42.5 & 61.0 & 16.2 & < 0.18 & 2.39+/-0.64 & 4.6+/-1.0 & 6.3+/-1.0 & 0.99972 & 276 \\
7238 & 2 & 53.146624 & -27.779476 & 3.33 & -- & -- & 9.7+/-1.5 & 42.01 & -- & 19.1 & 0.52+/-0.05 & 3.48+/-1.20 & 1.4+/-0.9 & 7.7+/-1.5 & 0.99335 & 365 \\
7243 & 2 & 53.127778 & -27.786485 & 4.45 & 28.3 & 17.92 & 6.9+/-0.9 & 42.16 & 648.4 & 12.9 & 0.22+/-0.03 & 1.66+/-0.73 & 1.7+/-0.9 & 3.0+/-0.8 & 0.97571 & 261 \\
7251 & 2 & 53.137663 & -27.782558 & 4.85 & 28.7 & 17.71 & 7.4+/-1.1 & 42.28 & 214.2 & 16.3 & 0.34+/-0.03 & 2.19+/-0.89 & 3.2+/-1.2 & 3.5+/-1.1 & 0.98062 & 206 \\
7269 & 2 & 53.156937 & -27.804686 & 4.5 & 28.4 & 17.88 & 9.5+/-1.1 & 42.31 & 165.2 & 21.0 & 0.49+/-0.02 & 4.93+/-2.26 & 5.2+/-0.4 & 3.8+/-1.0 & 0.97524 & 181 \\
7301 & 2 & 53.166439 & -27.807914 & 3.26 & 28.8 & 16.91 & 6.6+/-1.1 & 41.82 & 409.2 & 11.5 & 0.30+/-0.05 & 2.39+/-1.45 & 1.3+/-2.3 & 4.6+/-2.3 & 0.92578 & 151 \\
7325 & 2 & 53.144018 & -27.793844 & 4.34 & 29.9 & 16.28 & 4.2+/-0.8 & 41.92 & -- & 13.1 & 0.11+/-0.10 & 2.05+/-0.96 & 0.8+/-0.8 & 3.5+/-1.0 & 0.97833 & 253 \\
7359 & 2 & 53.151005 & -27.7952 & 3.67 & 29.9 & 16.08 & 19.3+/-3.0 & 42.41 & -- & 28.7 & 0.17+/-0.08 & 3.18+/-0.68 & 0.1+/-1.2 & 11.1+/-1.2 & 1.0 & 244 \\
\hline
\label{bigtab}
\end{longtable}
\end{landscape}
\end{small}
\end{center}

\end{appendix}

\end{document}